\documentclass[letterpaper]{article} 
\usepackage{aaai2026} 
\usepackage{times}
\usepackage{helvet}
\usepackage{courier}
\usepackage[hyphens]{url}
\usepackage{graphicx}
\usepackage{natbib}
\usepackage{caption}
\usepackage{algorithm}
\usepackage{algorithmic}
\usepackage{subcaption}
\usepackage{xcolor}
\usepackage{amsmath}
\usepackage{newfloat}
\usepackage{listings}
\DeclareCaptionStyle{ruled}{labelfont=normalfont,labelsep=colon,strut=off}
\floatstyle{ruled}
\newfloat{listing}{tb}{lst}{}
\floatname{listing}{Listing}
\nocopyright
\title{Generative AI Literacy Training Improves Intelligence Analysts' Discrimination of Real and AI-Generated Images}
\author{
    Negar Kamali\textsuperscript{\rm 1,4},
    Candice Rockell Gerstner\textsuperscript{\rm 2},
    Jessica Hullman\textsuperscript{\rm 1},
    Matthew Groh\textsuperscript{\rm 1,3,4}
}
\affiliations{
    \textsuperscript{\rm 1}Department of Computer Science, Northwestern University\\
    \textsuperscript{\rm 2}Research Directorate and AI Security Center, National Security Agency\\
    \textsuperscript{\rm 3}Kellogg School of Management, Northwestern University\\
    \textsuperscript{\rm 4}Ryan Institute on Complexity, Northwestern University\\
    negar.kamali@u.northwestern.edu, crgerst@uwe.nsa.gov, matthew.groh@kellogg.northwestern.edu
}

\begin{document}
\maketitle
\begin{abstract}
Across social and online platforms, people are increasingly exposed to AI-generated images. As a consequence, the task of distinguishing AI-generated from authentic images is becoming a central challenge for information ecosystems. While humans perform better than chance, accuracy falls short of many operational needs. Initial evidence shows that visually oriented training can improve deepfake detection but does not improve participants' ability to identify real images as real. Here, we investigate the efficacy of a brief training intervention for intelligence analysts employed by the United States government in 2024 to help them improve at distinguishing real images from AI-generated images. We conducted a counterbalanced within-subject randomized experiment in which we showed participants real and AI-generated images varying in pose complexity and scene context and asked them whether each image was real or AI-generated, both before and after an expert delivered a 30-minute training that pointed out patterns in seven real and 50 AI-generated images. We collected 2,544 image-level judgments from 32 intelligence analysts. We find training increased overall accuracy by 9 percentage points (95\% CI: [2.7, 15.4]) from a baseline of 72\%. Unlike prior work where the effects of training were driven by improvement on detecting deepfakes, we find the improvement is driven by a 14.2 percentage point increase in accuracy for real images (95\% CI: [0.7, 27.7]). Through a careful experimental setup that curated matched pairs of real and AI-generated images across pose complexity categories, we reveal how these trainings influence people with different levels of digital forensics and generative AI experience and identify the kind of image-based content where this training intervention appears to be most effective. Ultimately, these results provide causal evidence that a brief, structured training can improve human judgment across a diverse array of real and AI-generated images, informing organizational responses to AI-generated visual misinformation.
\end{abstract}
\section{Introduction}
Synthetic media can seamlessly move from online platforms into operational decision pipelines, where a single misclassification can disrupt critical infrastructure. For example, in December 2025, after a minor earthquake in England, an account posted to social media an AI-generated image showing the Carlisle Bridge in Lancaster looking severely damaged. In response, Network Rail halted rail services until safety inspections could be carried out, leading to 32 services being delayed. Incidents like this illustrate how visual misinformation propagates from online platforms into human decision pipelines, where misclassification can cascade through sociotechnical systems.
Throughout this paper, ``AI-generated images'' refers to photorealistic still images produced by contemporary text-to-image systems based on diffusion models such as Stable Diffusion, Midjourney, and Adobe Firefly; our study does not examine AI-generated video or audio. Across many variations of image-based deepfake detection, both humans and automated classifiers are far from perfect but better than random guessing. Leading open-source detection models achieve F1-scores between 0.41 and 0.77 on the Deepfake-Eval-2024 benchmark of in-the-wild diffusion-generated and GAN-generated imagery~\cite{chandra2025deepfake}, and their accuracy degrades under common transformations such as compression, cropping, and resizing~\citep{9879575, corvi2023intriguingpropertiessyntheticimages}. Human accuracy varies widely with scene complexity, generative model, and viewing context~\citep{nightingale2022ai, groh2024human, kamali2025, roca2025}, and professional experience with forensics or facial recognition is not on its own sufficient to confer a reliable advantage~\citep{Ramon2024, gray2025training}. Recent work has shown that targeted training on visual artifacts can improve detection of AI-generated images~\citep{chen2025training, geissler2026}. However, in Chen et al., improvements on AI-generated images come at the cost of increased false positives on real images, leaving open the question of whether training can boost overall accuracy without simply increasing skepticism toward authentic content. Existing evaluations have also relied on GAN-generated face portraits or small stimulus sets of diffusion-model images, leaving unresolved whether training effects generalize to the diverse, complex scenes characteristic of real-world synthetic media. Finally, few studies have evaluated whether such training works for professional populations whose verification decisions carry operational weight, such as intelligence analysts, journalists, or content moderators.
We address these gaps by studying intelligence analysts employed by the United States government. Intelligence analysts encounter synthetic media as part of their daily workflows, which involve building assessments that are presented to senior decision-makers~\citep{DHS2025}. Intelligence analysts, like journalists, content moderators, and trust-and-safety teams, make judgments about visual authenticity that have downstream consequences in information ecosystems. We study a professional analyst population because their verification decisions carry operational weight. This design also lets us examine how training interacts with prior forensics experience and generative AI familiarity, and whether improvements are concentrated on particular image types or distributed broadly across stimuli.
Our work addresses the following research questions:
\begin{itemize}
 \item \textbf{RQ1}: To what extent does a brief, 30-minute generative AI literacy training improve intelligence analysts' ability to distinguish real from AI-generated images as measured by overall accuracy and three subsets of the data: real images, AI-generated images, and pairs of real and AI-generated images with similar content?
 \item \textbf{RQ2}: How do the effects of training vary across intelligence analyst experience with digital forensics and generative AI?
 \item \textbf{RQ3}: How much of the effect of training is influenced by a subset of images?
\end{itemize}
To answer these questions, we combine an intervention study with a stimulus-level analysis of performance across a diverse set of images varying in pose complexity and scene context. The intervention is a 30-minute training built around documented diffusion-model artifacts~\citep{aiforensics2025, kamali2024, NYTdeepfake2025, nytimes2024deepfake}. Did it work? We measure accuracy as our main outcome. We then ask who benefited and on what kinds of images. As a robustness check, we re-run the analysis after dropping the handful of stimuli that the training deck itself happened to show. No prior study has run this kind of evaluation with intelligence analysts.
\section{Related Work}
\subsection{Synthetic Media and Operational Risk}
Synthetic media is not just a technical problem. A fabricated image can shift what people believe before any classifier has weighed in, and that gap matters most for organizations whose analysts are expected to react quickly~\citep{hancock2021social}. Two harms are now well documented: deepfakes used to defraud businesses and individuals~\citep{mink2022deepphish}, and deepfakes used in coordinated political influence operations~\citep{ruffin2024impact}. Prior work documents the growing presence of synthetic media in governmental and security contexts~\citep{bateman2022deepfakes}. Although automated detectors can screen content at scale, they remain brittle in practice. Across a wide range of approaches, detection accuracy degrades sharply under simple and common transformations such as cropping, compression, resizing, or blurring (operations that routinely occur on social media and messaging platforms)~\citep{9879575, corvi2023intriguingpropertiessyntheticimages, cozzolino2024raisingbaraigeneratedimage}. More fundamentally, detectors often fail to generalize across generative paradigms: models trained to recognize GAN artifacts frequently miss diffusion-generated images, and diffusion-specific signatures themselves are not stable under post-processing or architectural change~\citep{xi2023aigeneratedimagedetectionusing, ojha2024universalfakeimagedetectors, ricker2024detectiondiffusionmodeldeepfakes}. As generative models evolve, detection systems must continually chase shifting statistical regularities, making long-term robustness difficult to sustain~\citep{Mirsky2021, lin2024}. In practice, no detector verdict is sufficient on its own. Accuracy drops once an image has been compressed or resized in the way that social-media pipelines do by default, and a model trained on one generator's output does not necessarily catch the next. Analysts end up making the call. Visual literacy education for people who evaluate online visual content is needed alongside automated detection, given the persistent limitations of computer-vision tools~\citep{kim2025unmasking}.
\subsection{Human Detection of AI-Generated Visual Content}
Research on human detection of AI-generated media reveals substantial variability in performance across individuals and contexts. Detection accuracy depends on media modality, scene complexity, viewing conditions, and social context and not just a single stable visual cue~\citep{nightingale2022ai, Lago_2022, groh2024human, kamali2025}.
Early studies found GAN-generated faces particularly difficult to distinguish from real photographs, yet performance improves for video deepfakes, likely because viewers are sensitive to temporal inconsistencies and dynamic facial cues~\citep{groh2022deepfake}.
Social context shapes both what diffusion models generate~\citep{luccioni2024stable} and how people form beliefs about the resulting images. For instance, detection ability is influenced by shared identity between the viewer and the subject of the content~\citep{mink2024s}.
This variability raises the question of whether domain expertise can improve detection. Prior research suggests that cognitive flexibility better explains individual differences than occupational background~\citep{diel2024cognitive}, and that ability or experience alone is insufficient to confer a reliable detection advantage on state-of-the-art synthetic media~\citep{Ramon2024, gray2025training}.
\subsection{Literacy, Training, and Warning Interventions}
AI literacy refers to a broad set of competencies necessary for effectively interacting with AI systems. This involves the ability to critically evaluate AI technologies, communicate and collaborate effectively with AI systems, and use them as a tool across a wide range of contexts~\citep{long2020}. However, a recent meta-analysis of surveys for measuring AI literacy shows that these surveys are mostly based on self-reported questions, have limited evidence of validity or responsiveness, and have undergone little testing for cross-cultural validity~\citep{linter2024}. This body of work primarily treats AI literacy as a general-purpose concept and operationalizes it accordingly.
General media-literacy programs that teach broad evaluative skills such as checking sources, investigating URLs, and reading laterally across sites have shown limited effectiveness. In an evaluation of Facebook's ``Tips to Spot False News'' campaign, researchers found that while the intervention improved discernment by 26.5\% in the US, the effects did not persist over time and failed entirely in populations with low social media use~\citep{guess2020exposure}. More recently, \citet{aslett2024online} show that online searching to verify misinformation, a strategy promoted by such literacy programs, can paradoxically increase its perceived veracity. Moreover, misinformation often continues to influence thinking even after people accept a correction, and even a single prior exposure to false content increases its perceived accuracy~\citep{lewandowsky2020misinformation, pennycook2020prior}.
More targeted interventions have shown greater promise. Warning labels can reduce sharing of flagged content and correct misbeliefs, though their effectiveness depends on the source of the warning and how labels are framed~\citep{horne2025warnings, lee2025partisan, jose2025label}. These approaches, however, remain reactive because they flag content after exposure instead of equipping people to evaluate it themselves. Inoculation theory offers a preemptive alternative by exposing people to weakened manipulative techniques paired with refutational explanations, which can build resistance to future misinformation~\citep{traberg2023prebunking, hameleers2024state, king2025promoting}.
With respect to AI literacy focused on synthetic images, researchers have found evidence that a variety of trainings can improve human ability to tell real images from AI-generated. Chen et al.\ demonstrated that two different kinds of trainings, one focused on pointing out visible artifacts and the other on showing how faces are generated, can improve participants' accuracy at identifying GAN face portraits by nearly 20 percentage points but decrease participants' accuracy by 5 percentage points on real faces~\citep{chen2025training}. Notably, in Chen et al.'s stimulus set, the StyleGAN3 images contained more easily detectable artifacts than the StyleGAN2 images. These patterns highlight both the promise of perceptual training and the importance of evaluating such interventions on more subtle generative AI systems such as the diffusion models that power many of the most widely used text-to-image commercial products, including Google Nano Banana Pro and Midjourney. Geissler et al.\ examined five digital literacy interventions focused on 30 diffusion-model-generated images and found that the most effective intervention is pointing out the kinds of artifacts that emerge in diffusion-model-generated images~\cite{geissler2026}. Bray et al.\ further showed that text-only checklists of visual cues, without accompanying examples or feedback, produce no significant improvement~\citep{bray2023}. Artifact-based perceptual training on face portraits, including evaluations with crowdworkers~\citep{rehman2025} and with trained super-recognizers~\citep{gray2025training}, similarly improves fake-image detection but consistently decreases real-image accuracy. These studies demonstrated that targeted perceptual training can improve human discrimination of AI-generated images with the caveat that training often increases skepticism toward real images. Few studies have evaluated AI-image-detection training on professionals whose verification decisions carry operational weight, nor on stimulus sets that extend beyond face portraits to diverse, complex scenes. Our study addresses both gaps and demonstrates that brief, expert-led artifact training can improve real-image accuracy as opposed to only increasing skepticism toward authentic content.
\section{Methods}
\subsection{Overview}
We conducted a randomized experiment to evaluate the effect of a brief, 30-minute generative AI literacy training on intelligence analysts' ability to distinguish real from AI-generated images. The full experiment lasted 90 minutes and involved three stages. In the first 30 minutes, we showed participants 40 images each and asked them to judge whether the image was real or generated by AI (see Figure~\ref{fig:interface} for a screenshot of the interface). In the second 30 minutes, an expert in distinguishing AI-generated images from real images presented a slide deck drawn from a written guide on how to distinguish AI-generated images from real images~\cite{kamali2024}. In the final 30 minutes, we showed participants another 40 images each. We assigned images to participants based on a counterbalanced within-subject design presented in Figure~\ref{fig:counterbalance}. Based on this design, we address RQ1 and estimate the effect of training while controlling for image difficulty and order effects, examining changes in overall accuracy as well as performance on real images, AI-generated images, and matched real-fake image pairs. Next, we address RQ2 by examining how training outcomes vary across analysts' prior experience with digital forensics and generative AI tools. Finally, we examine the robustness of our results across image stimuli posed in RQ3 by analyzing whether improvements are distributed broadly across stimuli or concentrated in a subset of images.
\subsection{Image Stimuli}
\label{sec:imagestimuli}
Drawing on a dataset of 149 verified real photographs and 450 photorealistic AI-generated images used in previous research on deepfake detection~\cite{kamali2025}, we initially selected 100 images to serve as the stimulus set for this experiment. These images are balanced with respect to authenticity (50 are real, 50 are AI-generated), scene complexity (25 are portraits, 25 are full-body shots, 25 are posed group shots, and 25 are candid shots), and image-to-text descriptions (50 text descriptions describe 50 pairs of real and AI-generated images). We balanced the stimulus set along these three dimensions in order to enable a controlled experiment and bound the generalizability of the effects of the training intervention. Each participant evaluated 80 of these images (40 in each experimental phase) drawn from the deployed stimulus set of 97 images. Furthermore, the stimulus pairing enables within-condition comparisons that reduce confounding from differences in content or baseline difficulty.
\subsection{Training}
The expert presented the training remotely to intelligence analysts employed by the government. Based on a guide for distinguishing AI-generated images from authentic photographs~\citep{kamali2024}, the 30-minute training focused on helping analysts answer two questions: (1) Does an image show any evidence of generative AI? And, (2) Does an image show anything generative AI tends to be incapable of? The training consisted of 32 slides with 50 AI-generated images and 7 real images. Following the guide for distinguishing AI-generated images~\citep{kamali2024}, the training was structured around five categories of artifacts: anatomical implausibilities (e.g., hands, eyes, teeth, body structure, identity inconsistencies, celebrity overfitting), stylistic artifacts (e.g., cinematization, plastic textures, hyper-real detail), functional implausibilities (e.g., dysfunctional objects, incorrect mechanical structure, malformed text and logos), violations of physical constraints (e.g., inconsistent lighting, shadows, reflections, or perspective), and sociocultural implausibilities (e.g., historically or culturally incongruent scenarios).
For each artifact category, the training presented concise explanations paired with multiple visual examples. Examples were drawn from a mixture of real photographs and AI-generated images produced using contemporary diffusion models (Midjourney, Adobe Firefly, and Stable Diffusion), with visual callouts highlighting regions of interest.
The training materials also emphasized limits and failure modes of visual inspection. Analysts were explicitly cautioned against over-reliance on any single cue, and were encouraged to consider how artifact visibility varies with scene complexity, pose, image resolution, and viewing time. Interactive checkpoints throughout the session prompted participants to apply the framework to novel examples and reflect on sources of uncertainty in their judgments.
The impact of this training intervention on detection accuracy is evaluated using a counterbalanced within-subject design described in the following section.
\subsection{Experimental Design}
\label{sec:design}
\begin{figure}[!t]
 \centering
 \includegraphics[width=\linewidth]{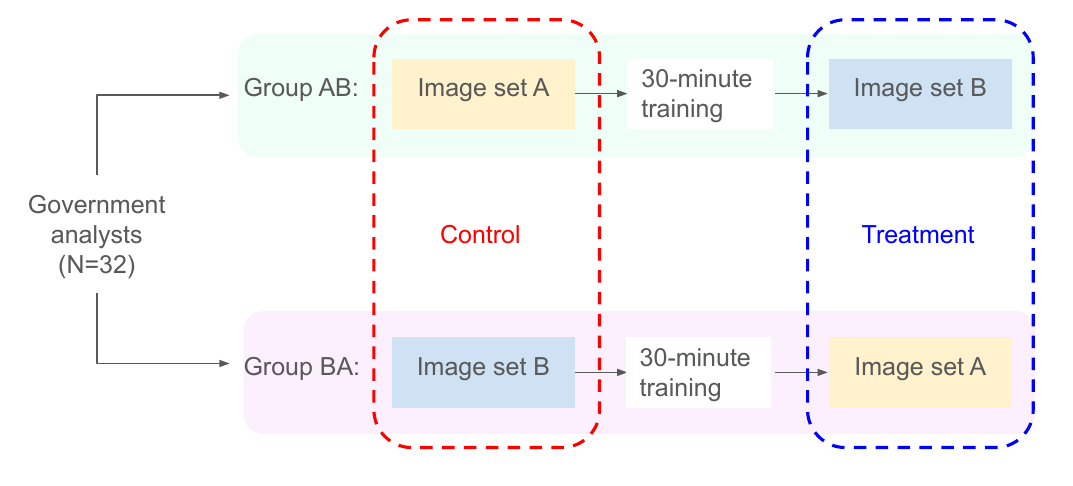}
 \caption{\textbf{Counterbalanced experimental design.} The stimulus set was divided into two halves, A and B. Participants were randomly assigned to one of two groups: Group~AB viewed Set~A before training and Set~B after training, while Group~BA viewed Set~B before training and Set~A after training. This design creates a built-in control: Set~A in the AB condition serves as the control for Set~A in the BA condition (and likewise for Set~B).}
 \label{fig:counterbalance}
\end{figure}

Figure~\ref{fig:counterbalance} illustrates our counterbalanced within-subject design. We split the stimulus set into two disjoint pools (Set A and Set B). Each participant evaluated 40 images from their assigned set in each phase.
Assignment of images to Set A versus Set B was randomized, with the constraint that each half preserved the same scene-complexity composition as the full set (25\% portrait, 25\% full body, 25\% posed group, 25\% candid group). Participants were then randomly assigned to one of two experimental sequences. In the AB condition, participants evaluated Set A \emph{before} training and Set B \emph{after} training. In the BA condition, participants evaluated Set B before training and Set A after training. Thus, every participant evaluated both sets, but the training occurred at the midpoint and the order of sets was counterbalanced across groups.
This counterbalancing creates a built-in control for the training effect. For example, performance on Set A in the AB condition provides a pre-training baseline for that set, while performance on the same Set A in the BA condition reflects post-training performance. The analogous comparison holds for Set B. By comparing accuracy on the same images when they appear pre- versus post-training (across groups), we isolate the effect of the training intervention from stimulus difficulty and idiosyncratic set effects, strengthening the causal interpretability of observed pre--post changes.
\subsection{Experimental Interface}
\begin{figure}[!t]
 \centering
 \includegraphics[width=\linewidth]{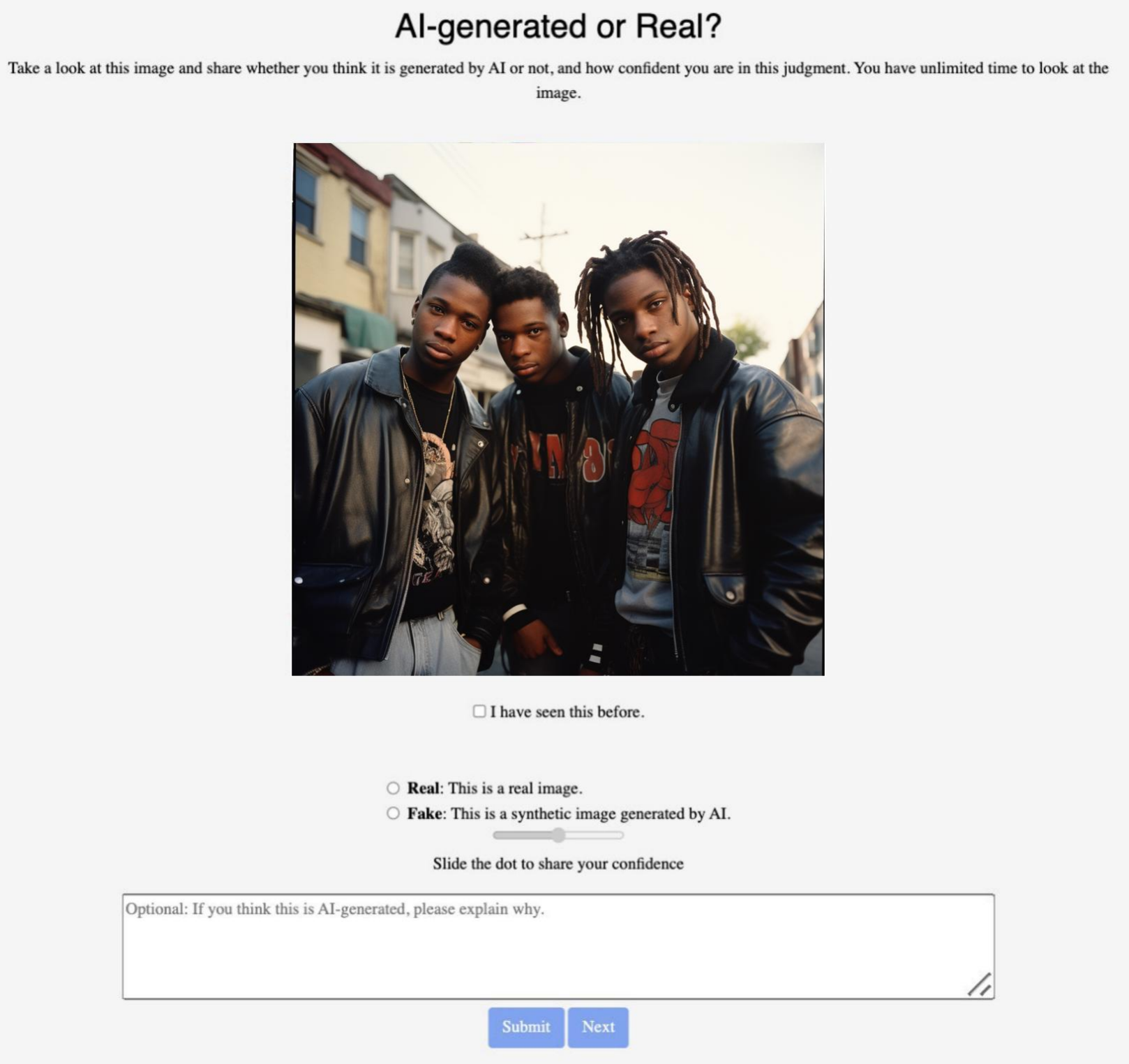}
 \caption{Experiment interface for image evaluation. Participants viewed one image at a time and judged whether it was AI-generated or real, reported confidence, and optionally provided a textual explanation.}
 \label{fig:interface}
\end{figure}

The experiment was administered through a custom web-based interface (Figure~\ref{fig:interface}). To ensure consistency in visual presentation, the study was conducted on desktop devices for all participants. On each trial, participants were shown a single image and asked to judge whether it was AI-generated or a real photograph. Participants were given unlimited time to examine each image, reflecting realistic analytic conditions.

Below the image, participants selected a binary authenticity judgment (``Real'' or ``AI-generated'') and reported their confidence on a 5-point Likert scale (1 = not at all confident, 5 = very confident). An optional free-text field allowed participants to explain their reasoning when they believed an image was AI-generated, enabling later qualitative analysis of detection strategies and cues. Participants could also indicate whether they had previously seen the image; responses marked as ``seen before'' were excluded from analysis.
To support careful inspection, the interface allowed people to zoom into the image using standard browser interactions. After submitting a response, participants advanced to the next image using a ``Next'' button.
\subsection{Participants}
We recruited 32 intelligence analysts through the North Carolina State Laboratory for Analytic Sciences (LAS). All participants were adults and completed the study as part of a professional research engagement.
Prior to the experiment, participants completed a brief pre-survey capturing background information, including professional experience and familiarity with generative AI tools.
Participants varied in their prior experience with digital forensics and AI-generated media. Approximately one third reported no prior professional experience with digital forensics tasks, one third reported less than one year of experience, and one third reported more than one year of experience. Familiarity with generative AI tools followed a similar distribution: roughly one third reported rare use, one third occasional use, and one third frequent use of generative AI systems.
Because the study was conducted in a supervised research setting with trained analysts, we did not include attention-check questions or time-based exclusions.
Field experiments with specialized professional populations are resource-intensive but offer essential insight into how trained practitioners make consequential decisions in the context of their organizations. Recent work in this tradition includes Bashardoust et al.'s prompt-engineering study with 29 professional journalists~\citep{bashardoust2024}, Senoner et al.'s explainable-AI study with 48 factory workers~\citep{senoner2024}, and Li et al.'s generative AI perceptions study with 20 UX design professionals~\citep{li2024ux}. Our study contributes a counterbalanced randomized within-subject experiment with 32 intelligence analysts, yielding 2,544 image-level judgments across the design.
\section{Pre-Registered Analysis}
We pre-registered\footnote{\url{https://aspredicted.org/rh2q-8gv5.pdf}} ordinary least squares (OLS) regression models for accuracy and pair accuracy to analyze the two aggregate-level outcomes, using individual trial responses as the dependent variable with standard errors clustered at the participant level. The model specification included three predictors: (1) assignment to BA versus AB order, (2) whether images were seen post-training, and (3) the interaction between these factors. The interaction term provides the critical test of training effectiveness by comparing performance on the same image sets before versus after training while controlling for order effects.
Formally, the model is:
\begin{equation}
\begin{split}
Y_{ij} = {}& \beta_0 + \beta_1 BA_i + \beta_2 Post_{ij} \\
&+ \beta_3 (BA_i \times Post_{ij}) + \varepsilon_{ij},
\end{split}
\end{equation}
where $Y_{ij}$ is the 0/1 accuracy of participant $i$ on image $j$, $BA_i$ indicates whether the participant was assigned to BA order, $Post_{ij}$ indicates whether the image was shown post-training, and $\varepsilon_{ij}$ is the error term clustered at the participant level.
The intercept $\beta_0$ represents the baseline accuracy for AB participants on pre-training images. The coefficient $\beta_1$ captures any baseline difference in accuracy between BA and AB groups. The coefficient $\beta_2$ estimates the average difference between pre- and post-training performance across groups. Finally, $\beta_3$ provides the critical test of training effectiveness by comparing performance on the same image sets before versus after training while controlling for order effects.
We report overall accuracy, defined as the proportion of correctly classified individual trials, and pair accuracy, a stricter measure coded as correct only if both images within a real--fake pair were correctly classified. Although participants evaluated each image individually, the stimulus set was constructed from matched real--fake pairs depicting similar scenes. Pair accuracy is computed post hoc by checking whether each participant correctly classified both members of a pair, reducing the likelihood that partial or chance-level guessing inflates measured performance. We also report a third ancillary outcome: the false positive rate on real images (mislabeling real images as fake). This tests whether training improves performance without merely making participants more skeptical of real images.
Our analysis strictly follows the pre-registered plan, but our experiment has two operational deviations from what we pre-registered. First, each participant evaluated 40 images before training and 40 after, instead of the pre-registered 50 and 50. Second, our final stimulus set contained 97 images instead of the pre-registered 100. Both deviations were applied uniformly across all participants and conditions, and preserved the counterbalanced design, stimulus composition, and analysis plan. These changes were made during deployment on a secure research server, after submitting the pre-registration but prior to the start of data collection, in response to technical difficulties.
\section{Results}
\subsection{Overall Training Effect}
In total, we collected 2,544 image-level responses. Because random guessing in this task yields 50\% accuracy, baseline performance provides an important reference point for interpreting training effects. Prior to training, analysts achieved an overall accuracy of 72\% ($\beta_0 = 0.72$), indicating performance reliably above chance but still short of operational ideals. This baseline is broadly consistent with results from large-scale public samples reported in prior work~\cite{kamali2025}, where non-expert participants achieved approximately 75\% accuracy on the larger image pool from which our stimulus set was drawn. The training intervention produced a meaningful improvement in detection performance: overall accuracy increased by 9 percentage points (95\% CI: [2.7, 15.4]), corresponding to a predicted post-training accuracy of 82\%.
This effect size represents a meaningful improvement given both the brief, 30-minute duration of the intervention and the challenging nature of the discrimination task. Results were consistent across outcome measures: pairwise accuracy increased by 11 percentage points (95\% CI: [1.8, 21.0]). Training did not induce a conservative shift toward labeling more images as fake. Instead, accuracy on real images increased by 14.2 percentage points (95\% CI: [0.7, 27.7]), corresponding to a 14.2 percentage point reduction in false-positive errors and indicating sharper discrimination capabilities. As expected, neither the order of image set presentation (BA versus AB) nor the overall difference between pre- and post-training images, independent of training, showed meaningful effects.
\begin{table*}[!t]
 \caption{Pre-registered OLS estimates of the training effect on overall, real-image, AI-generated-image, and pair accuracy. Standard errors clustered at the participant level appear in parentheses.}\label{tab:ols_results}
 \centering
 \scriptsize
 \resizebox{\linewidth}{!}{
 \begin{tabular}{lcccc}
 \hline
 & Overall (1) & Real image (2) & AI-generated image (3) & Pair (4) \\
 \hline
 Constant & 0.72$^{***}$ (0.03) & 0.67$^{***}$ (0.04) & 0.78$^{***}$ (0.04) & 0.58$^{***}$ (0.05) \\
 Assignment to BA & -0.04 (0.04) & -0.06 (0.06) & -0.01 (0.05) & -0.05 (0.06) \\
 Post-training image & 0.01 (0.02) & -0.04 (0.04) & 0.06 (0.04) & -0.00 (0.03) \\
 BA $\times$ Post & 0.09$^{**}$ (0.03) & 0.14$^{*}$ (0.07) & 0.04 (0.05) & 0.11$^{*}$ (0.05) \\
 \hline
 Participants & 32 & 32 & 32 & 32 \\
 Images & 97 & 97 & 97 & 97 \\
 Observations & 2544 & 1248 & 1296 & 2544 \\
 \textit{p}-value on BA $\times$ Post& 0.0049 & 0.0392 & 0.3966 & 0.0204 \\
 \hline
 \multicolumn{5}{l}{\tiny Note: Standard errors in parentheses. $^{*}$\textit{p}$<$0.05; $^{**}$\textit{p}$<$0.01; $^{***}$\textit{p}$<$0.001}
 \end{tabular}
 }
\end{table*}
\subsection{Robustness Check with Logistic Regression}
We replicate our pre-registered OLS analysis (Table~\ref{tab:ols_results}) with logistic regression, which is naturally suited to the binary nature of the image-level accuracy outcome ($Y_{ij} \in \{0, 1\}$). Linear regression yields unbiased estimates of average treatment effects on binary outcomes and produces directly interpretable coefficients in probability units~\citep{gomila2021}, motivating our pre-registered OLS choice; the logit specification serves as a robustness check. The model preserves the same structure as the OLS specification but links the linear predictor to the probability of a correct response through the logit function:
\begin{equation}
\begin{split}
\mathrm{logit}\!\left(P(Y_{ij}=1)\right) = {}& \beta_0 + \beta_1 BA_i + \beta_2 Post_{ij} \\
&+ \beta_3 (BA_i \times Post_{ij}),
\end{split}
\end{equation}
where $\mathrm{logit}(p) = \log(p/(1-p))$ denotes the log-odds, $Y_{ij}$ is the accuracy (0/1) of participant $i$ on image $j$, $BA_i$ indicates assignment to the BA order (versus AB), and $Post_{ij}$ indicates whether the image was shown in the post-training block. The coefficient of interest, $\beta_3$, captures the training effect through the BA $\times$ Post interaction. Standard errors are clustered at the participant level.
The logistic regression confirms the results of the OLS analysis reported in Table~\ref{tab:ols_results}. Baseline accuracy in the untrained reference condition (AB, Pre) was $0.725$, closely matching the OLS estimate. Training produced a statistically significant improvement in detection performance, increasing the log-odds of a correct response by $\beta_3 = 0.479$ (SE $= 0.170$, $p = 0.0047$), which corresponds to a marginal effect of approximately $9.5$ percentage points, similar to the $9.1$ percentage-point effect estimated by OLS (Table~\ref{tab:ols_results}, column 1). The subgroup pattern also mirrors the OLS findings (Table~\ref{tab:ols_results}, columns 2--3): the training effect is larger and statistically significant on real images ($\beta = 0.631$, $p = 0.040$), whereas the effect on AI-generated images is positive but not significant ($\beta = 0.331$, $p = 0.330$), consistent with the interpretation that training sharpened discrimination without inducing a conservative bias toward labeling images as fake.
\subsection{Signal Detection Analysis}
In the training session, we showed participants 50 AI-generated images and 7 real images. This asymmetry reflects the structure of the artifact taxonomy: the training covered five categories of artifacts with multiple examples each, and AI-generated images carried the cues the training was designed to teach. The real images shown during training contained non-generative AI artifacts (e.g., compression and focus inconsistencies) so that analysts would learn that the presence of an artifact does not by itself imply that an image is AI-generated. To address whether the asymmetric training composition could explain the post-training accuracy gain, we conducted a signal detection analysis. Signal detection theory decomposes performance into two components: discrimination sensitivity ($d'$), which captures the ability to distinguish AI-generated from real images, and response criterion ($c$), which reflects the threshold for labeling an image as AI-generated. Formally, these measures are defined as:
\begin{equation}
d' = Z(\text{Hit Rate}) - Z(\text{False Alarm Rate})
\end{equation}
\begin{equation}
c = -\frac{1}{2}\left[Z(\text{Hit Rate}) + Z(\text{False Alarm Rate})\right]
\end{equation}
where $Z(\cdot)$ denotes the inverse of the standard normal cumulative distribution function (i.e., the probit function). Higher $d'$ indicates better discrimination independent of bias; negative values of $c$ indicate a greater tendency to classify images as AI-generated, whereas positive values indicate a greater tendency to classify images as real.
The signal-detection analysis aggregates each participant's Hit Rate ($P(\text{``AI''} \mid \text{AI-generated})$) and False-Alarm Rate ($P(\text{``AI''} \mid \text{real})$) from their raw pre-training and post-training trials, averaging across the two counterbalance orderings (AB and BA) so that set-difficulty differences cancel in the aggregate. Training produced a substantial improvement in discrimination sensitivity ($\Delta d' = +0.36$, $p = 0.003$) alongside a small, non-significant criterion shift toward classifying images as AI-generated ($\Delta c = -0.12$, $p = 0.230$).
\subsection{Stimulus-Level Heterogeneity}
\begin{figure*}[t]
 \centering
 \begin{subfigure}[t]{0.32\textwidth}
 \centering
 \includegraphics[width=\linewidth]{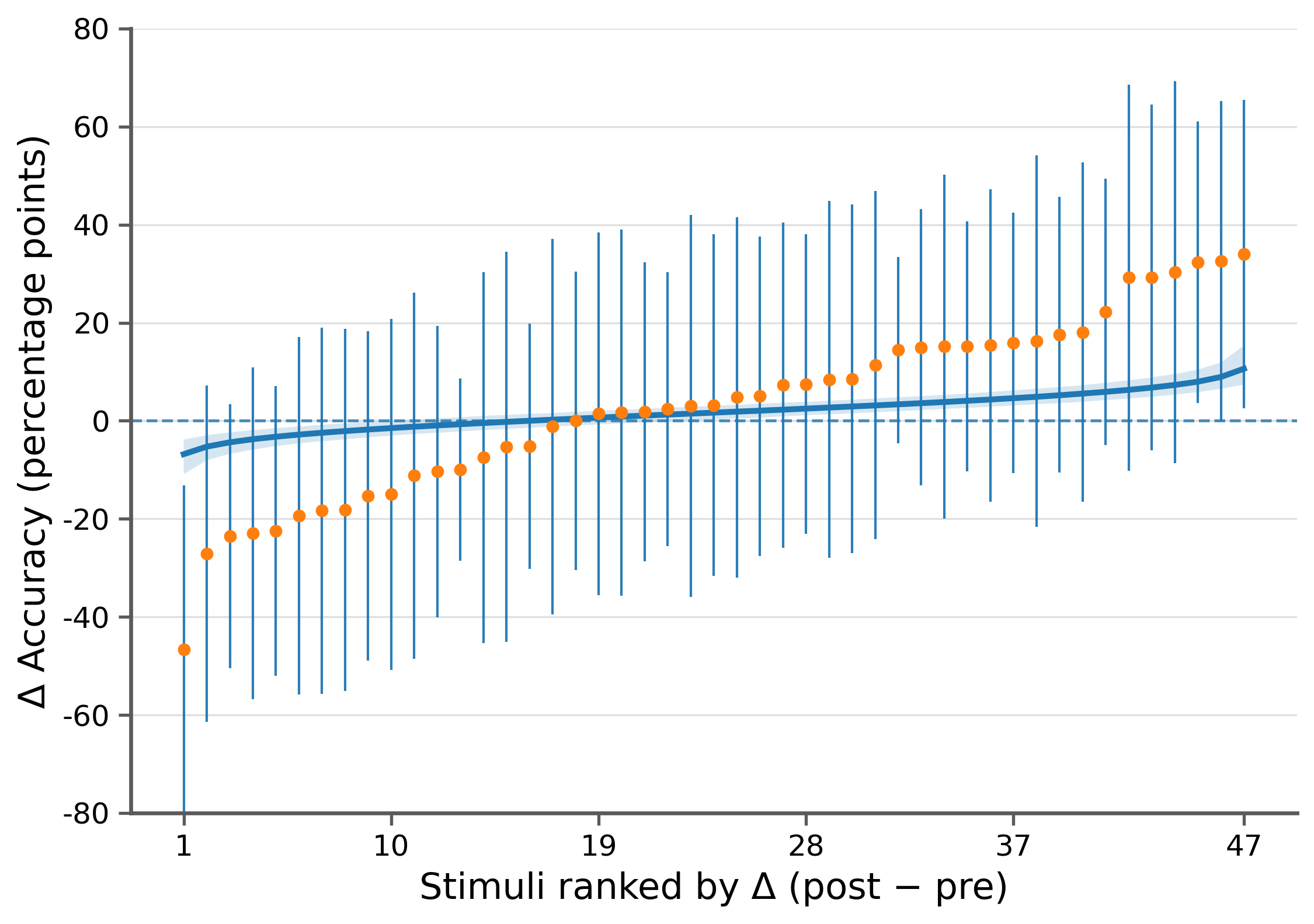}
 \caption{Real images}
 \label{fig:stimulus-real}
 \end{subfigure}\hfill
 \begin{subfigure}[t]{0.32\textwidth}
 \centering
 \includegraphics[width=\linewidth]{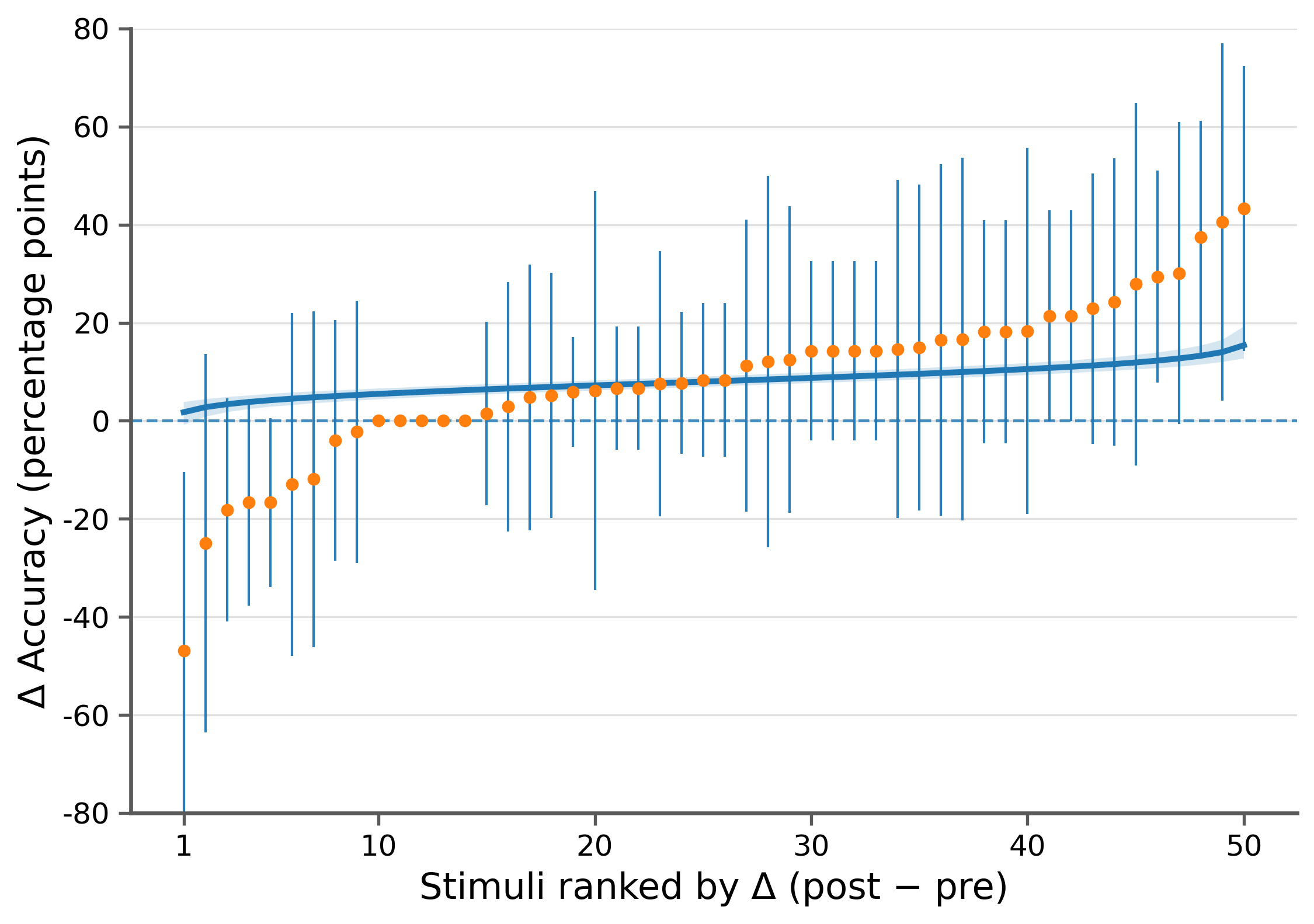}
 \caption{AI-generated images}
 \label{fig:stimulus-fake}
 \end{subfigure}\hfill
 \begin{subfigure}[t]{0.32\textwidth}
 \centering
 \includegraphics[width=\linewidth]{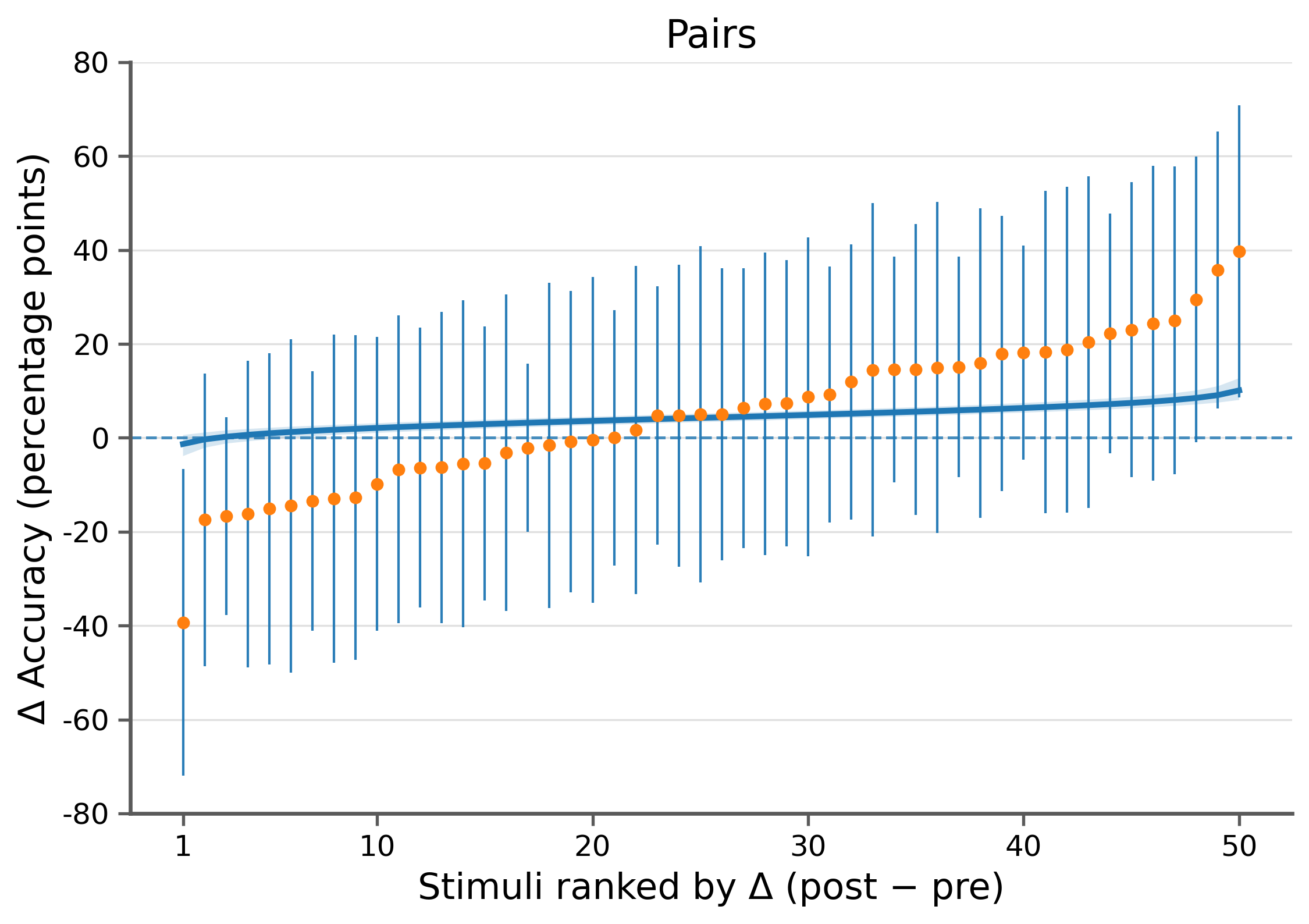}
 \caption{Real--fake pairs}
 \label{fig:stimulus-pair}
 \end{subfigure}
 \caption{Ranked stimulus- and pair-level training effects. (A) Real images, (B) AI-generated images, and (C) real--fake image pairs. Each point shows the change in accuracy from pre- to post-training ($\Delta=\mathrm{post}-\mathrm{pre}$), with stimuli/pairs ordered by $\Delta$. Vertical whiskers show 95\% confidence intervals for each point estimate. The solid curve and shaded band show the expected ranked pattern under a null model with no stimulus-specific effects (i.e., sampling-driven heterogeneity only).}
 \label{fig:stimulus-plots}
\end{figure*}

To assess whether the observed training gains reflect broad learning as opposed to improvements driven by a small number of easy or extreme images, we examined stimulus-level changes in accuracy across the full image set. We find that the training effect is distributed across images and is not concentrated in a narrow subset, which indicates that performance gains do not hinge on a few highly responsive stimuli.
To support this claim, we plot image-level changes in accuracy following the visualization approach proposed by Simonsohn et al.~\citep{simonsohn2025stimulus} (Fig.~\ref{fig:stimulus-real}, Fig.~\ref{fig:stimulus-fake}, Fig.~\ref{fig:stimulus-pair}). Each point represents a single image, ordered by its change in accuracy from pre- to post-training. Each point includes a 95\% confidence interval (vertical whiskers) that reflects uncertainty in that stimulus' estimated $\Delta$, which varies with the number of observations in the pre- and post-training blocks. The shaded band denotes the range of variation in accuracy expected for individual images under a null model in which training has no effect, accounting for sampling variability. Images falling outside this 95\% expected band exhibit changes unlikely to arise from noise alone.
Two insights emerge from these plots. First, many images lie outside the expected band, reflecting substantial heterogeneity in image difficulty and confirming that the stimulus set spans a diverse range of perceptual challenges. Second, positive changes appear throughout the ranked distribution as opposed to being confined to a small number of images; the aggregate training effect is not driven by isolated outliers. These patterns indicate that training yields robust improvements that generalize across heterogeneous stimuli.
To characterize what distinguishes the most training-responsive images from the least responsive ones, we cross-referenced stimulus-level changes with image pose category. The largest positive deltas concentrate among portrait stimuli, while full-body and posed-group images populate the lower tail of the ranked-delta distribution. The Heterogeneity by Image Pose Category subsection below interprets this pattern. Other image-level features (resolution, lighting conditions, scene complexity) likely contribute additional variation; we leave a fuller covariate analysis of stimulus-level training response to future work.
\subsection{Heterogeneity by Digital-Forensics Expertise}
The following subsections present exploratory heterogeneity analyses across digital-forensics expertise, generative AI familiarity, and image pose category. These analyses were pre-registered as secondary analyses, alongside the primary pre-registered hypothesis test of overall training effectiveness. Given that each subgroup contained roughly ten participants, we present them as exploratory rather than confirmatory and caution that interaction estimates from small subgroups can be unstable.
Before turning to these comparisons, we verified that the BA and AB sequences were balanced on prior digital-forensics experience ($\chi^2$ test $p=0.238$), self-reported familiarity with generative AI tools ($\chi^2$ test $p=0.459$), and pre-training accuracy (Welch $t$-test $p=0.429$).
We next examined whether the effect of training differed across groups defined by prior digital-forensics experience. Before training, accuracy levels differed with participants who reported forensics experience achieving an average accuracy of 62.7\% (95\% CI: [57.9, 67.5]), compared to 74.5\% (95\% CI: [68.8, 80.3]) among those with less than one year of experience and 73.9\% (95\% CI: [66.4, 80.7]) among those with more than one year of experience (Fig.~\ref{fig:beeswarm-forensics-pre}). False-positive rates on real images were highest in the no-experience group (43.9\%; 95\% CI: [21.2, 47.7]) and lower in the other groups (28.5--32.0\%).
Our training increased accuracy within all forensics-experience groups. The training effect, estimated by the BA~$\times$~Post interaction, was largest among participants with no prior forensics experience, corresponding to a 9.2 percentage point increase in accuracy (95\% CI: [3.5, 14.2]). Participants with less than one year and more than one year of experience showed smaller estimated gains of 5.4 percentage points (95\% CI: [0.5, 10.0]) and 2.0 percentage points (95\% CI: [-3.9, 8.3]), respectively (Fig.~\ref{fig:beeswarm-forensics-delta}).
\subsection{Heterogeneity by Generative AI Familiarity}
\begin{figure*}[!t]
 \centering
 \begin{subfigure}[t]{0.23\linewidth}
 \centering
 \includegraphics[width=\linewidth]{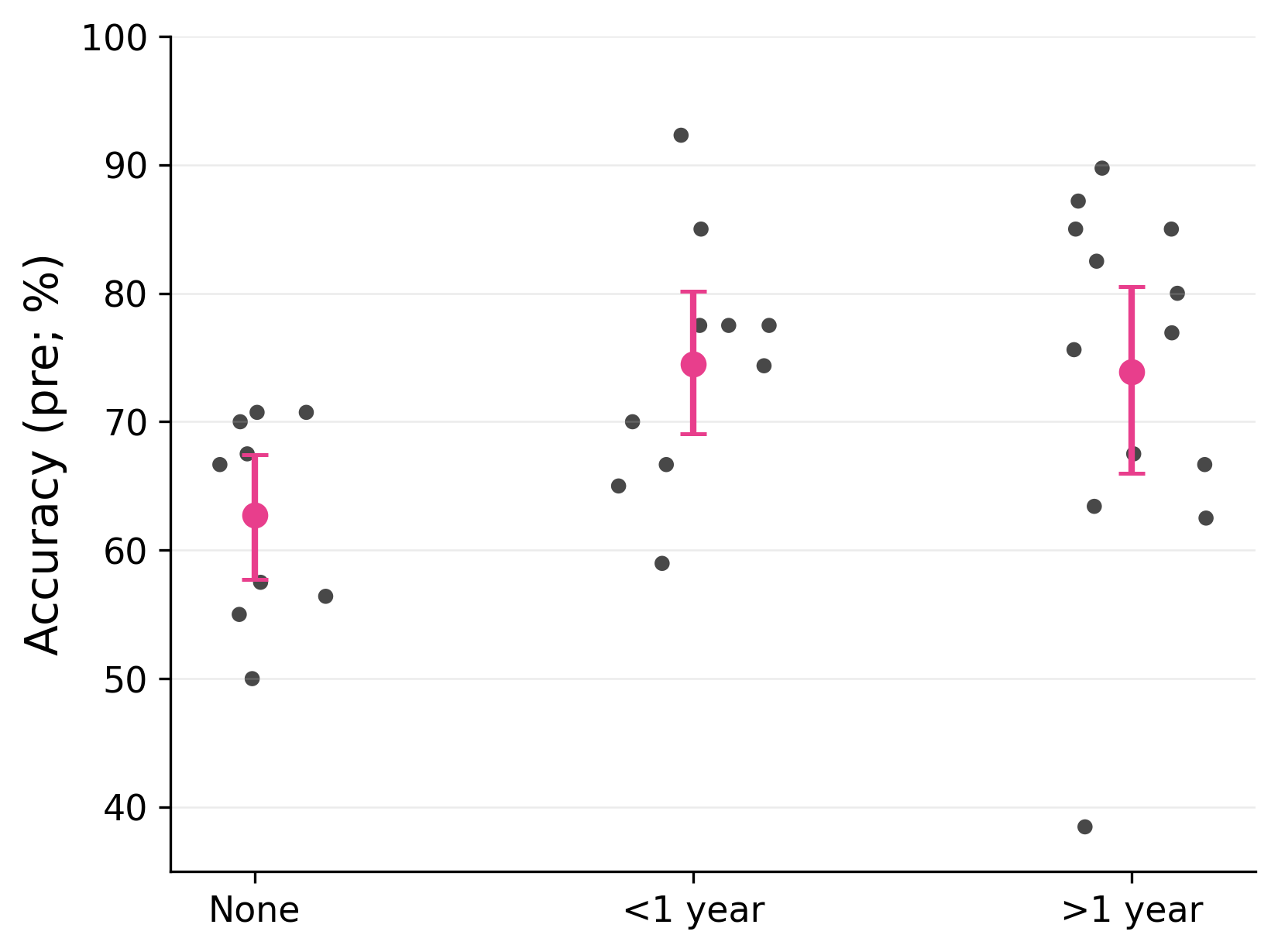}
 \caption{}
 \label{fig:beeswarm-forensics-pre}
 \end{subfigure}\hfill
 \begin{subfigure}[t]{0.23\linewidth}
 \centering
 \includegraphics[width=\linewidth]{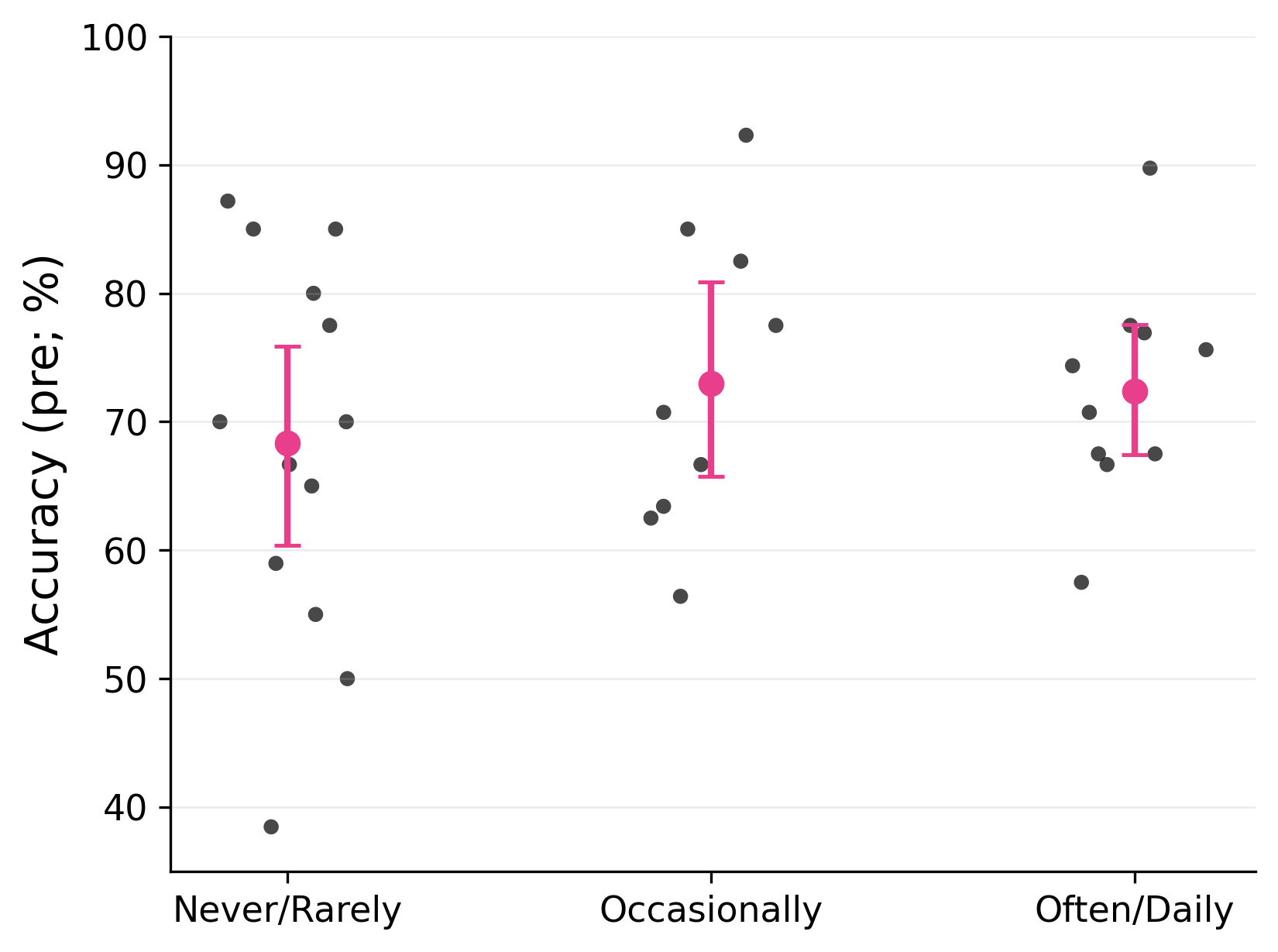}
 \caption{}
 \label{fig:beeswarm-genai-pre}
 \end{subfigure}\hfill
 \begin{subfigure}[t]{0.23\linewidth}
 \centering
 \includegraphics[width=\linewidth]{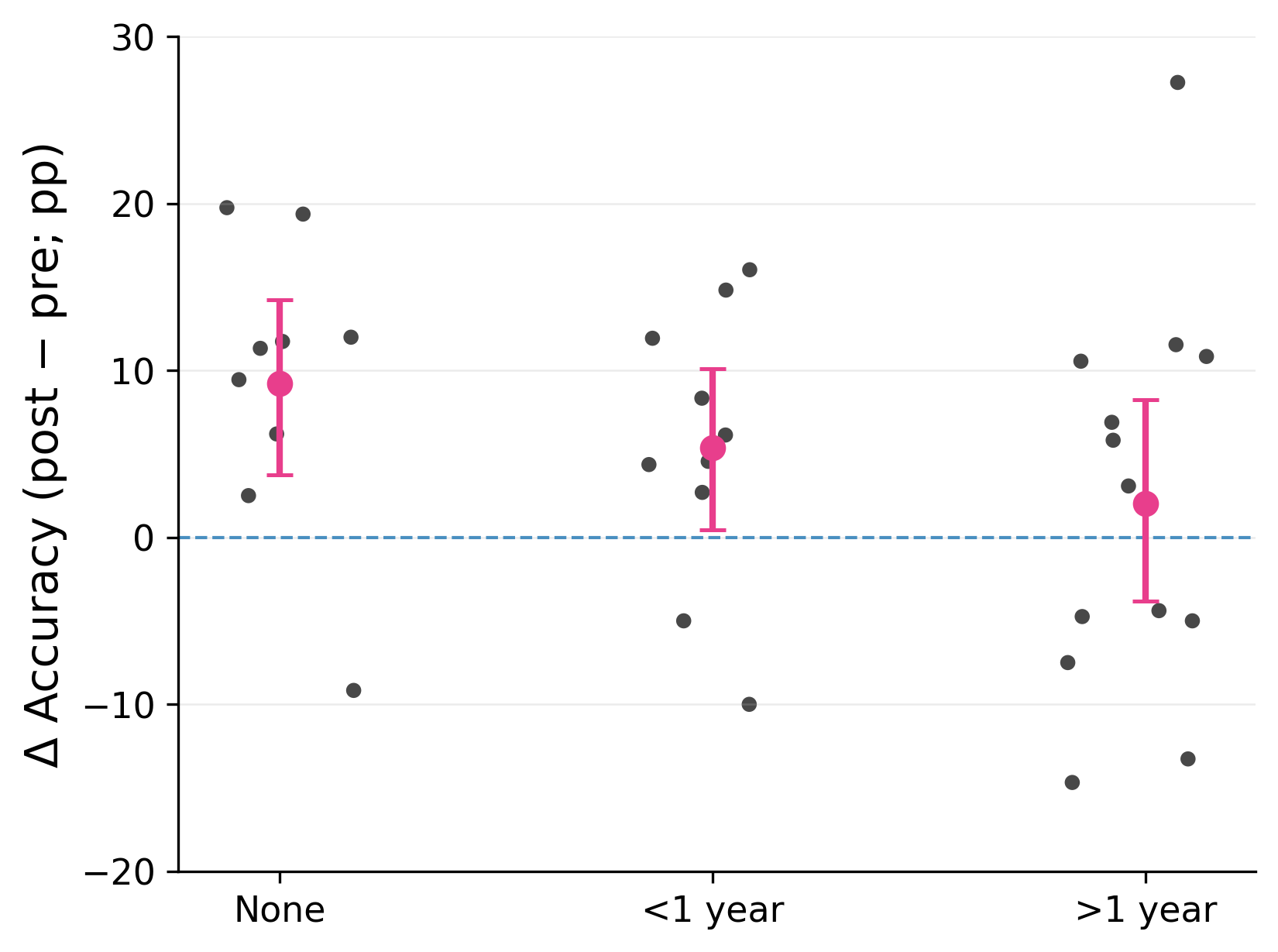}
 \caption{}
 \label{fig:beeswarm-forensics-delta}
 \end{subfigure}\hfill
 \begin{subfigure}[t]{0.23\linewidth}
 \centering
 \includegraphics[width=\linewidth]{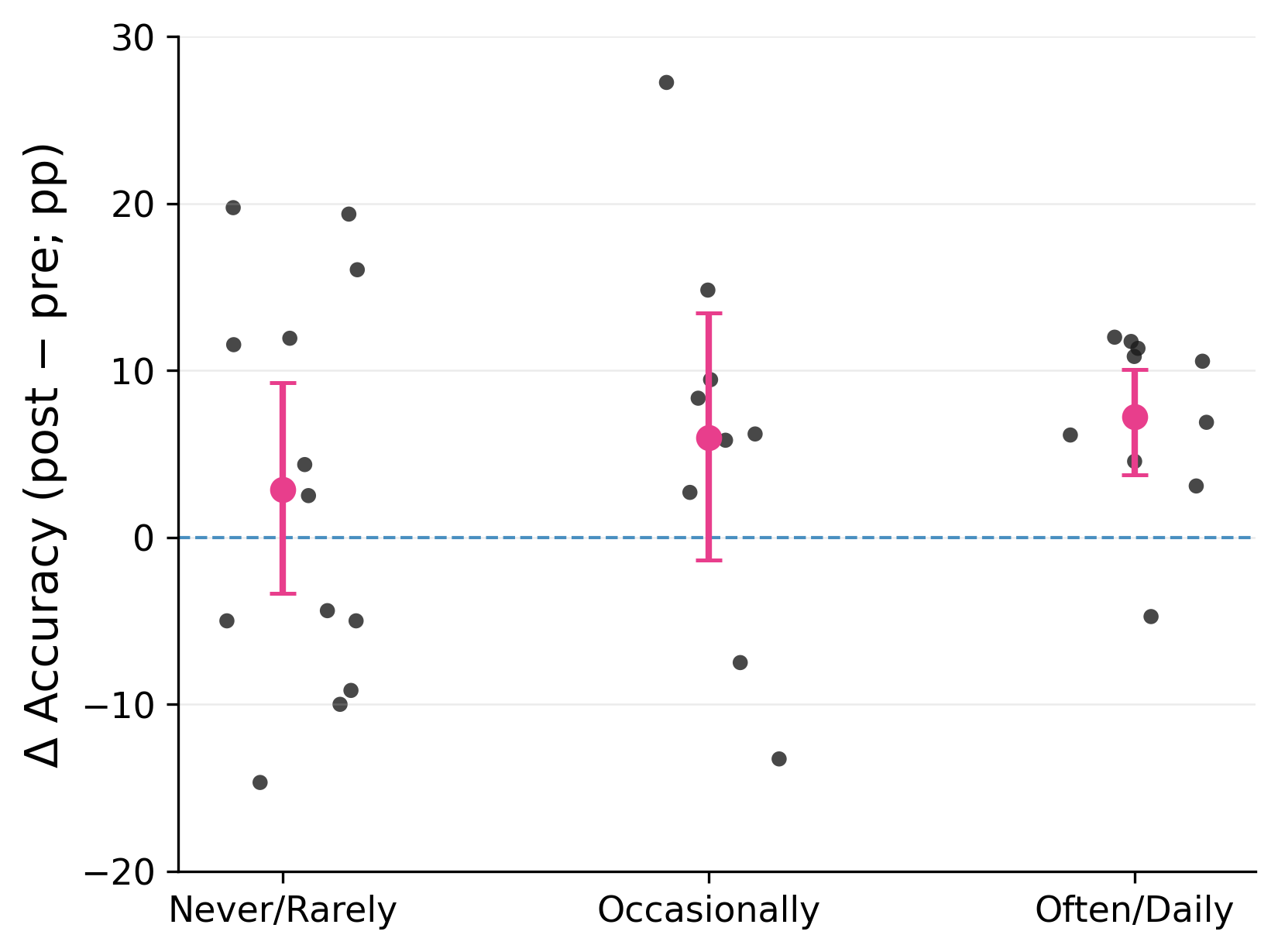}
 \caption{}
 \label{fig:beeswarm-genai-delta}
 \end{subfigure}
 \caption{\textbf{Participant-level accuracy and training effects by expertise and tool use.} Each point represents one participant. Pink points indicate group means, with error bars showing 95\% bootstrapped confidence intervals within each group. Panels (a--b) show baseline (pre-training) accuracy grouped by self-reported digital forensics expertise and generative AI tool use. Panels (c--d) show participant-level training effects, computed as the change in accuracy from pre- to post-training.}
 \label{fig:beeswarm-subgroups}
\end{figure*}

Accuracy patterns also differed across groups defined by participants' self-reported familiarity with generative AI tools. Before training, baseline accuracy ranged from 68.4\% (95\% CI: [60.5, 76.1]) among participants reporting rare use to 73.0\% (95\% CI: [65.7, 80.7]) among those reporting occasional use and 72.4\% (95\% CI: [67.4, 77.5]) among those reporting frequent or daily use (Fig.~\ref{fig:beeswarm-genai-pre}). False-positive rates on real images likewise differed across these groups, with higher rates among participants reporting less frequent use (40.5\% among people who rarely use generative AI) and lower rates among those reporting daily use (18.5\%). Estimates for participants reporting no prior use of generative AI tools are not interpreted due to the small sample size in this category.
Our training improved participant accuracy across most generative AI familiarity groups. The estimated training effect was largest among participants reporting frequent or daily prior use of generative AI tools, corresponding to a 7.2 percentage point increase in accuracy (95\% CI: [3.9, 10.0]). Participants reporting occasional and rare use exhibited smaller estimated gains of 6.0 percentage points (95\% CI: [-1.3, 13.2]) and 2.9 percentage points (95\% CI: [-3.3, 9.0]), respectively (Fig.~\ref{fig:beeswarm-genai-delta}).
\subsection{Heterogeneity by Image Pose Category}
We further examined whether training effects differed across image pose categories (portraits, full-body images, posed groups, and candid scenes). Pre-training accuracy varied across pose categories: 69.3\% (95\% CI: [65.0, 73.5]) on portraits, 64.6\% (95\% CI: [58.4, 70.5]) on full-body images, 74.0\% (95\% CI: [68.4, 79.5]) on posed groups, and 75.6\% (95\% CI: [70.0, 81.2]) on candid scenes. Portraits and full-body images had the lowest baselines, leaving the most room for training-driven improvement.
Training effects, estimated by the BA $\times$ Post interaction within each pose category, varied accordingly. The estimated training effect was largest for portrait images, corresponding to a statistically significant 39.8 percentage point increase in accuracy (95\% CI: [27.5, 52.0]). Full-body images and posed group images exhibited directional non-significant estimated effects of -11.6 percentage points (95\% CI: [-24.2, 1.0]) and -5.2 percentage points (95\% CI: [-18.6, 8.2]), respectively. Candid images showed a modest gain of 8.7 percentage points, though this effect was not statistically significant (95\% CI: [-1.7, 19.1]). Given the small per-category sample size, we treat the negative directional estimates for full-body and posed-group images as exploratory; their confidence intervals include zero, so they should not be interpreted as evidence that training harmed performance on these categories.
We speculate the portrait effect reflects how training shifted analysts' attention. Before training, analysts tended to rely on anatomical cues such as hand, finger, and body-proportion anomalies, cues that are mostly not visible in close-up portrait framing. The training broadened analysts' detection vocabulary to include stylistic artifacts (waxy or plastic textures, hyper-real detail, color and resolution inconsistencies) that are highly legible in AI-generated portraits because the face fills the frame. Pre-training reasoning therefore relied on cues that were largely unavailable in portraits, while the training introduced cues that are particularly visible there. Full-body images did not show a comparable gain, plausibly because the anatomical cues that analysts were already attending to remain visible in full-body framing, leaving smaller marginal value from the additional artifact categories the training introduced.
\subsection{Robustness to Training-Stimulus Overlap}
Five stimuli (four AI-generated images and one real photograph) that appeared in the post-training evaluation were also shown to analysts during the 30-minute training. Excluding all 129 trials involving these five stimuli does not alter our findings: BA $\times$ Post $= +11.3$ percentage points (95\% CI: [4.7, 17.8], $p = 0.0007$) on overall accuracy and $+13.4$ percentage points (95\% CI: [3.5, 23.2], $p = 0.008$) on pair accuracy.
\subsection{Qualitative Analysis of Detection Reasoning}
To explore how training shifted analysts' reasoning, we conducted an exploratory analysis of 365 free-text comments explaining detection decisions from 24 intelligence analysts. Of these, 250 (68.5\%) were collected pre-training and 115 (31.5\%) post-training; 11 analysts contributed both pre- and post-training comments, enabling a within-subject view of reasoning evolution. Each comment was coded along the five-category artifact taxonomy from \citet{kamali2024} (anatomical, stylistic, functional, physics, and sociocultural implausibilities). Coding was performed with LLM-assisted classification using Claude (Opus 4.6). This is an exploratory analysis that complements our quantitative results.

\begin{figure}[h]
 \centering
 \includegraphics[width=0.9\linewidth]{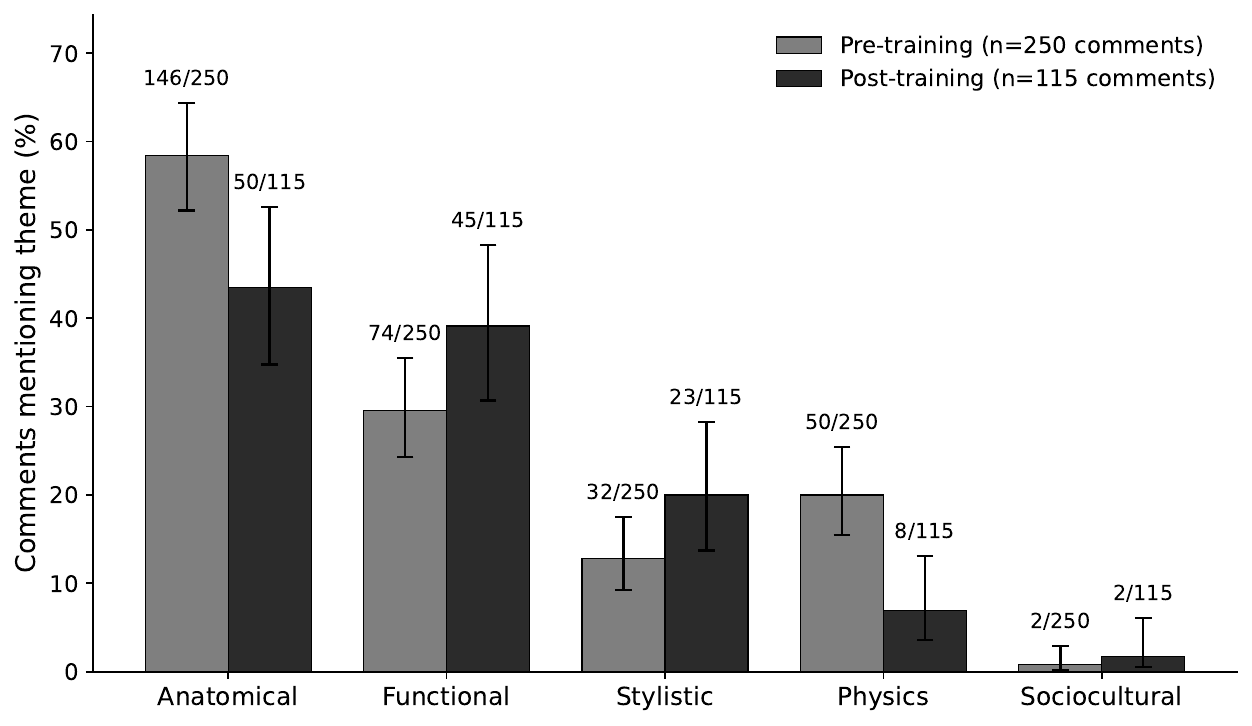}
 \caption{Artifact themes mentioned in analyst free-text comments before and after training. Bars show the proportion of comments referencing each artifact category, with 95\% Wilson confidence intervals. Counts above each bar ($k/n$) indicate the number of comments mentioning the theme out of total comments in that phase. Pre-training: $n=250$ comments from 24 analysts; post-training: $n=115$ comments. Themes are coded with the artifact taxonomy from \citet{kamali2024}.}
 \label{fig:qualitative-artifacts}
\end{figure}

Training was associated with shifts in the artifact categories analysts referenced (Figure~\ref{fig:qualitative-artifacts}). References to physics violations \emph{decreased} from 20.0\% (95\% CI: [15.5, 25.4]) pre-training to 7.0\% (95\% CI: [3.6, 13.1]) post-training, the only between-phase difference for which the 95\% Wilson confidence intervals do not overlap. The share of comments mentioning anatomical artifacts decreased from 58.4\% to 43.5\% ($-14.9$ percentage points), consistent with analysts moving beyond body-part anomalies that were already prevalent in pre-training reasoning. Functional artifact mentions (implausible objects, malformed text, illogical placements) shifted from 29.6\% to 39.1\% (+9.5 percentage points), and stylistic mentions (waxy textures, hyper-real detail, color and resolution inconsistencies) shifted from 12.8\% to 20.0\% (+7.2 percentage points). These directional increases are consistent with broader vocabulary use after training but should be interpreted as exploratory given overlapping confidence intervals.

Comments also became more specific. High-specificity comments, those naming a concrete cue instead of a general impression, rose from 33.2\% pre-training to 41.7\% post-training (+8.5 percentage points). The character of comments shifted accordingly. Pre-training comments tended to name a body part or a vague impression: \emph{``hair''}, \emph{``fingers''}, \emph{``something looks off''}. Post-training comments referenced specific concepts taught in the session: \emph{``uniformity of dirt''}, \emph{``chair seats don't levitate, usually''}, \emph{``doors don't open unto empty space''}, \emph{``JPEG artifacts in clothes but not hair, at the same focal plane''}, and \emph{``lateral chromatic aberrations''}. A small number of post-training comments (7 of 115, 6.1\%) showed meta-cognitive reasoning about the generative process itself, including attempts to reverse-engineer the stock-photo prompts a generator might have used.

Among the 11 analysts with both pre- and post-training comments, individual comment patterns generally mirrored the aggregate shift toward functional and stylistic categories and toward higher specificity. Given the small paired subset, we present these observations as a descriptive companion to the quantitative within-subject design.

Together, these patterns suggest that the training transferred not only a set of visual cues but also a vocabulary for articulating detection reasoning. We caveat that this analysis is exploratory: comments were not collected as a primary outcome measure, comment density is uneven across participants, and LLM-based coding may carry systematic biases relative to expert human coding. We frame the qualitative findings as consistent with the quantitative training effects.

\section{Discussion}

We find that a brief generative AI literacy training improved intelligence analysts' accuracy at distinguishing between real and AI-generated images. Our counterbalanced within-subject randomized experiment reveals that the training increased overall accuracy by 9 percentage points (95\% CI: [2.7, 15.4]), with a 14.2-percentage-point reduction in false-positive errors on real images (95\% CI: [0.7, 27.7]) and a 4-percentage-point increase in accuracy on AI-generated images (95\% CI: [$-5.8$, 13.8]). In contrast to prior trainings that mostly improved identification of AI-generated images at the cost of misidentifying real images~\citep{chen2025training}, we find a modest but not statistically significant increase in identifying AI-generated images and a large, statistically significant increase in participants' identification of real images.

The differences in effects between prior trainings and this training may partly arise from the image content and type of training. The generative AI literacy training and associated evaluation covered a diverse range of human-oriented imagery and state-of-the-art 2024 text-to-image systems including Midjourney, Stable Diffusion, and Firefly. The 30-minute training covered 50 AI-generated images and seven real images from a manual on how to spot AI-generated images~\cite{kamali2024}, showcasing examples across five artifact categories~\cite{kamali2025}. Importantly, the training highlighted that viewers can mistake a real image for AI-generated if they are not paying careful attention to the question: ``Does this image show anything generative AI tends to perform poorly on?'' The images in the training span many levels of pose and scene complexity; examples include a portrait-style image of a man eating a pizza, two images of couples posing on their wedding day, individuals and groups sitting, walking, and running, and more. Likewise, the images in the evaluation include close-up portraits of people stirring drinks, looking at insects, walking, running, and posing. While the evaluation in \citet{geissler2026} also considers a wide range of human imagery, many prior evaluations of training focused on GAN-generated face portraits~\cite{rehman2025, gray2025training, chen2025training}, which is a more constrained domain of images. 

Notably, \citet{geissler2026} provide experimental evidence that visual examples and text-based explanations have the strongest effect on improving detection among the interventions they tested relative to text tips, gamified spot-the-error, practice with feedback, and knowledge explanation. The visual-examples training in \citet{geissler2026} used five images from DALL-E 3, while our training drew on 50 AI-generated images from Midjourney, Stable Diffusion, and Firefly. We speculate that this difference in scale and model diversity contributes to the different outcomes across the two studies. With more visual examples, participants had the opportunity to build a more nuanced mental model of what generative AI tends to perform poorly on and how non-generative AI photography artifacts like compression or editing may appear. This combined exposure may help explain the improved identification of real images.

We did not observe a measurable advantage associated with professional experience as an intelligence analyst in distinguishing AI-generated from real images. Before training, intelligence analysts' accuracy was broadly comparable to large-scale public samples reported in prior work~\citep{kamali2025}, where average accuracy was approximately 72\% for real images and 76\% for AI-generated images. Our stimulus set is a balanced subset of the image pool used in that public study, which makes the two studies more comparable at the stimulus level. The alignment is consistent with prior evidence that professional experience and individual differences in face processing ability are insufficient on their own to confer a reliable advantage in deepfake detection~\citep{Ramon2024, gray2025training}. While participants with more than one year of digital forensics experience showed modestly higher baseline performance, training produced gains across all experience levels, with statistically significant gains among analysts with no prior experience. 

Stimulus-level analyses show that training effects are distributed broadly across images and are not driven by a small subset of easy cases. Of the 97 stimulus images, 65 (67\%) showed post-training accuracy gains and 32 (33\%) showed declines (Fig.~\ref{fig:stimulus-plots}). Of the 65 gain images, 47 fell above the null expectation under a uniform user-level effect, indicating gains larger than sampling variability alone would produce. The aggregate effect is strongly positive, but the underlying distribution is heterogeneous across stimuli. This pattern reflects the intentional stimulus heterogeneity of our design~\citep{simonsohn2025stimulus}.

Two features of the design strengthen the interpretation that the observed gains are evidence of analysts' classification improvement. First, the counterbalanced design with the same images evaluated both before and after training by two different groups of analysts precludes the possibility that the post-training block had differential difficulty. Second, the stimulus set is designed in pairs, which allows for a more conservative measure of accuracy. Each AI image is paired with a real photograph matched on subject, pose, and scene context. A pair is coded as correct only when both members are correctly classified, so chance-level guessing is less likely to produce a correct response on both than on a single image. Pair accuracy improved by 11 percentage points after training (95\% CI: [1.8, 21.0], $p=0.020$; Table~\ref{tab:ols_results}, column 4). The training effect persists under this more conservative outcome, consistent with the overall accuracy result. 

The design supports a practical claim that brief, artifact-focused training could complement automated detection. Prior interventions that deliver fixed rules or checklists of visual cues for spotting AI-generated content have not consistently improved classification accuracy~\citep{somoray2023, holmes2025, bray2023}. In contrast, our training does not teach fixed rules for classification. Instead, it offers many visual examples that guide analysts on where to look and what kinds of inconsistencies to consider when judging whether images are AI-generated. 

\paragraph{Limitations and scope.}
This study evaluates short-term training effects with 32 professional analysts and 2,544 image-level judgments. Each experimental phase allowed up to 30 minutes for 40 images. Analysts averaged 27 seconds per image, and most finished a phase in about 21 minutes. We do not assess long-term retention, and a follow-up assessment days or weeks after training would help characterize whether these gains persist and how rapidly they decay. These results should be interpreted as evidence of immediate post-training improvement under realistic time constraints. Future work could investigate whether these gains transfer to operational analyst workflows and how this training combines with automated detection tools.

\section{Conclusion}
This work provides evidence that brief, structured training can meaningfully improve intelligence analysts' ability to distinguish AI-generated images from real photographs, even when baseline performance mirrors that of the general public. Using a counterbalanced within-subject design and a stimulus set drawn from contemporary diffusion models across varied scene complexities, we show that a short intervention can sharpen discrimination and lead to a significant reduction in false-positive errors on real images.

Our findings suggest that strengthening human judgment does not necessarily require extensive retraining or domain-specific expertise but can be achieved through targeted guidance that helps analysts attend to recurring visual inconsistencies while accounting for uncertainty and image-level variability. At the same time, the persistence of heterogeneity across stimuli points to the limits of human judgment without access to the broader context associated with an image.

As generative image models continue to evolve, brief, lightweight interventions aimed at supporting analysts' detection of AI-generated images, such as the one studied here, offer a practical complement to automated tools in settings where visual authenticity remains ambiguous. Future work should examine how such training interacts with longer-term learning, real-world task integration, and assistive systems that provide just-in-time support.

\section{Broader Impact and Ethical Considerations}
This study was determined exempt from Northwestern University's Institutional Review Board review (IRB identification number STU00220627). All participants provided informed consent prior to participation.
The study posed minimal risk to participants, involving only standard image-classification tasks and survey questions. Participants were informed after the study that some of the images they viewed were AI-generated. All collected data were stored on secure, access-controlled servers. Participant identifiers were removed or replaced with randomized IDs prior to analysis, and no personally identifiable information was shared outside the research team.
\section{Acknowledgments}
The authors thank the North Carolina State Laboratory for Analytic Sciences (LAS) for facilitating the training session with intelligence analysts. We acknowledge funding and support from VeryAI, Robert Pozen, and the Department of Defense (DoD). Any opinions, findings, conclusions, or recommendations expressed are those of the authors and do not necessarily reflect the views of the DoD or any agency or entity of the United States Government. We used Claude to assist with the exploratory qualitative coding of analysts' text responses, which we described in the Qualitative Analysis of Detection Reasoning section.
\section{Code and Data Availability}
Code, de-identified data, and all stimulus images (50 AI-generated images and 47 real photographs) are available at \url{https://github.com/negarkamali/analyst-image-detection}.
\bibliography{aaai2026}
\onecolumn
\section{Appendix: Stimulus Pairs}
We display the 47 real photographs and 50 AI-generated images used in the stimulus set. The left column shows each real photograph and the right column shows its matched AI-generated image; three additional AI-generated images are shown at the end.

\begin{figure}[h!]
\centering
\renewcommand{\arraystretch}{0.6}
\begin{tabular}{cc}
\includegraphics[width=0.42\linewidth, height=0.21\textheight, keepaspectratio]{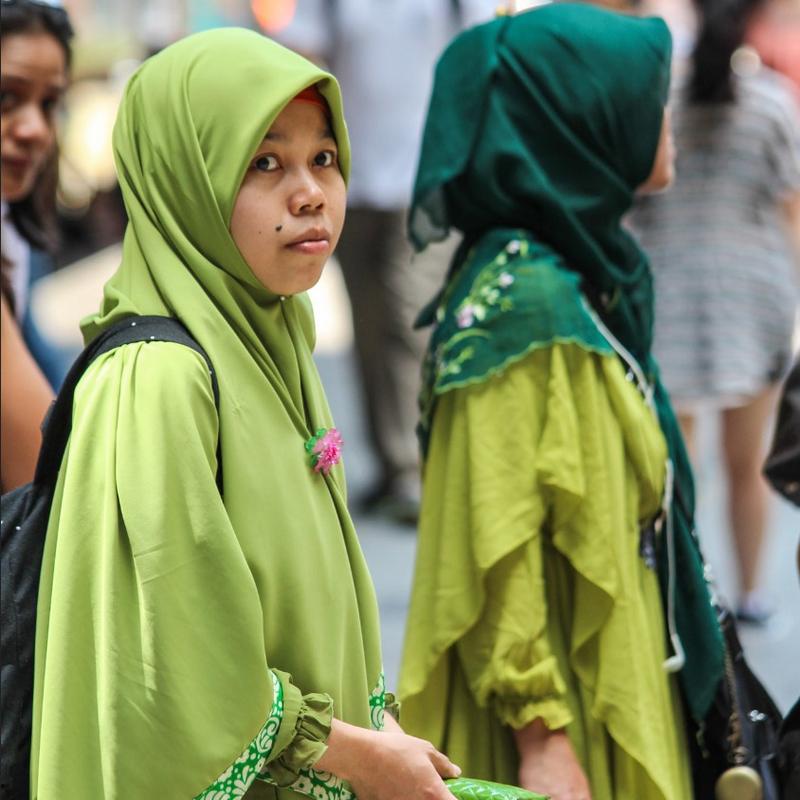} &
\includegraphics[width=0.42\linewidth, height=0.21\textheight, keepaspectratio]{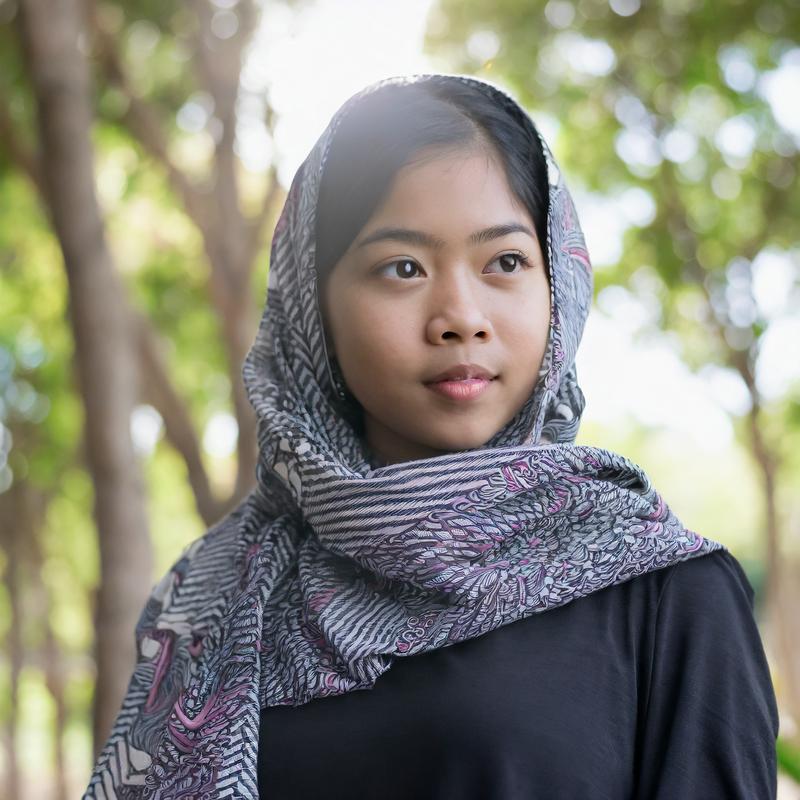} \\
\includegraphics[width=0.42\linewidth, height=0.21\textheight, keepaspectratio]{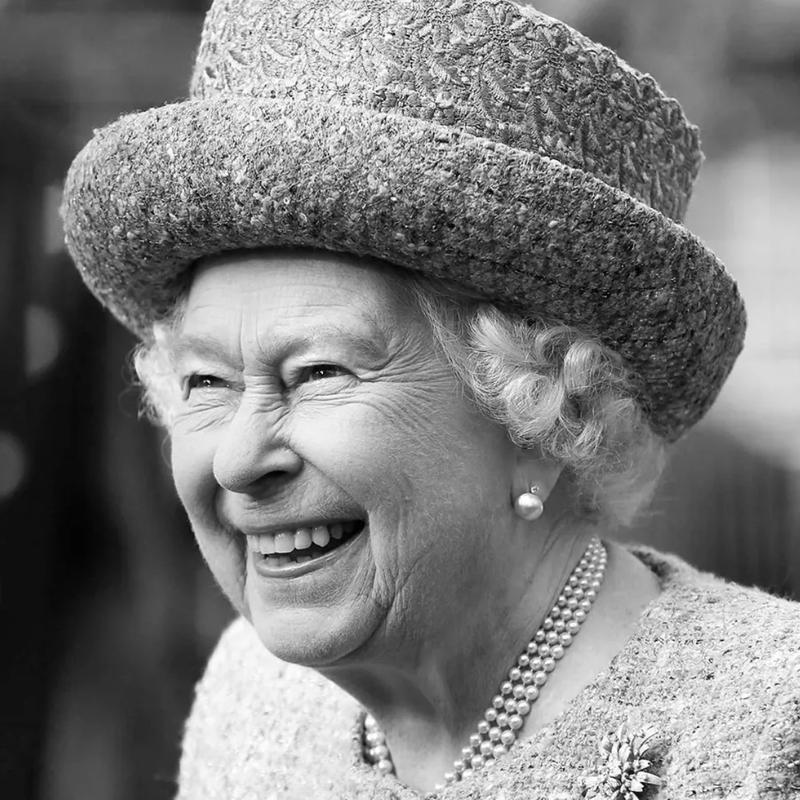} &
\includegraphics[width=0.42\linewidth, height=0.21\textheight, keepaspectratio]{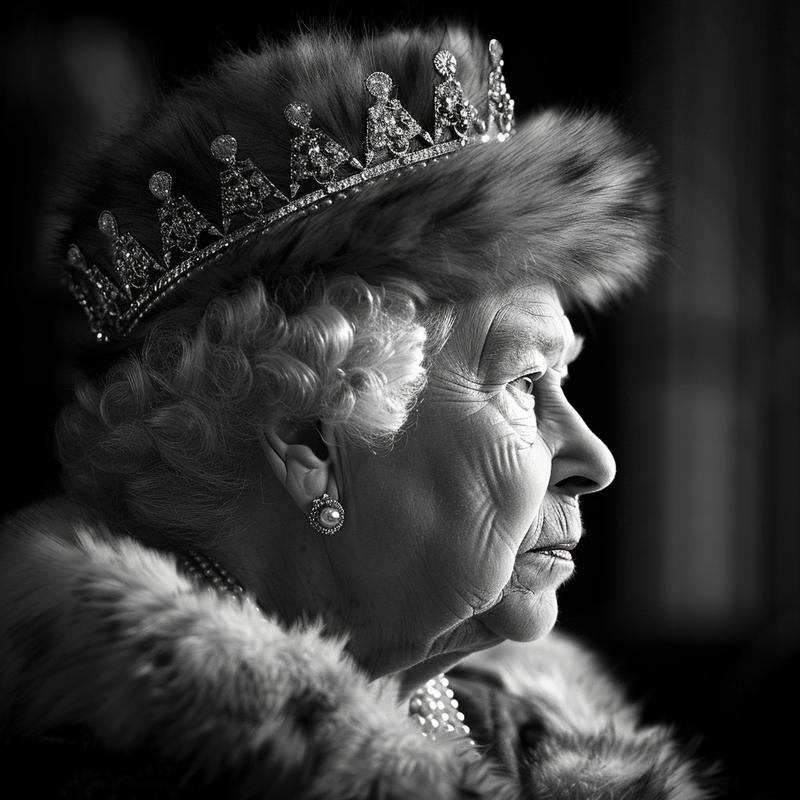} \\
\includegraphics[width=0.42\linewidth, height=0.21\textheight, keepaspectratio]{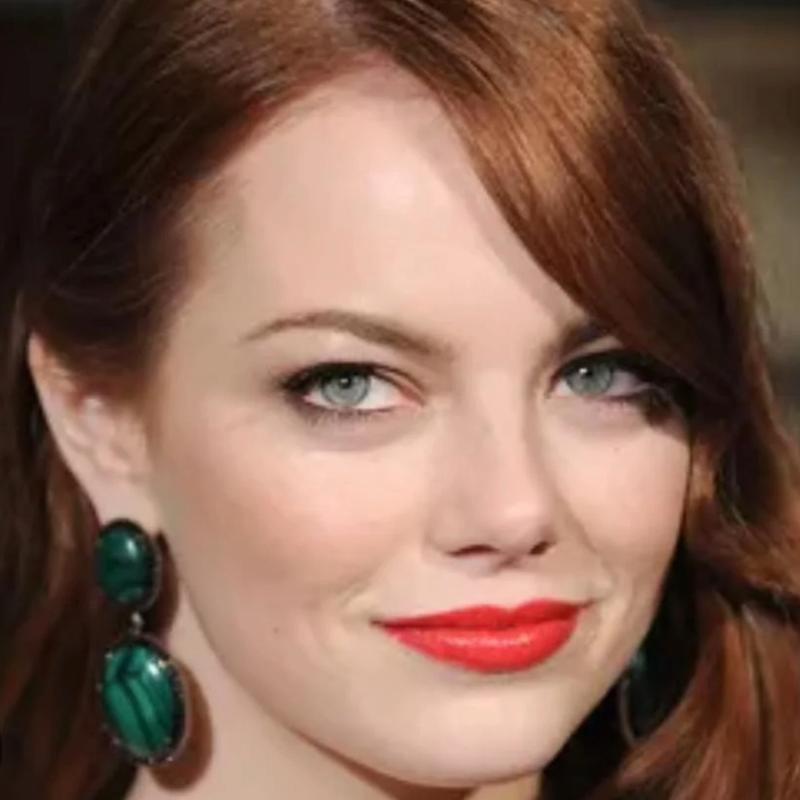} &
\includegraphics[width=0.42\linewidth, height=0.21\textheight, keepaspectratio]{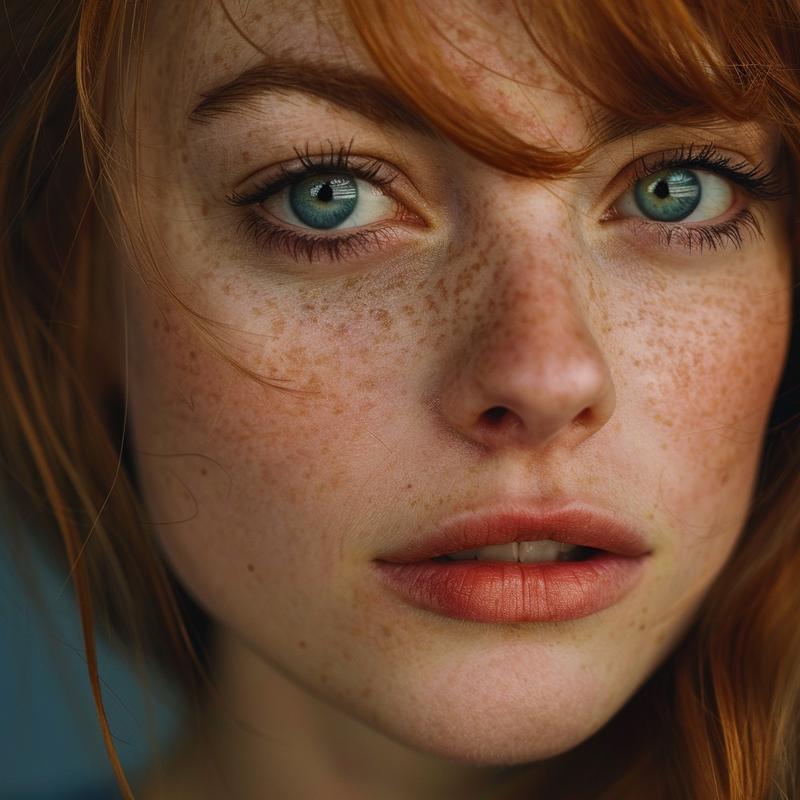} \\
\includegraphics[width=0.42\linewidth, height=0.21\textheight, keepaspectratio]{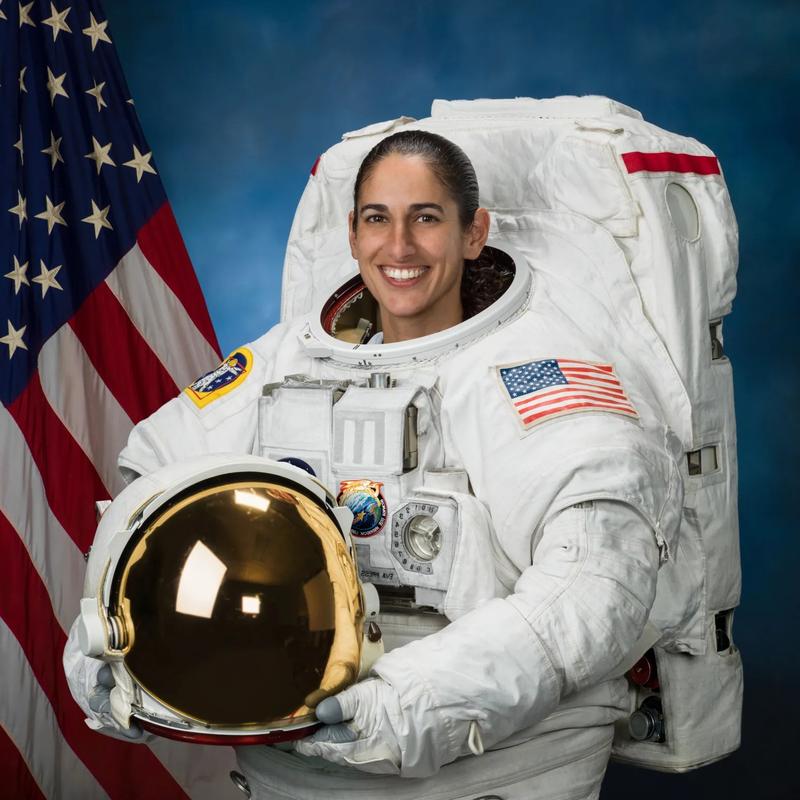} &
\includegraphics[width=0.42\linewidth, height=0.21\textheight, keepaspectratio]{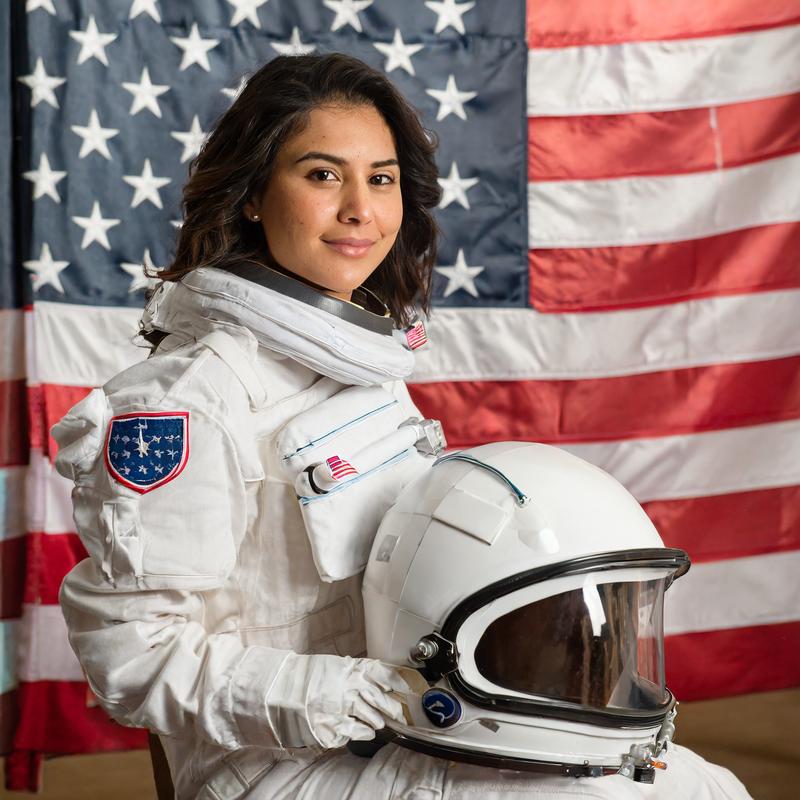} \\
\end{tabular}
\caption*{Pairs 1--4. Real photographs (left), matched AI-generated images (right).}
\end{figure}
\clearpage

\begin{figure}[h!]
\centering
\renewcommand{\arraystretch}{0.6}
\begin{tabular}{cc}
\includegraphics[width=0.42\linewidth, height=0.21\textheight, keepaspectratio]{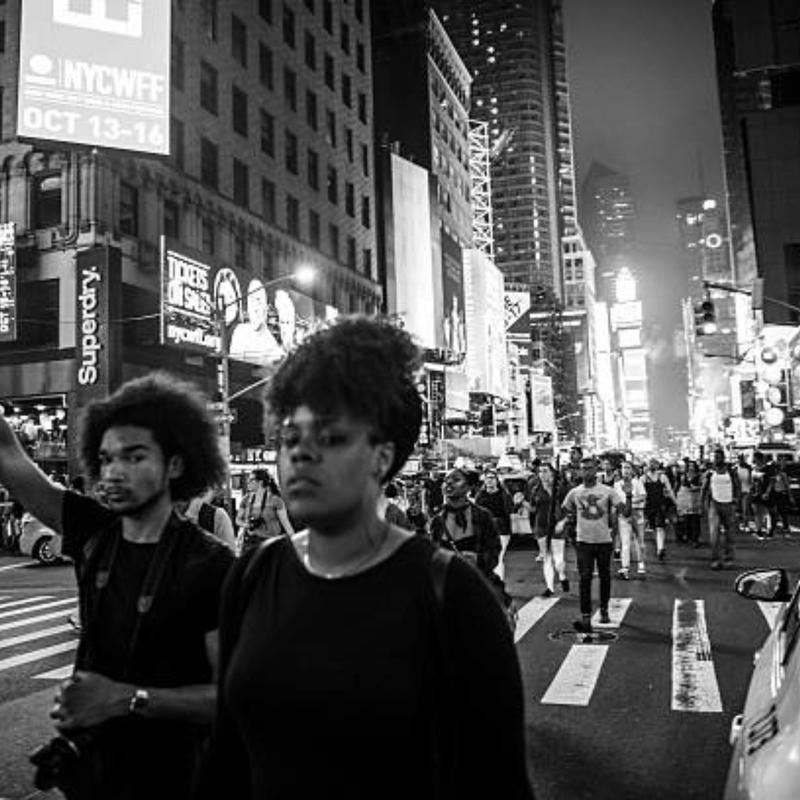} &
\includegraphics[width=0.42\linewidth, height=0.21\textheight, keepaspectratio]{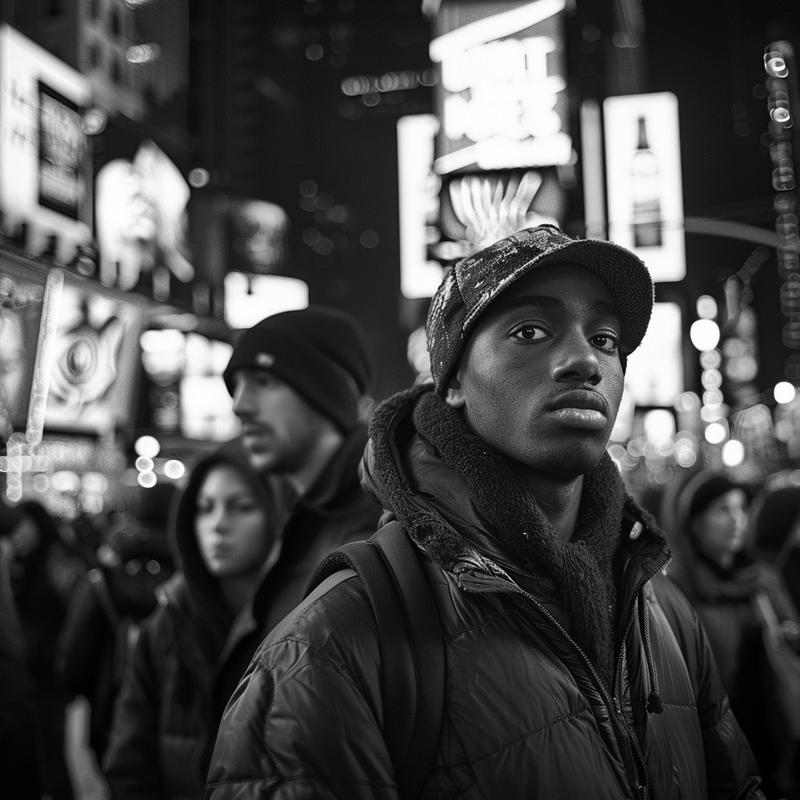} \\
\includegraphics[width=0.42\linewidth, height=0.21\textheight, keepaspectratio]{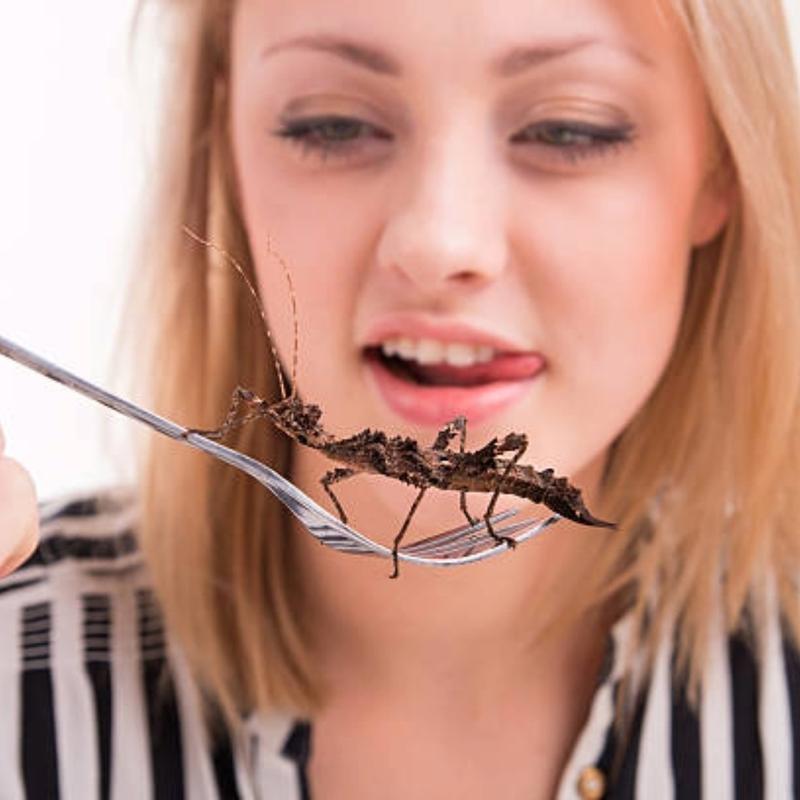} &
\includegraphics[width=0.42\linewidth, height=0.21\textheight, keepaspectratio]{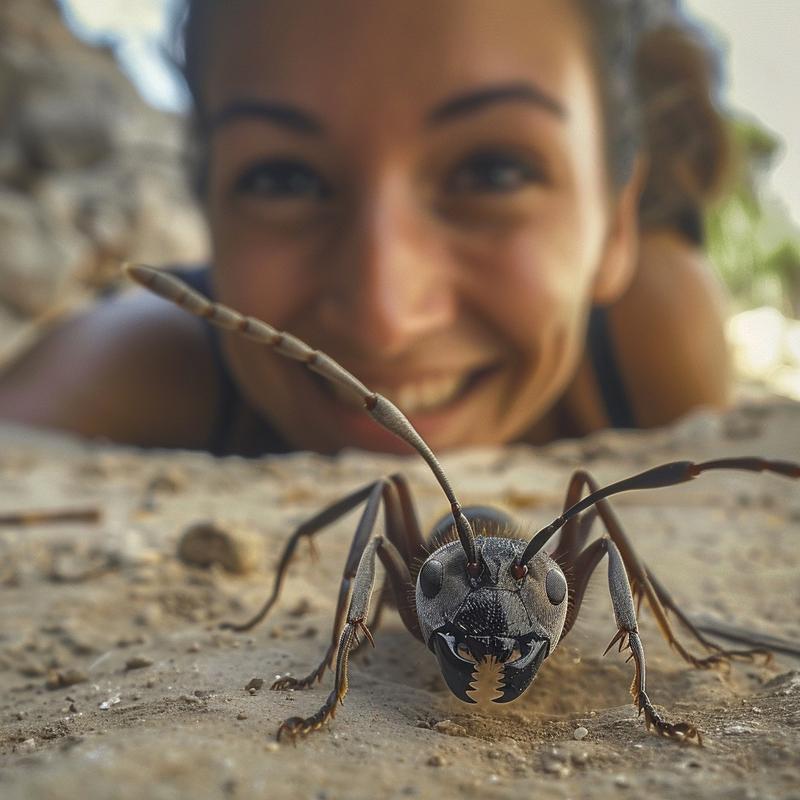} \\
\includegraphics[width=0.42\linewidth, height=0.21\textheight, keepaspectratio]{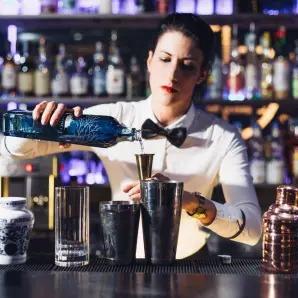} &
\includegraphics[width=0.42\linewidth, height=0.21\textheight, keepaspectratio]{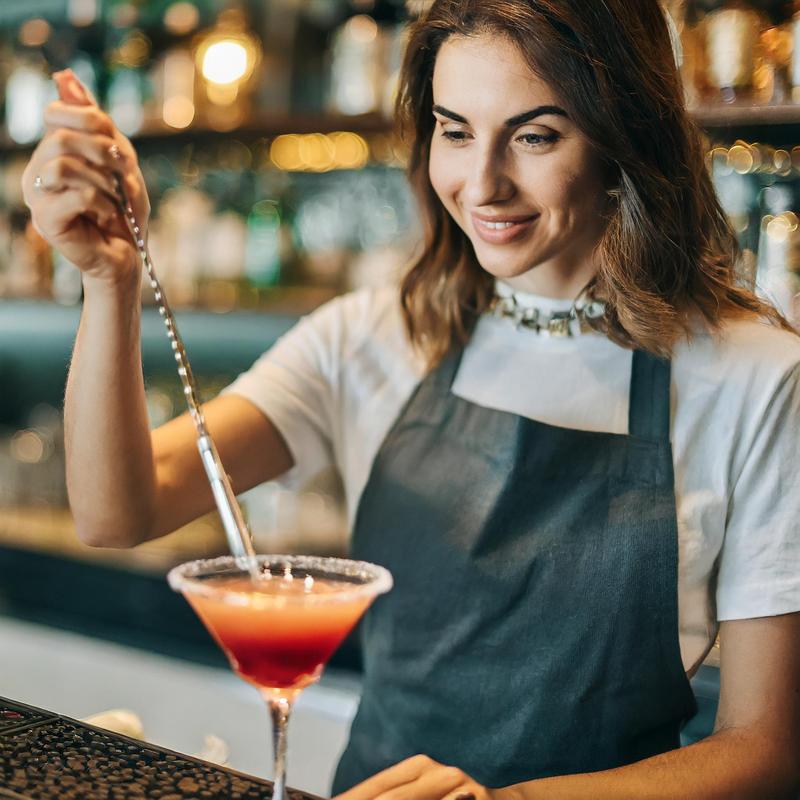} \\
\includegraphics[width=0.42\linewidth, height=0.21\textheight, keepaspectratio]{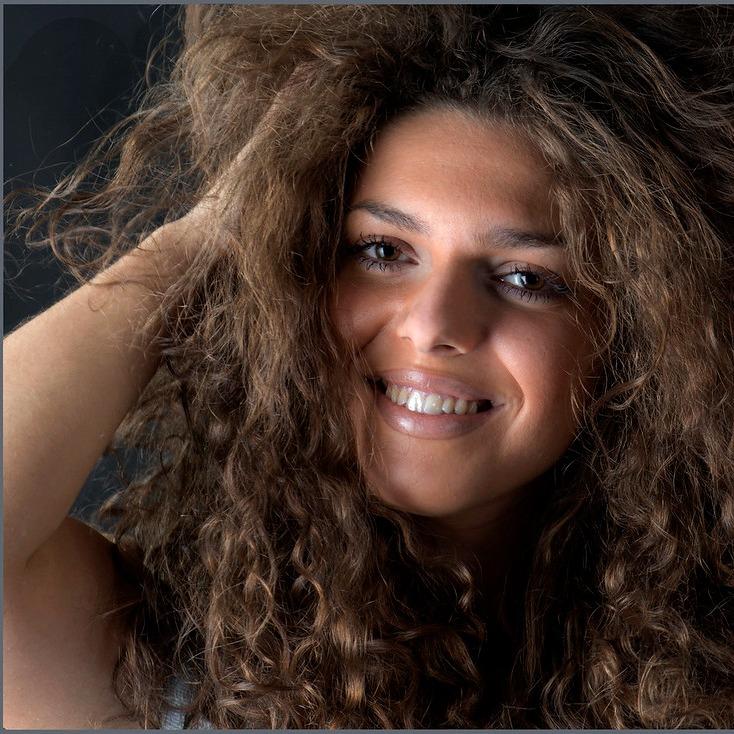} &
\includegraphics[width=0.42\linewidth, height=0.21\textheight, keepaspectratio]{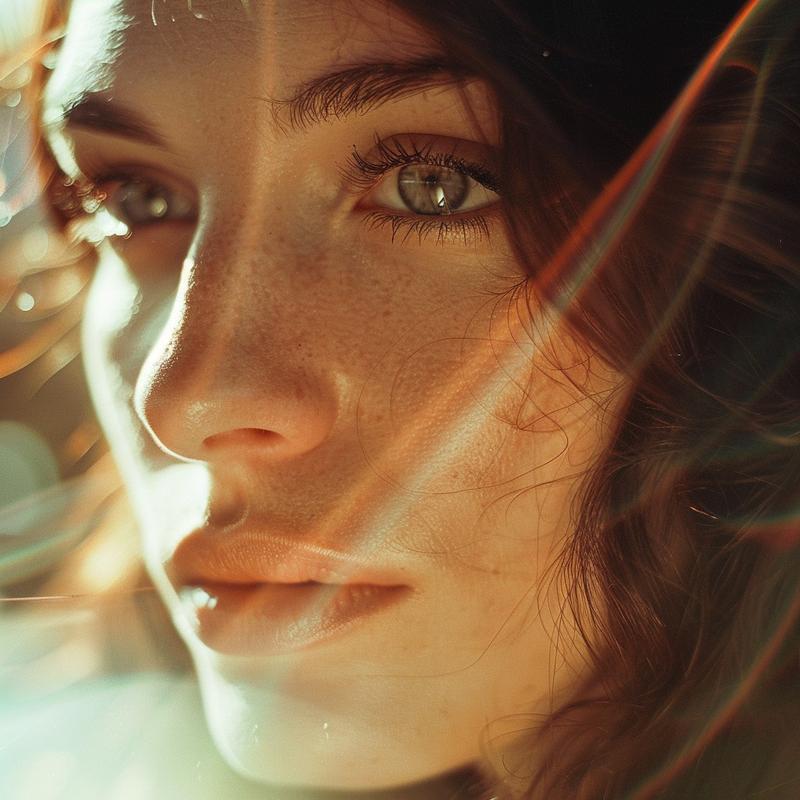} \\
\end{tabular}
\caption*{Pairs 5--8. Real photographs (left), matched AI-generated images (right).}
\end{figure}
\clearpage

\begin{figure}[h!]
\centering
\renewcommand{\arraystretch}{0.6}
\begin{tabular}{cc}
\includegraphics[width=0.42\linewidth, height=0.21\textheight, keepaspectratio]{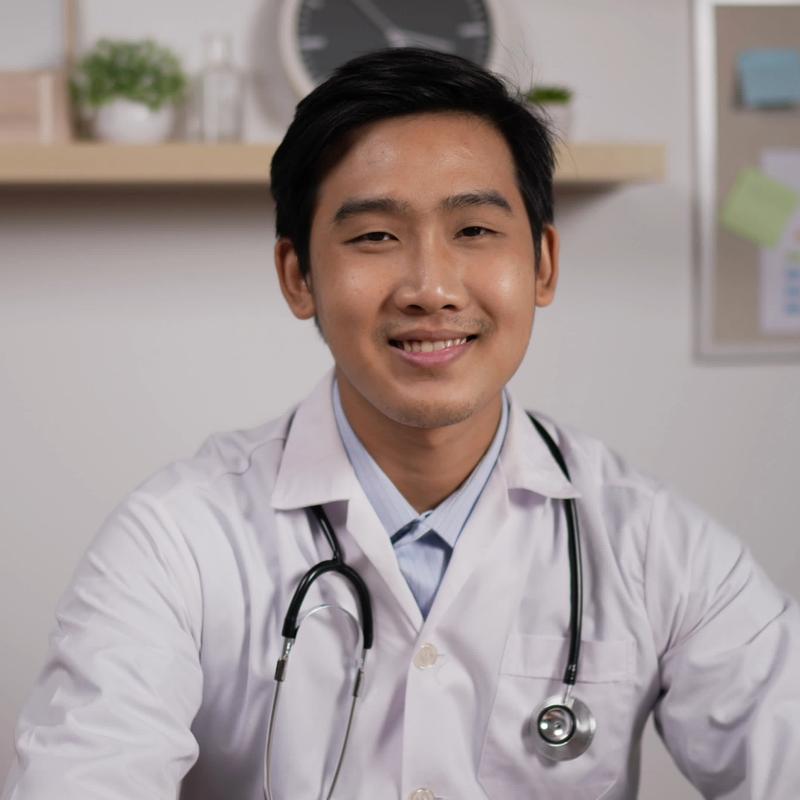} &
\includegraphics[width=0.42\linewidth, height=0.21\textheight, keepaspectratio]{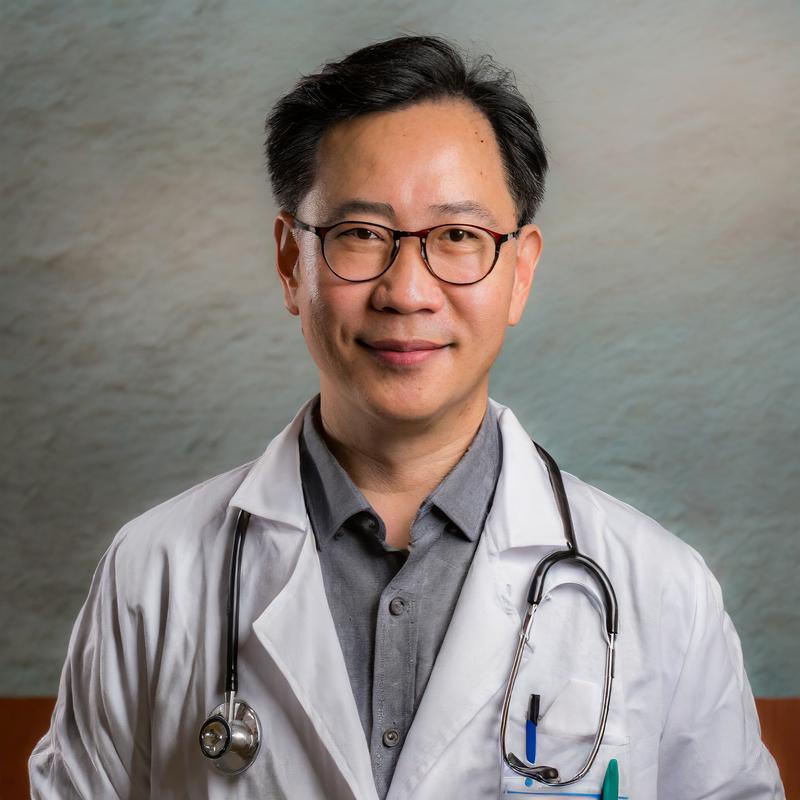} \\
\includegraphics[width=0.42\linewidth, height=0.21\textheight, keepaspectratio]{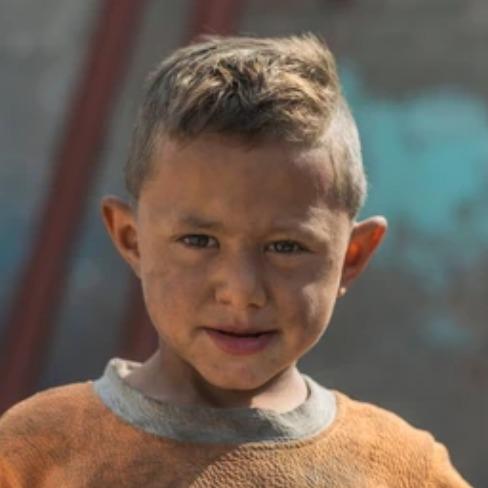} &
\includegraphics[width=0.42\linewidth, height=0.21\textheight, keepaspectratio]{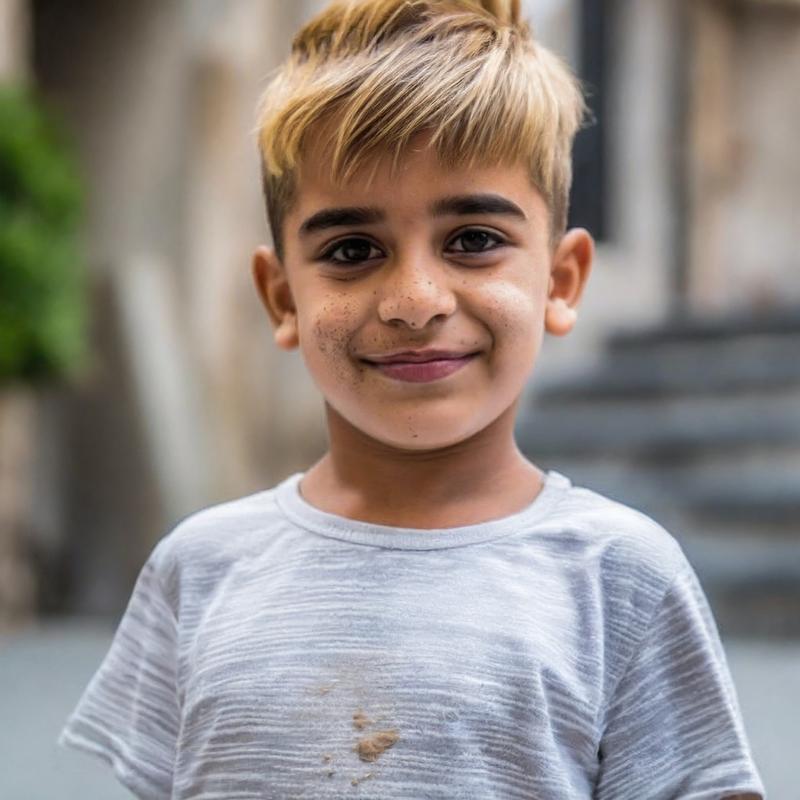} \\
\includegraphics[width=0.42\linewidth, height=0.21\textheight, keepaspectratio]{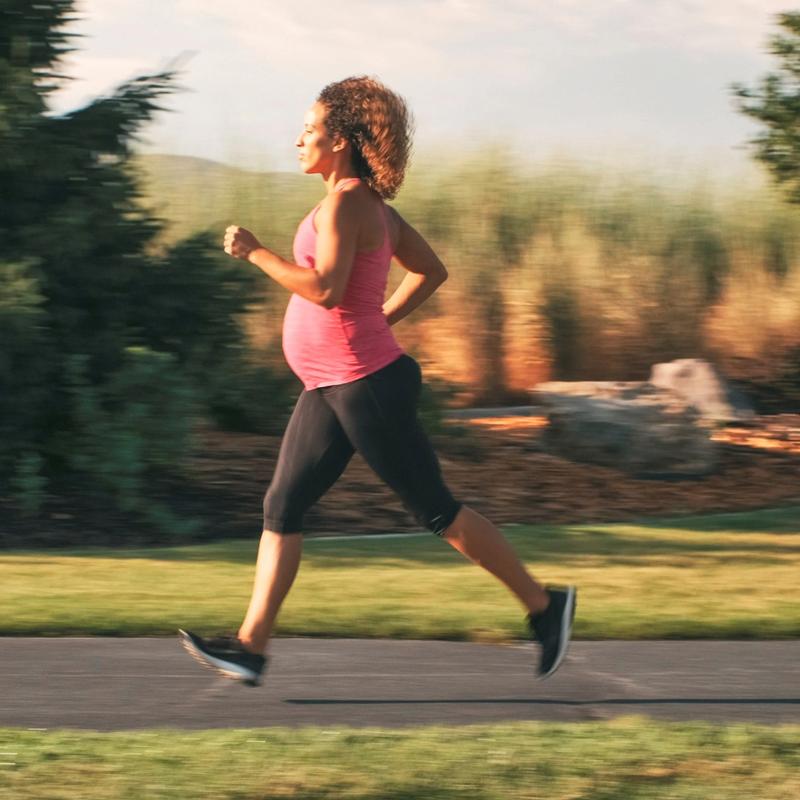} &
\includegraphics[width=0.42\linewidth, height=0.21\textheight, keepaspectratio]{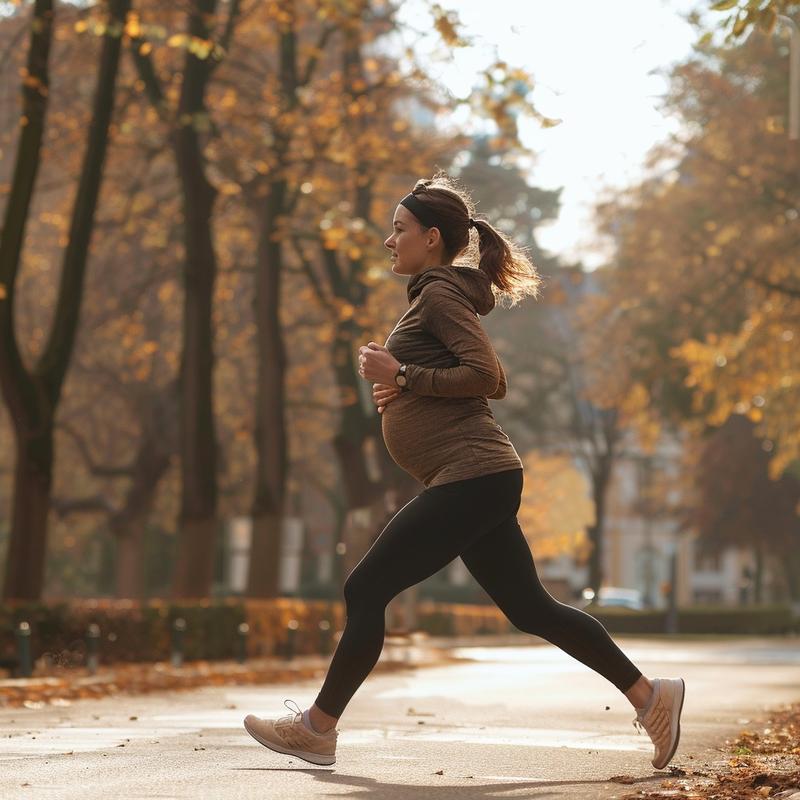} \\
\includegraphics[width=0.42\linewidth, height=0.21\textheight, keepaspectratio]{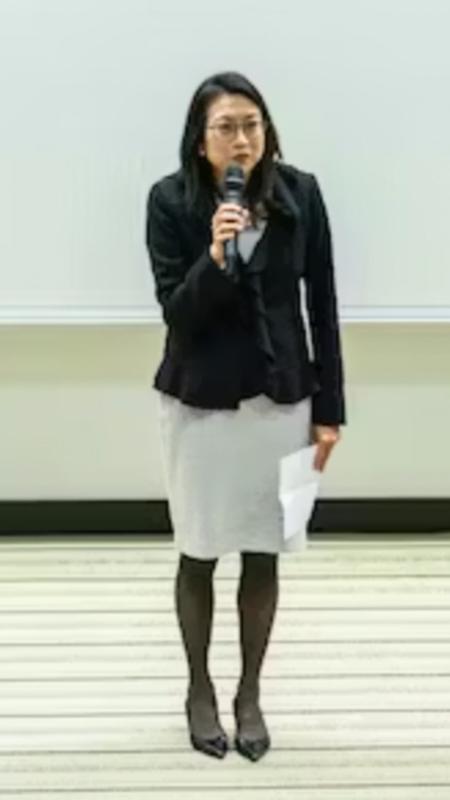} &
\includegraphics[width=0.42\linewidth, height=0.21\textheight, keepaspectratio]{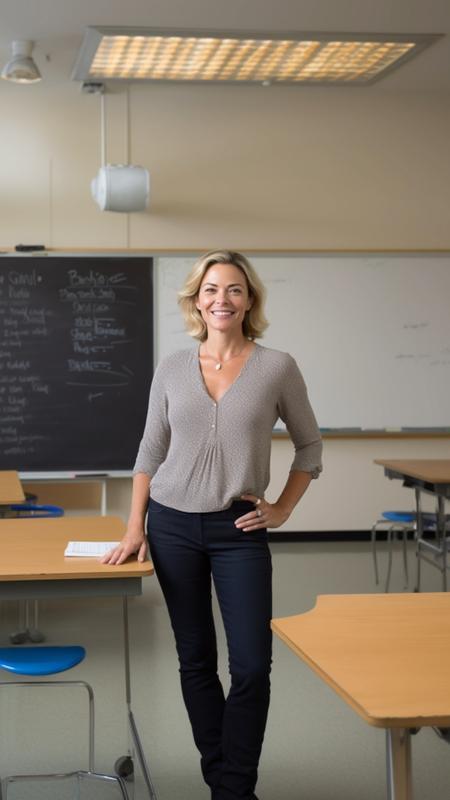} \\
\end{tabular}
\caption*{Pairs 9--12. Real photographs (left), matched AI-generated images (right).}
\end{figure}
\clearpage

\begin{figure}[h!]
\centering
\renewcommand{\arraystretch}{0.6}
\begin{tabular}{cc}
\includegraphics[width=0.42\linewidth, height=0.21\textheight, keepaspectratio]{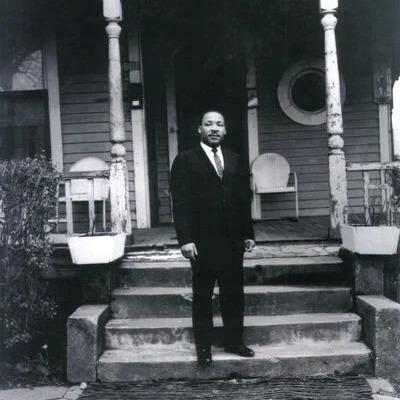} &
\includegraphics[width=0.42\linewidth, height=0.21\textheight, keepaspectratio]{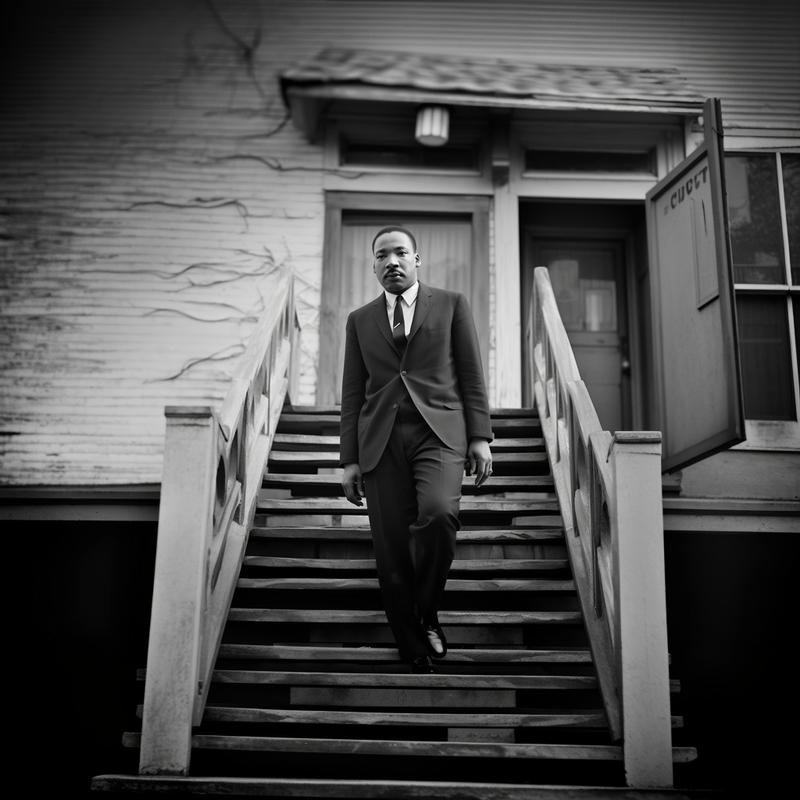} \\
\includegraphics[width=0.42\linewidth, height=0.21\textheight, keepaspectratio]{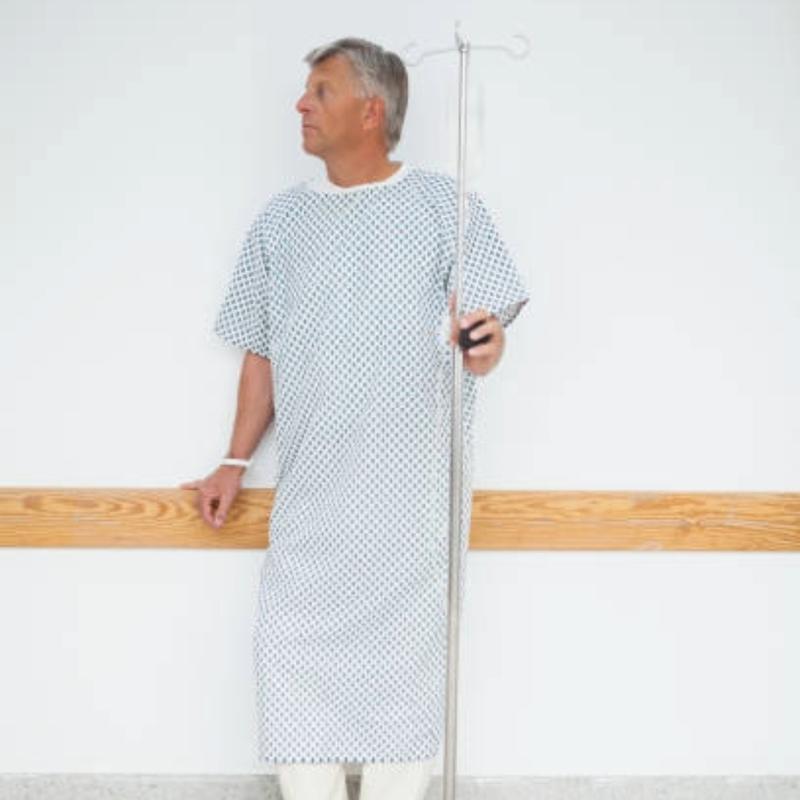} &
\includegraphics[width=0.42\linewidth, height=0.21\textheight, keepaspectratio]{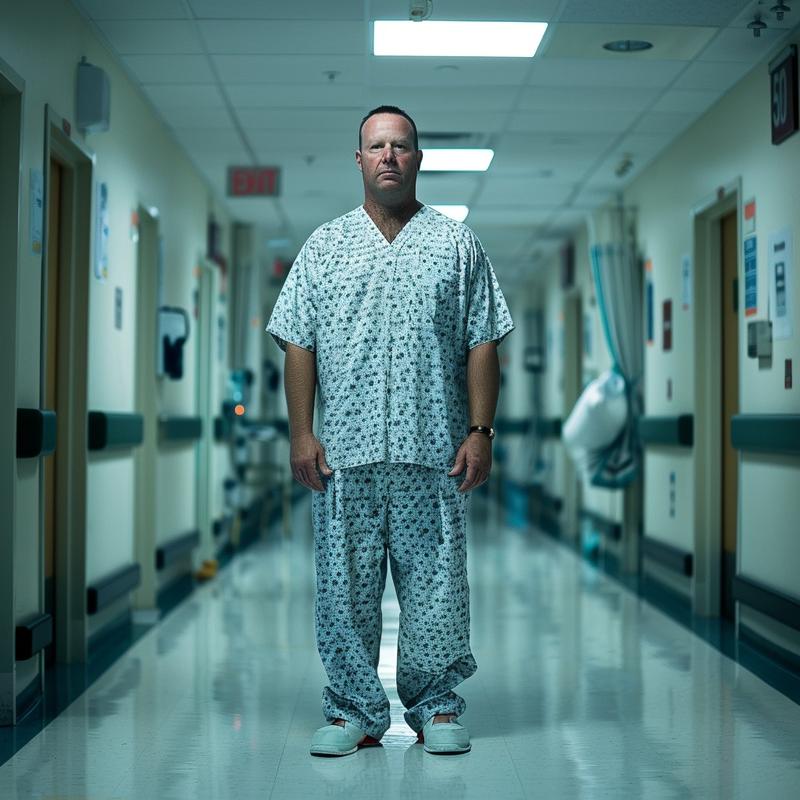} \\
\includegraphics[width=0.42\linewidth, height=0.21\textheight, keepaspectratio]{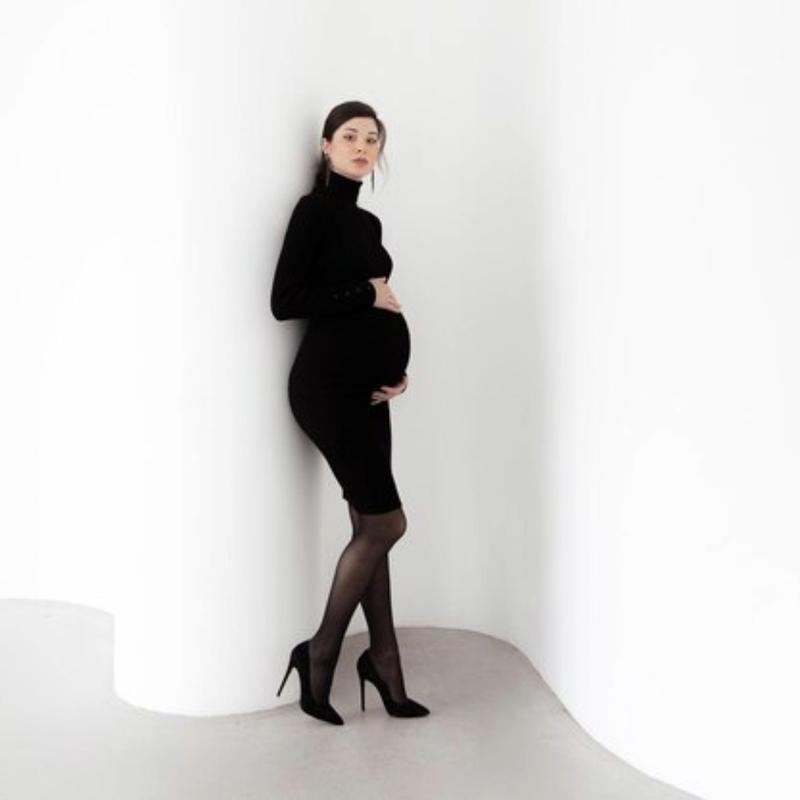} &
\includegraphics[width=0.42\linewidth, height=0.21\textheight, keepaspectratio]{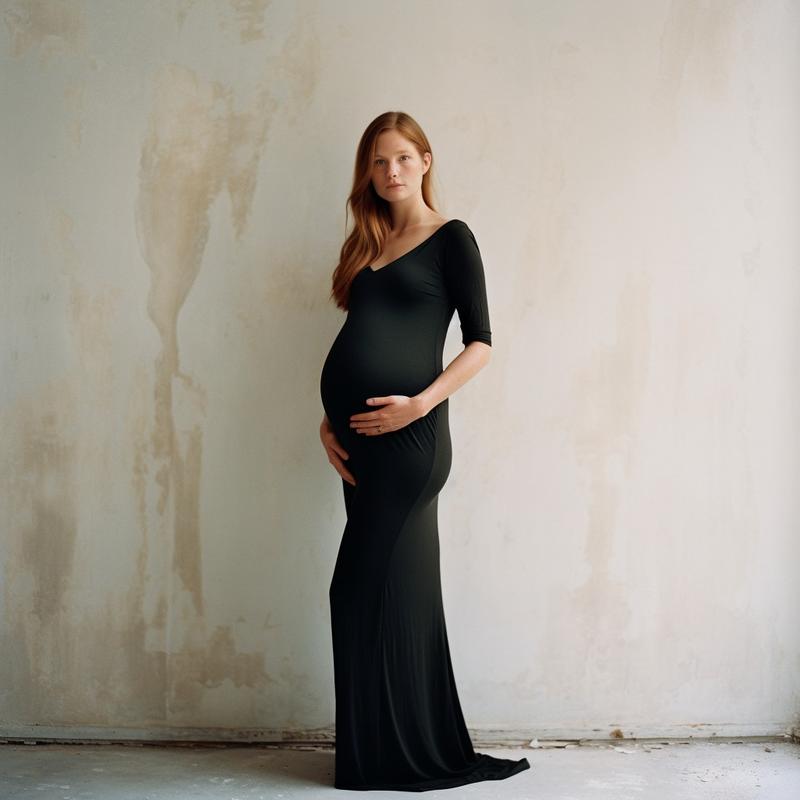} \\
\includegraphics[width=0.42\linewidth, height=0.21\textheight, keepaspectratio]{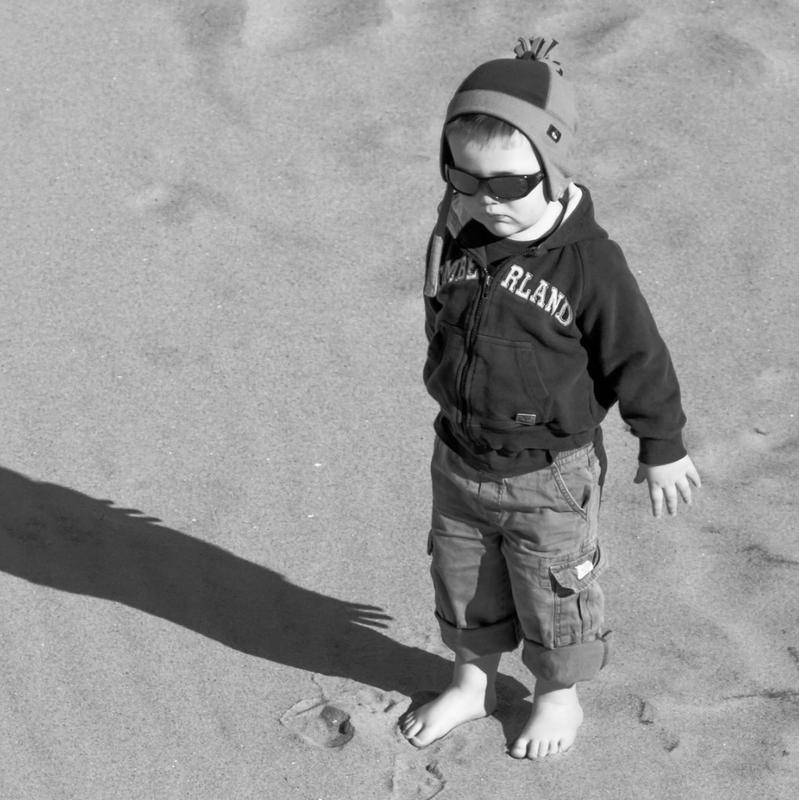} &
\includegraphics[width=0.42\linewidth, height=0.21\textheight, keepaspectratio]{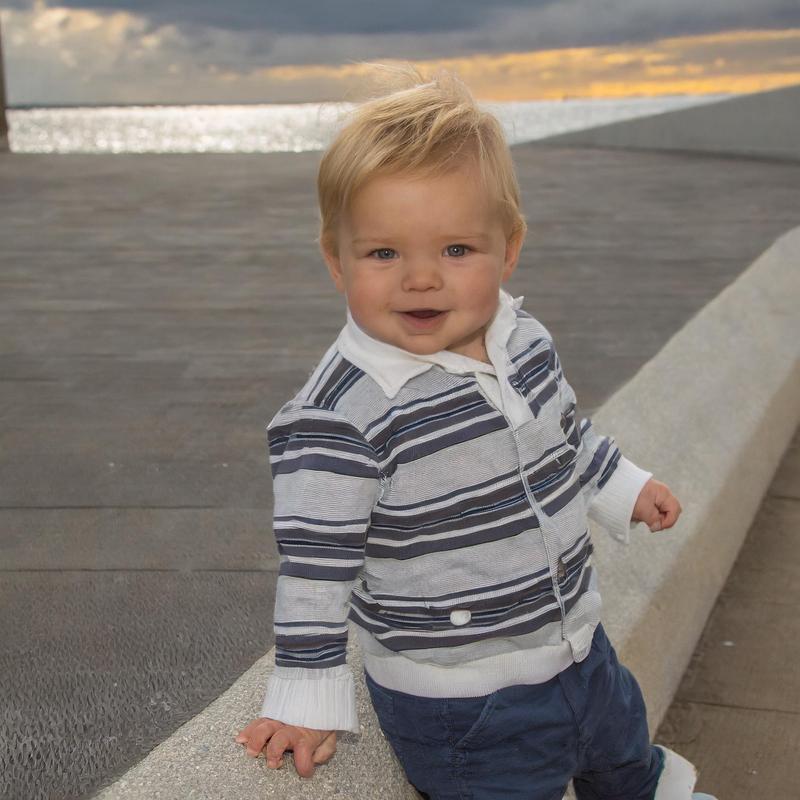} \\
\end{tabular}
\caption*{Pairs 13--16. Real photographs (left), matched AI-generated images (right).}
\end{figure}
\clearpage

\begin{figure}[h!]
\centering
\renewcommand{\arraystretch}{0.6}
\begin{tabular}{cc}
\includegraphics[width=0.42\linewidth, height=0.21\textheight, keepaspectratio]{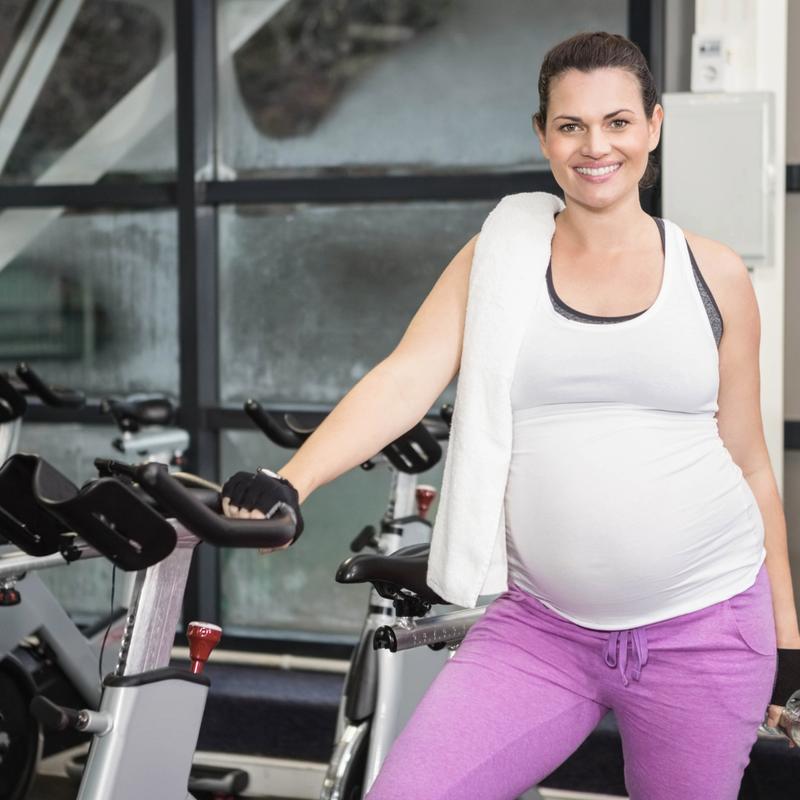} &
\includegraphics[width=0.42\linewidth, height=0.21\textheight, keepaspectratio]{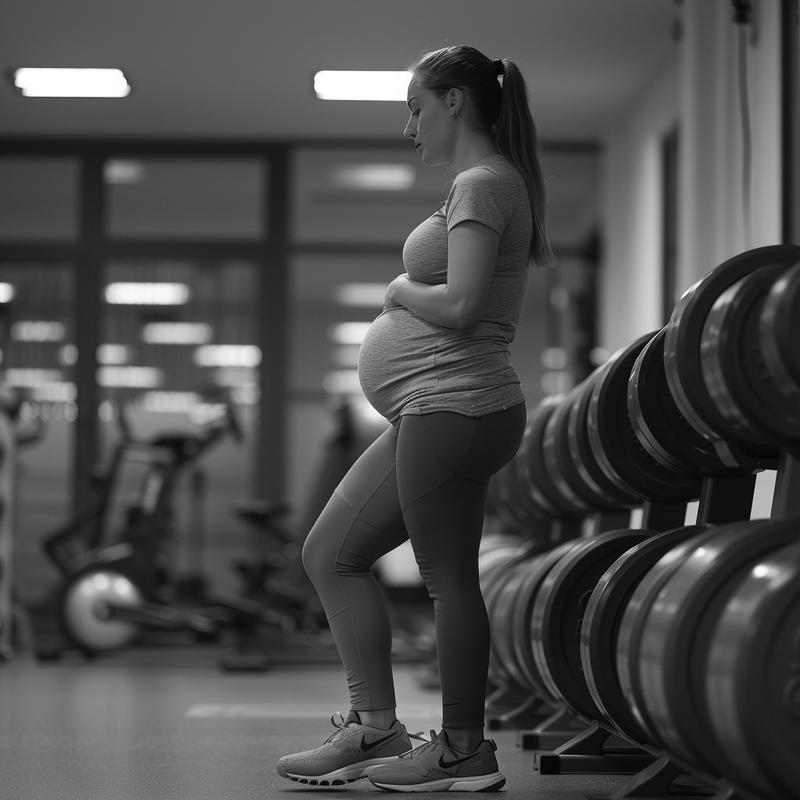} \\
\includegraphics[width=0.42\linewidth, height=0.21\textheight, keepaspectratio]{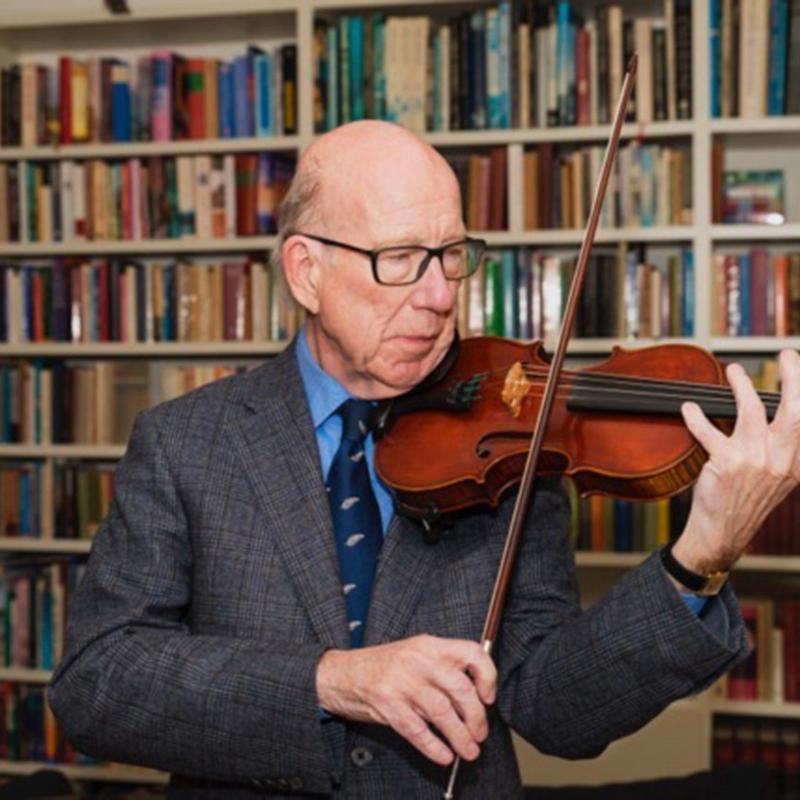} &
\includegraphics[width=0.42\linewidth, height=0.21\textheight, keepaspectratio]{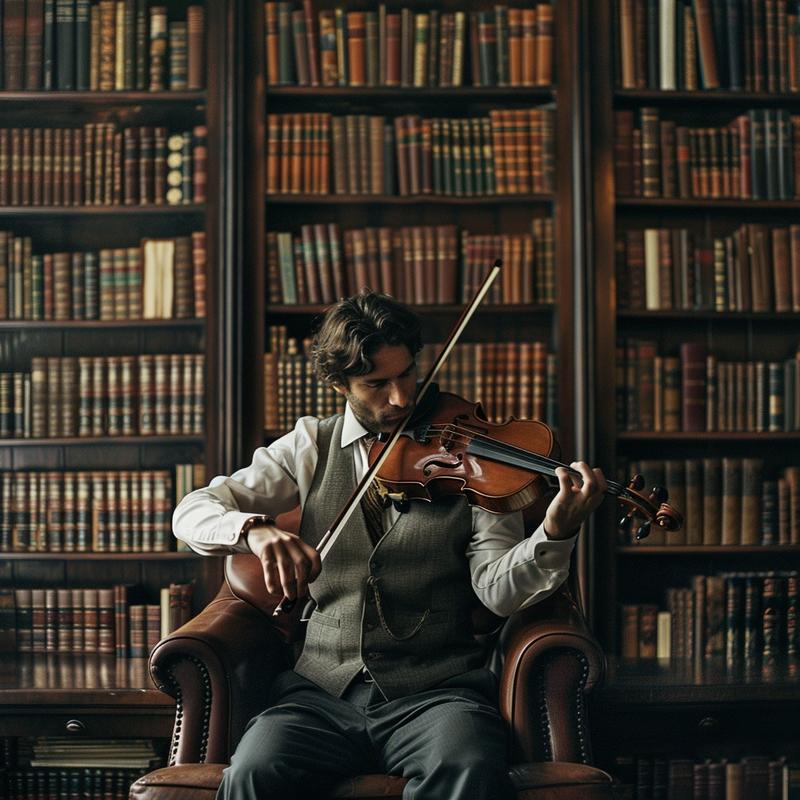} \\
\includegraphics[width=0.42\linewidth, height=0.21\textheight, keepaspectratio]{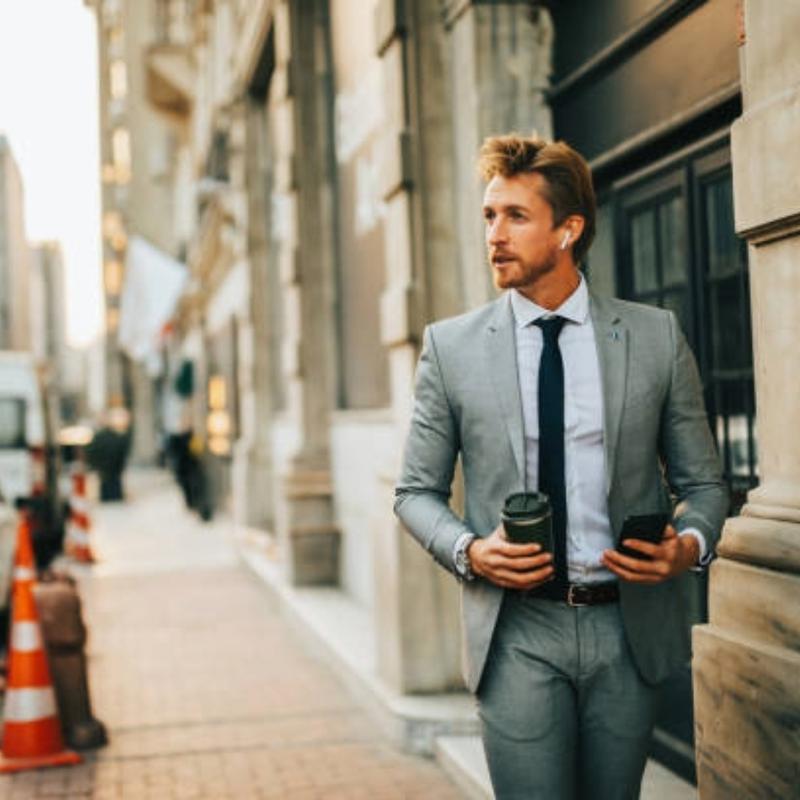} &
\includegraphics[width=0.42\linewidth, height=0.21\textheight, keepaspectratio]{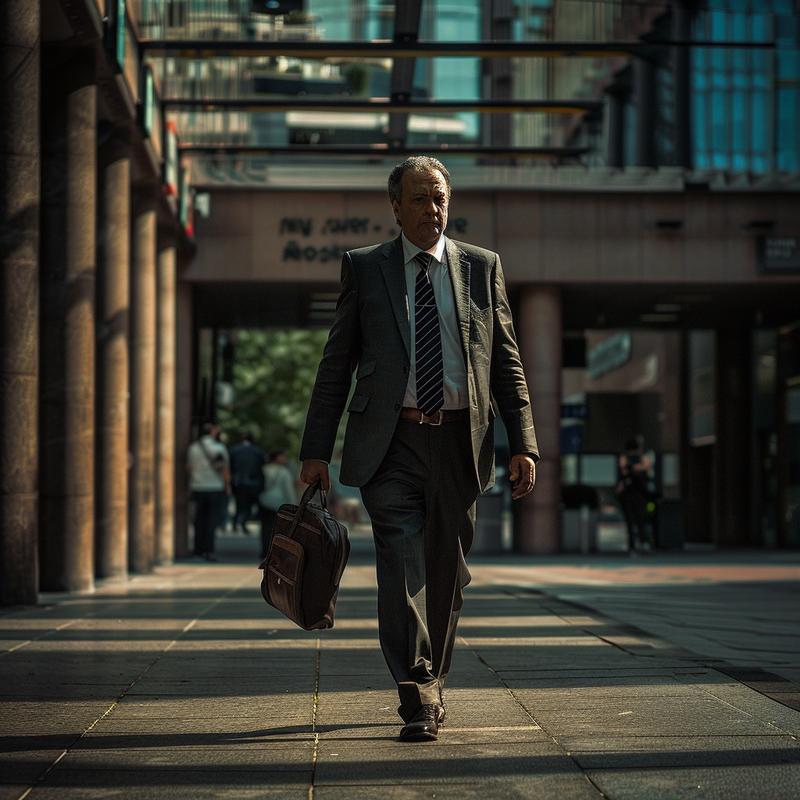} \\
\includegraphics[width=0.42\linewidth, height=0.21\textheight, keepaspectratio]{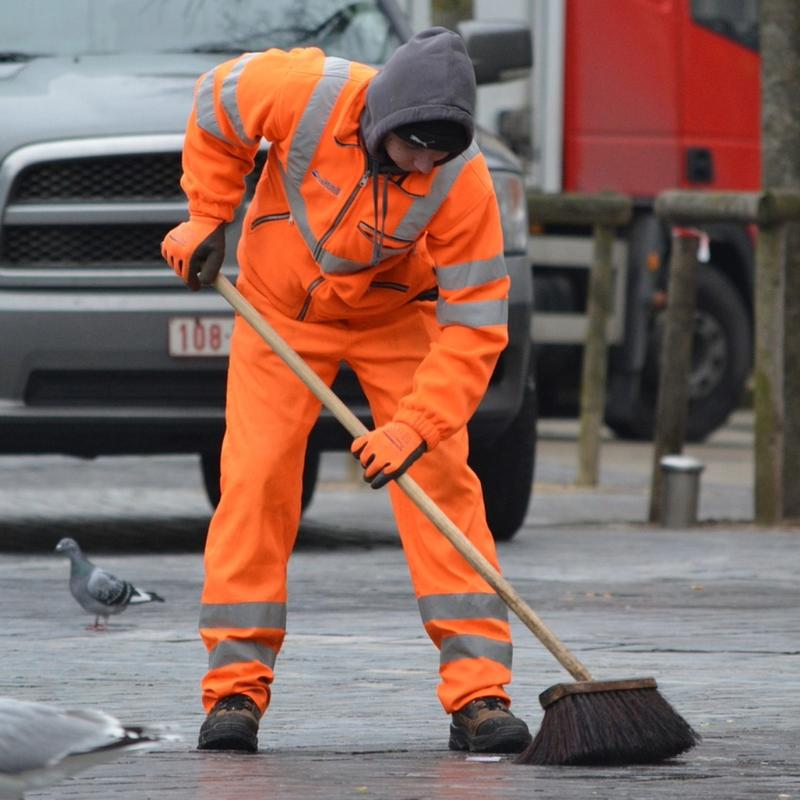} &
\includegraphics[width=0.42\linewidth, height=0.21\textheight, keepaspectratio]{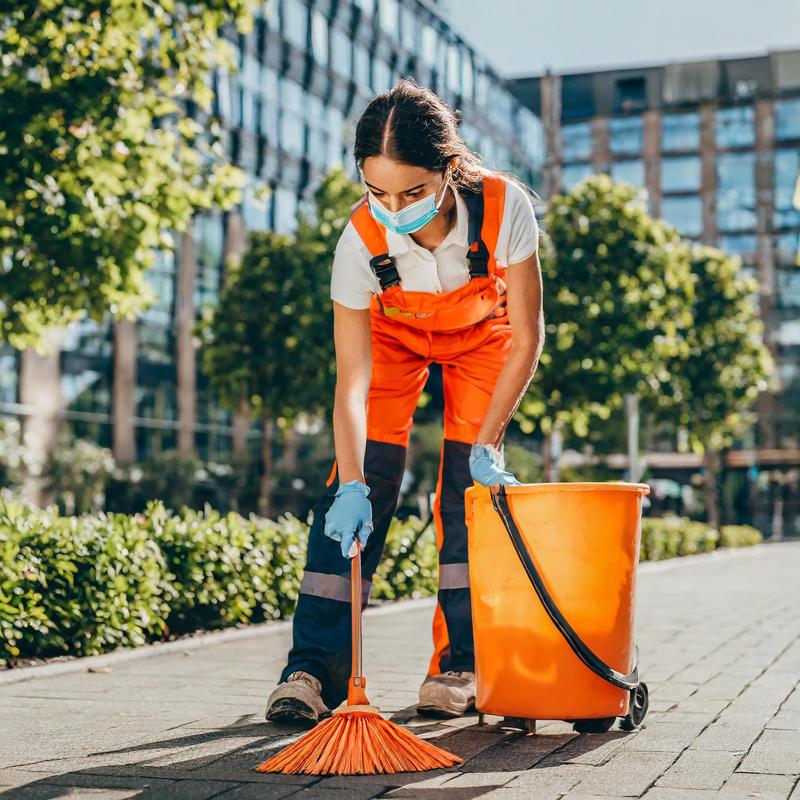} \\
\end{tabular}
\caption*{Pairs 17--20. Real photographs (left), matched AI-generated images (right).}
\end{figure}
\clearpage

\begin{figure}[h!]
\centering
\renewcommand{\arraystretch}{0.6}
\begin{tabular}{cc}
\includegraphics[width=0.42\linewidth, height=0.21\textheight, keepaspectratio]{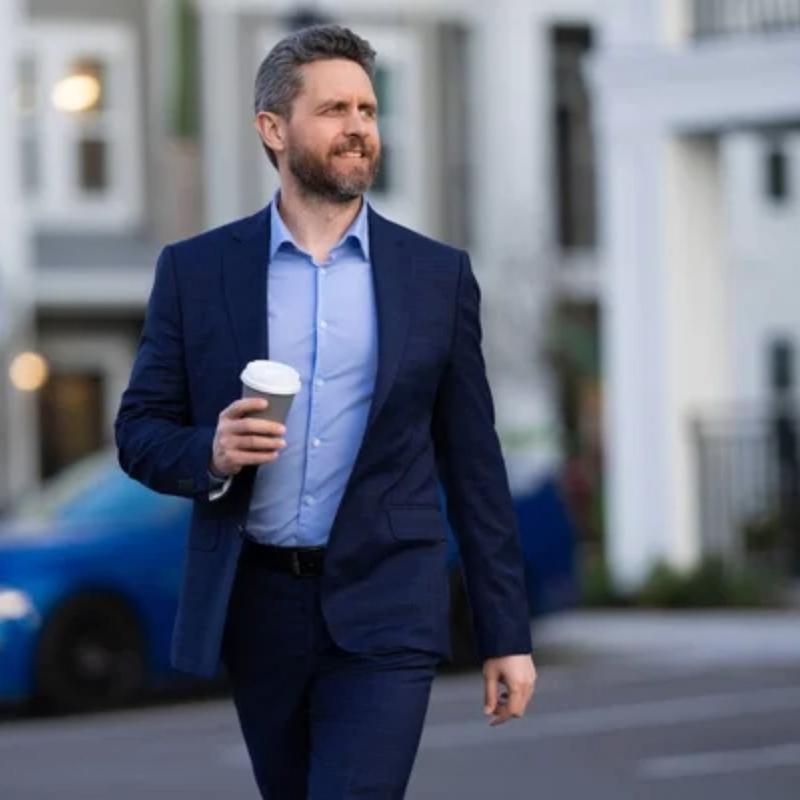} &
\includegraphics[width=0.42\linewidth, height=0.21\textheight, keepaspectratio]{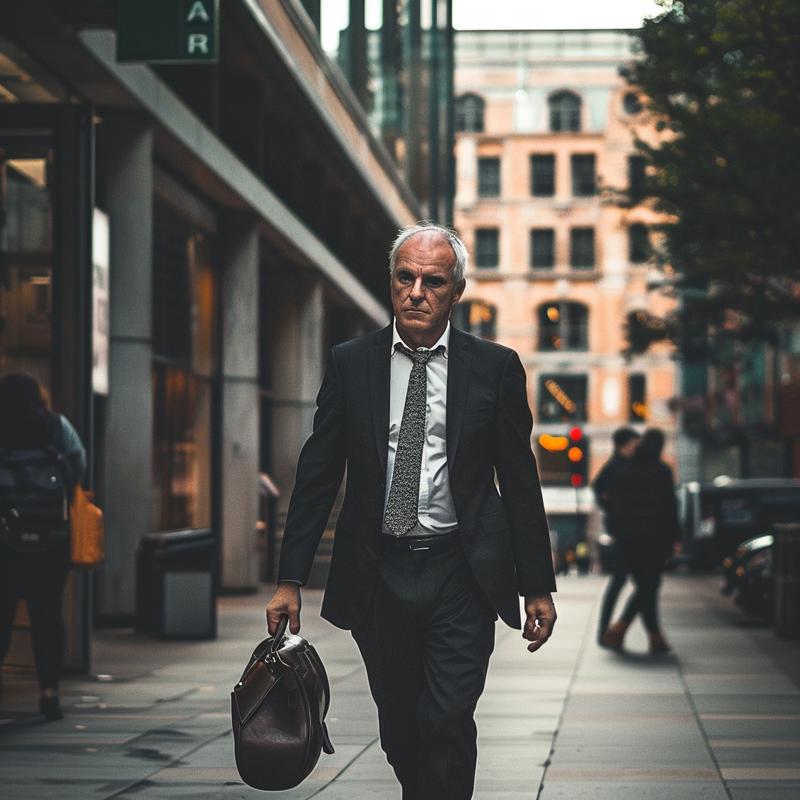} \\
\includegraphics[width=0.42\linewidth, height=0.21\textheight, keepaspectratio]{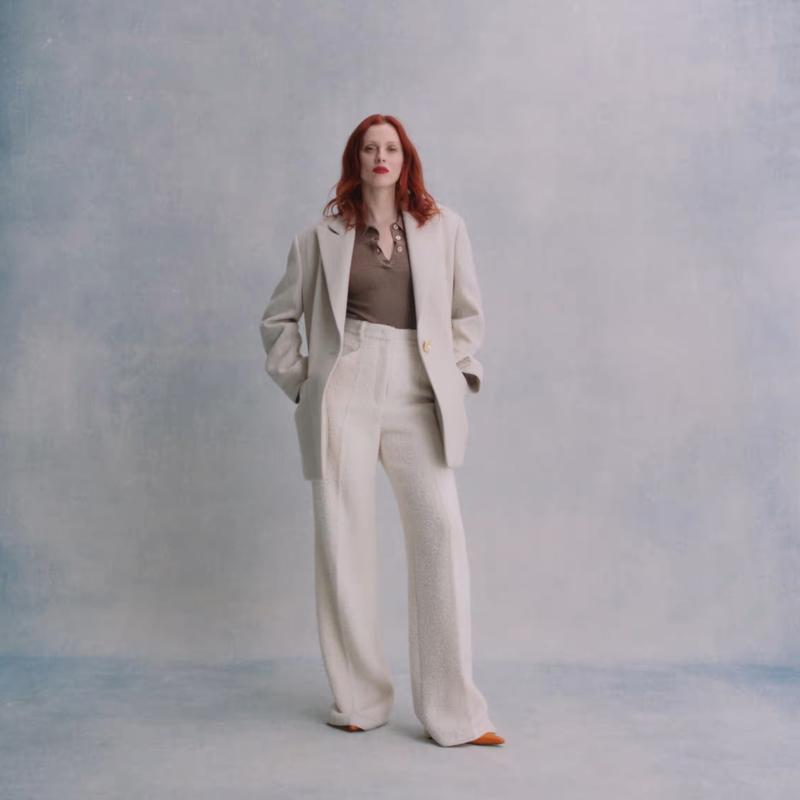} &
\includegraphics[width=0.42\linewidth, height=0.21\textheight, keepaspectratio]{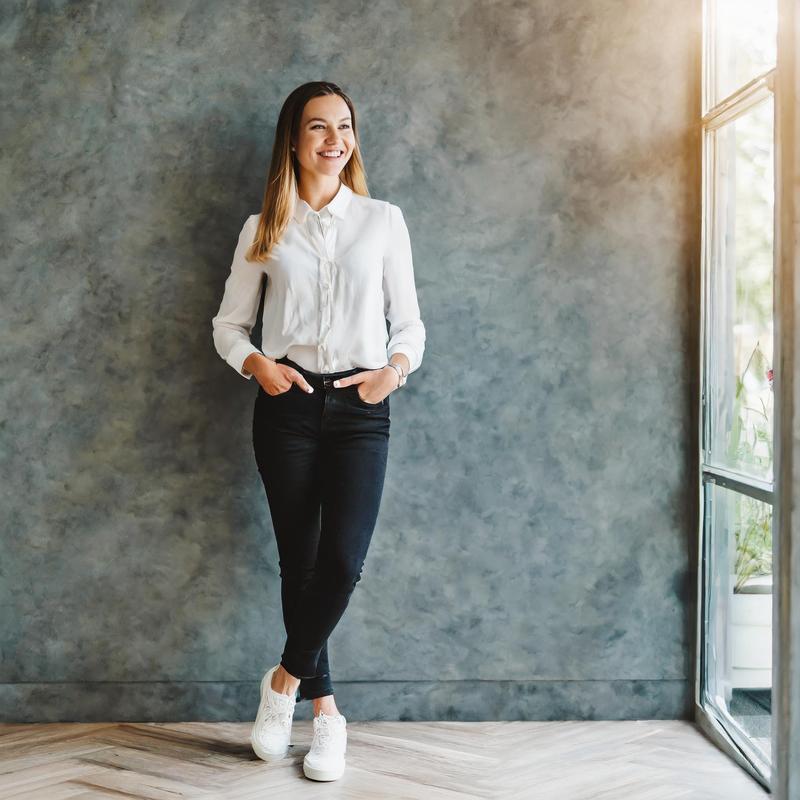} \\
\includegraphics[width=0.42\linewidth, height=0.21\textheight, keepaspectratio]{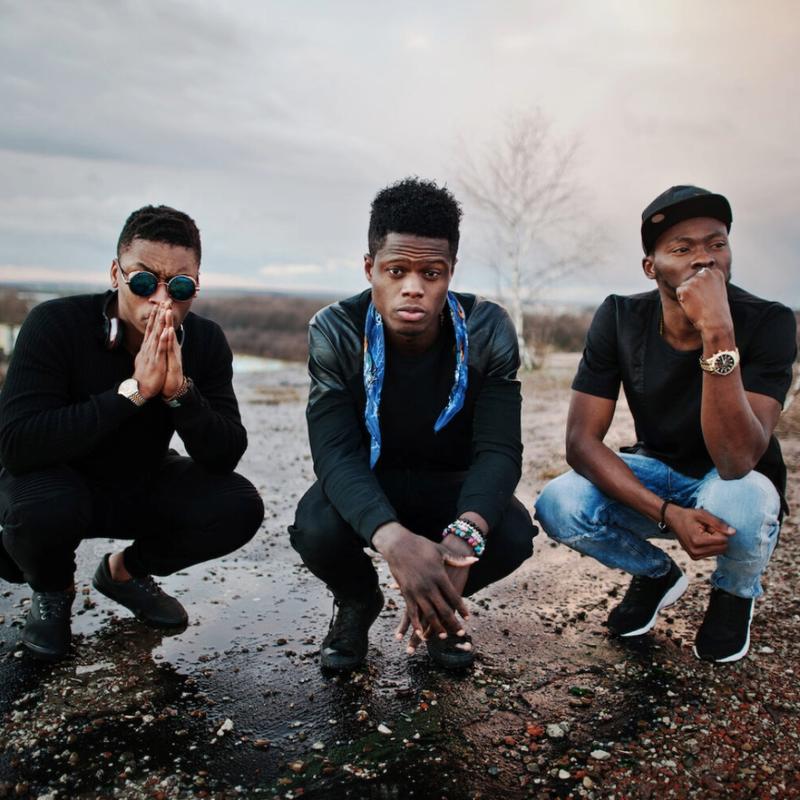} &
\includegraphics[width=0.42\linewidth, height=0.21\textheight, keepaspectratio]{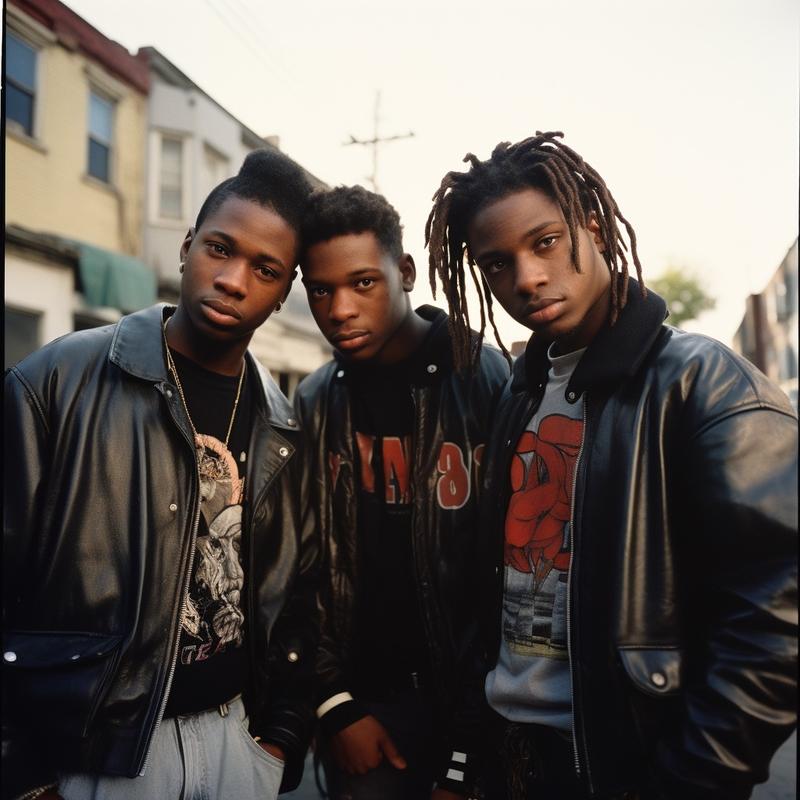} \\
\includegraphics[width=0.42\linewidth, height=0.21\textheight, keepaspectratio]{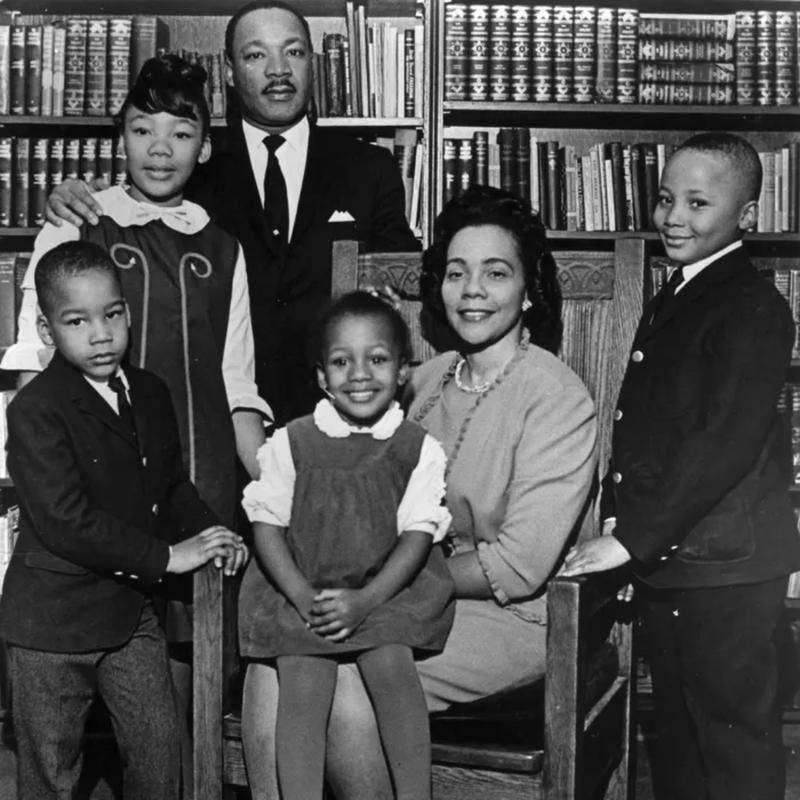} &
\includegraphics[width=0.42\linewidth, height=0.21\textheight, keepaspectratio]{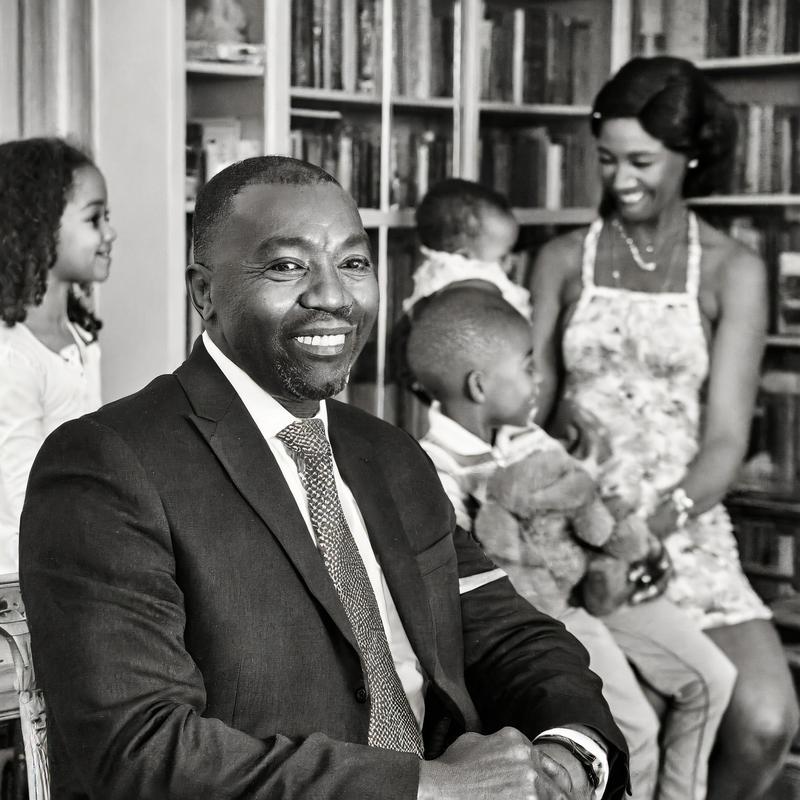} \\
\end{tabular}
\caption*{Pairs 21--24. Real photographs (left), matched AI-generated images (right).}
\end{figure}
\clearpage

\begin{figure}[h!]
\centering
\renewcommand{\arraystretch}{0.6}
\begin{tabular}{cc}
\includegraphics[width=0.42\linewidth, height=0.21\textheight, keepaspectratio]{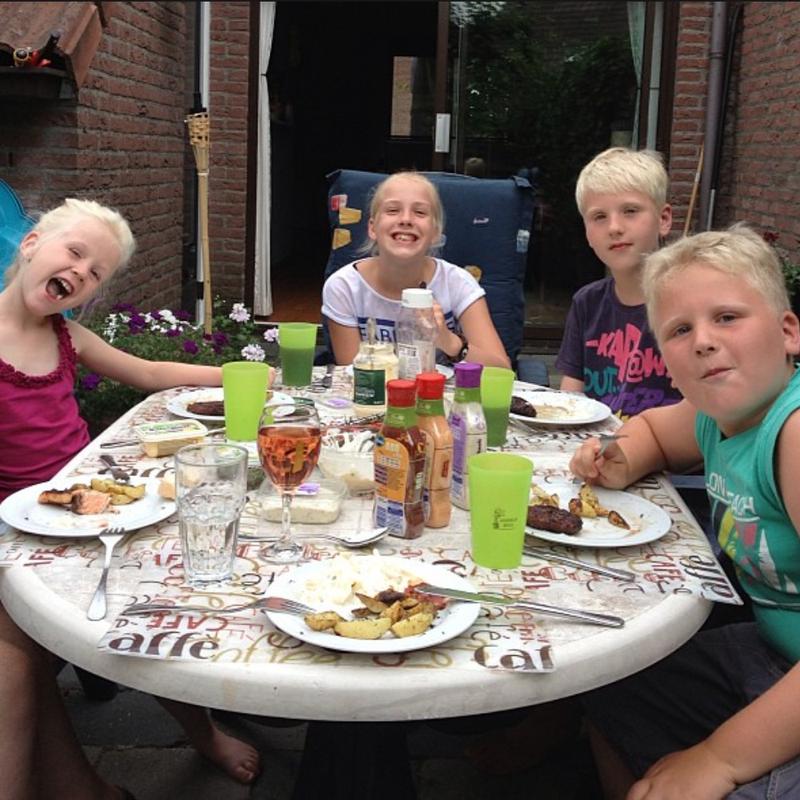} &
\includegraphics[width=0.42\linewidth, height=0.21\textheight, keepaspectratio]{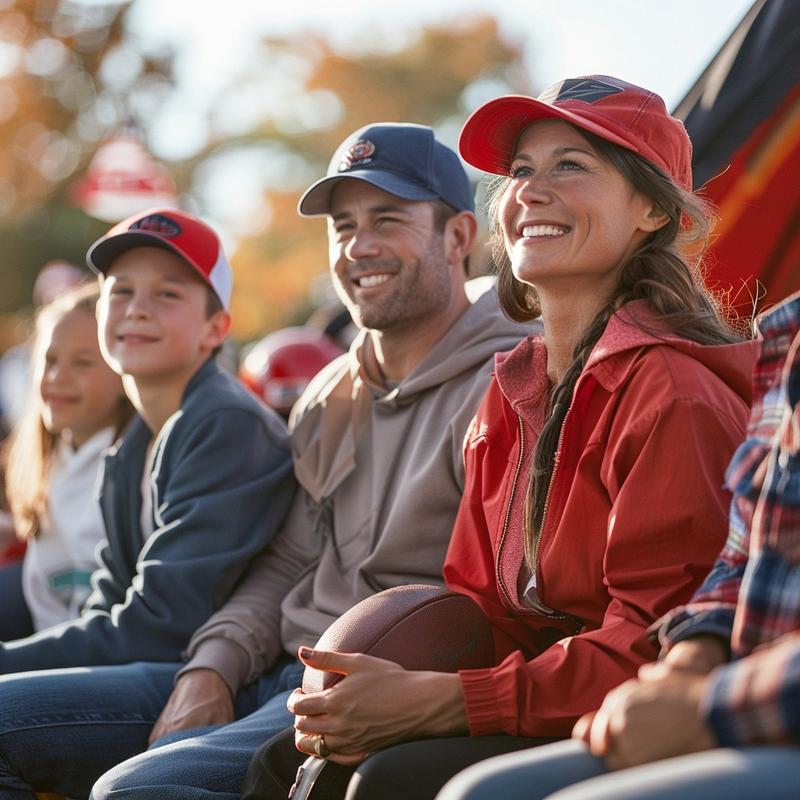} \\
\includegraphics[width=0.42\linewidth, height=0.21\textheight, keepaspectratio]{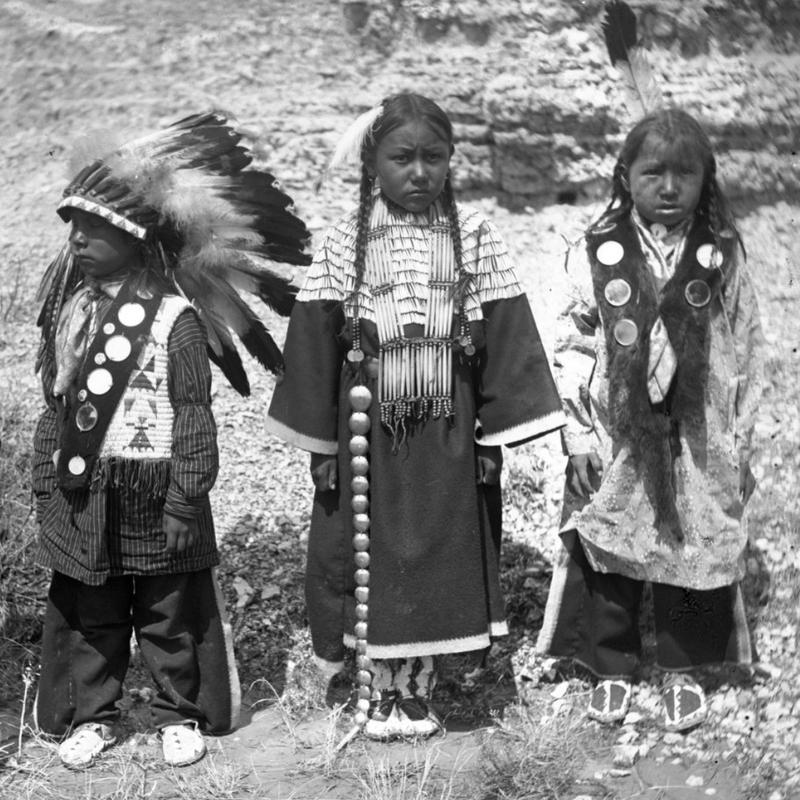} &
\includegraphics[width=0.42\linewidth, height=0.21\textheight, keepaspectratio]{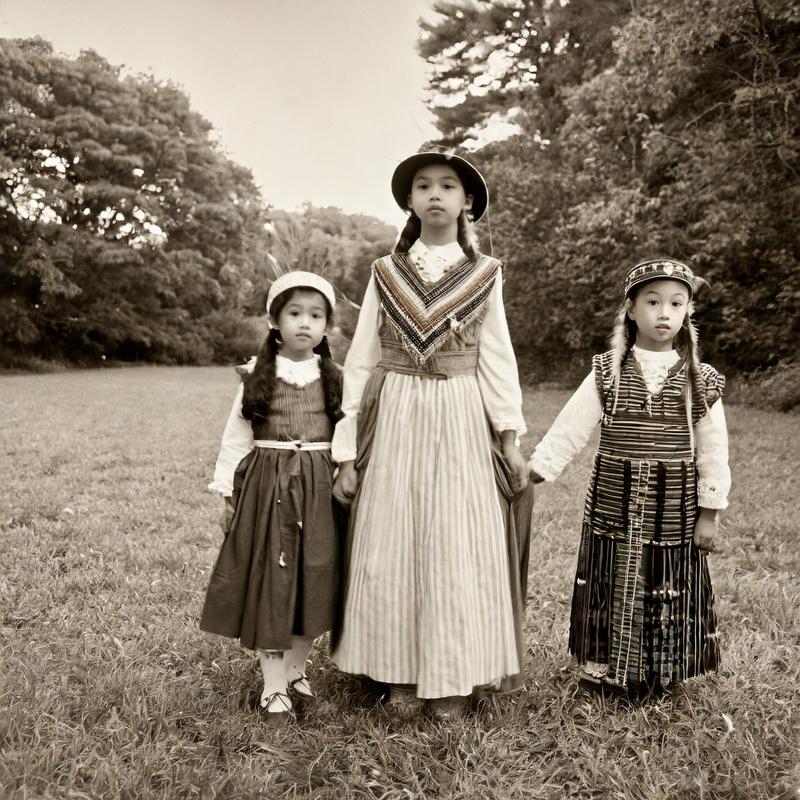} \\
\includegraphics[width=0.42\linewidth, height=0.21\textheight, keepaspectratio]{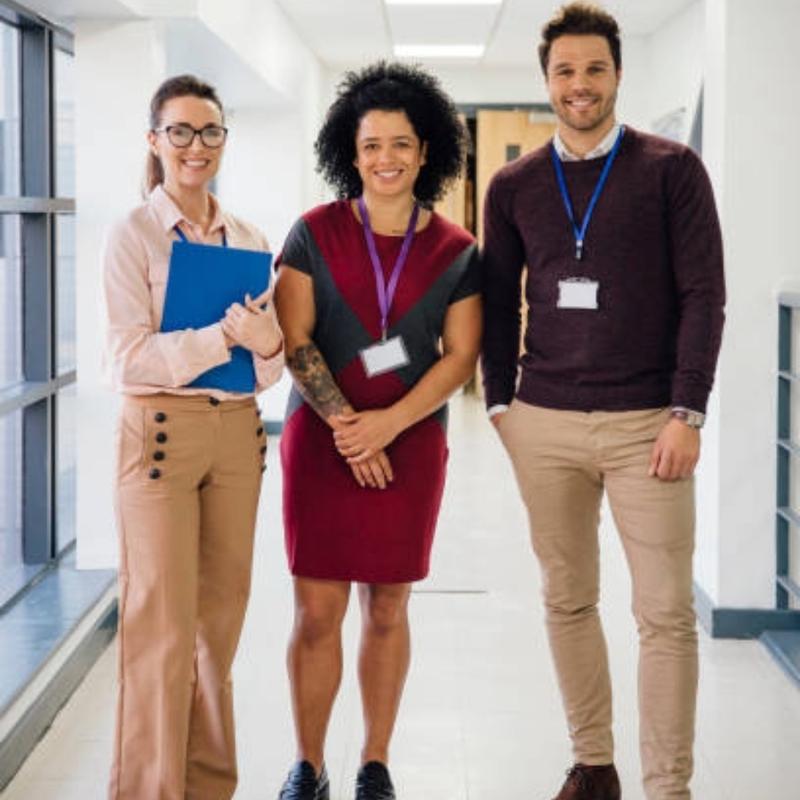} &
\includegraphics[width=0.42\linewidth, height=0.21\textheight, keepaspectratio]{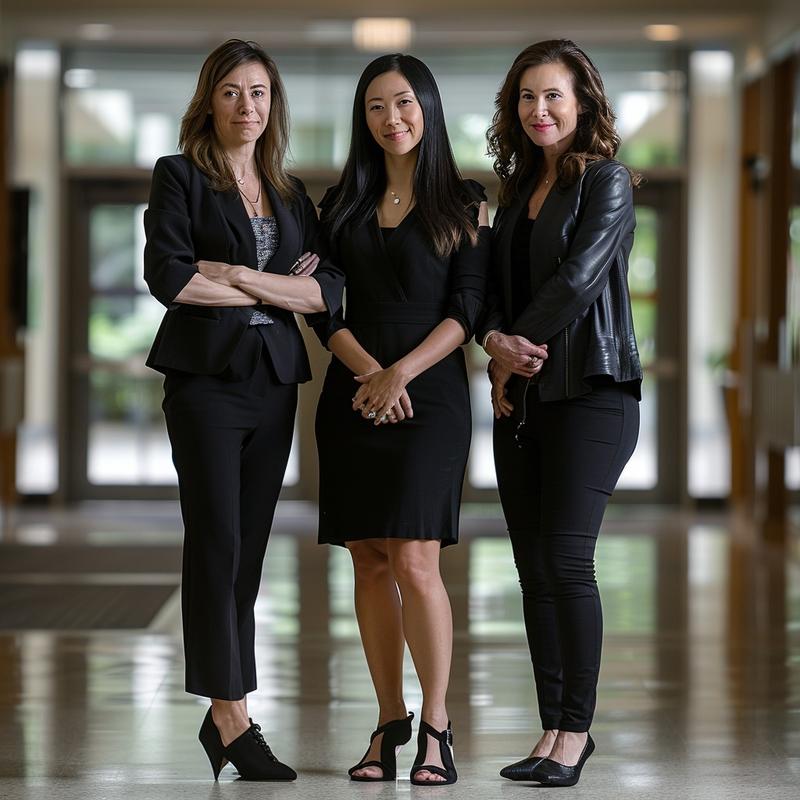} \\
\includegraphics[width=0.42\linewidth, height=0.21\textheight, keepaspectratio]{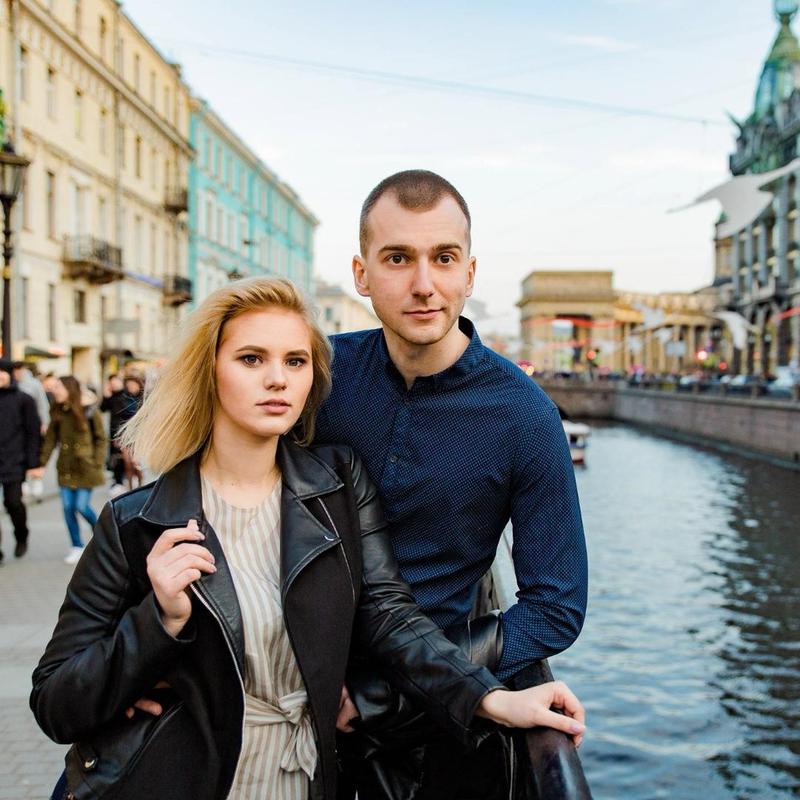} &
\includegraphics[width=0.42\linewidth, height=0.21\textheight, keepaspectratio]{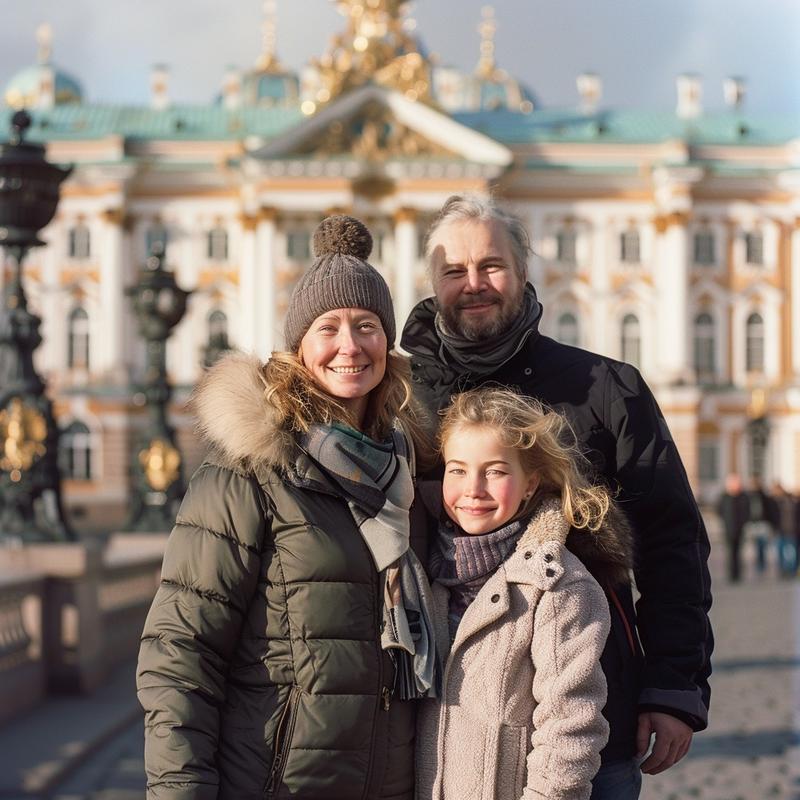} \\
\end{tabular}
\caption*{Pairs 25--28. Real photographs (left), matched AI-generated images (right).}
\end{figure}
\clearpage

\begin{figure}[h!]
\centering
\renewcommand{\arraystretch}{0.6}
\begin{tabular}{cc}
\includegraphics[width=0.42\linewidth, height=0.21\textheight, keepaspectratio]{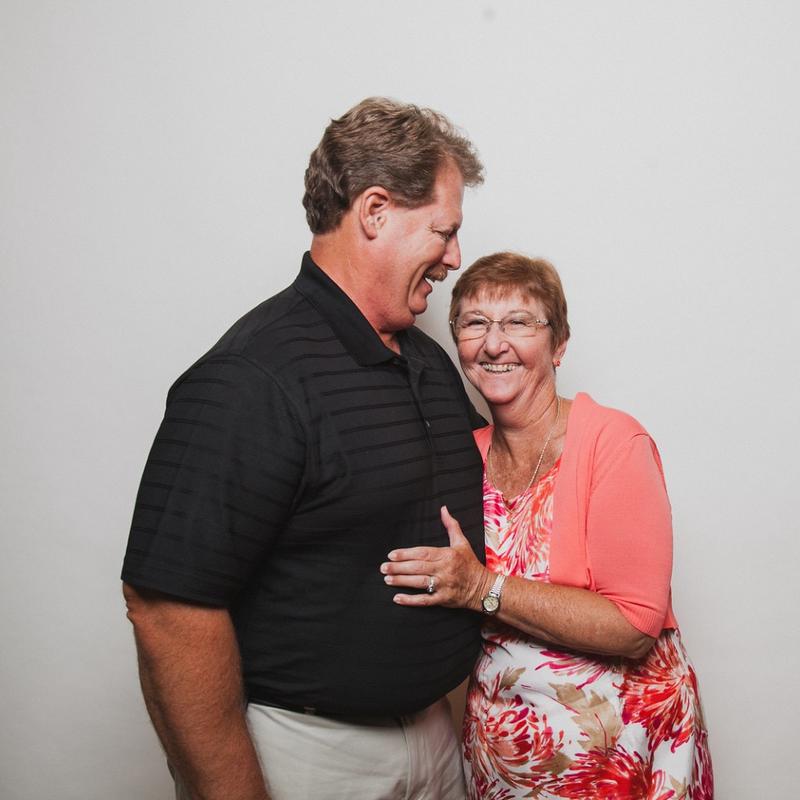} &
\includegraphics[width=0.42\linewidth, height=0.21\textheight, keepaspectratio]{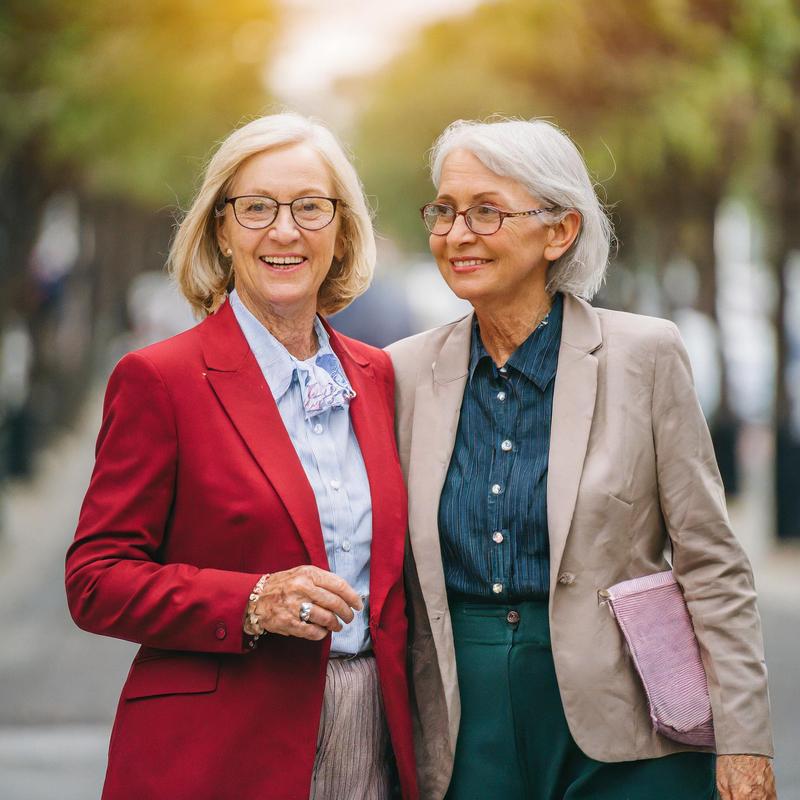} \\
\includegraphics[width=0.42\linewidth, height=0.21\textheight, keepaspectratio]{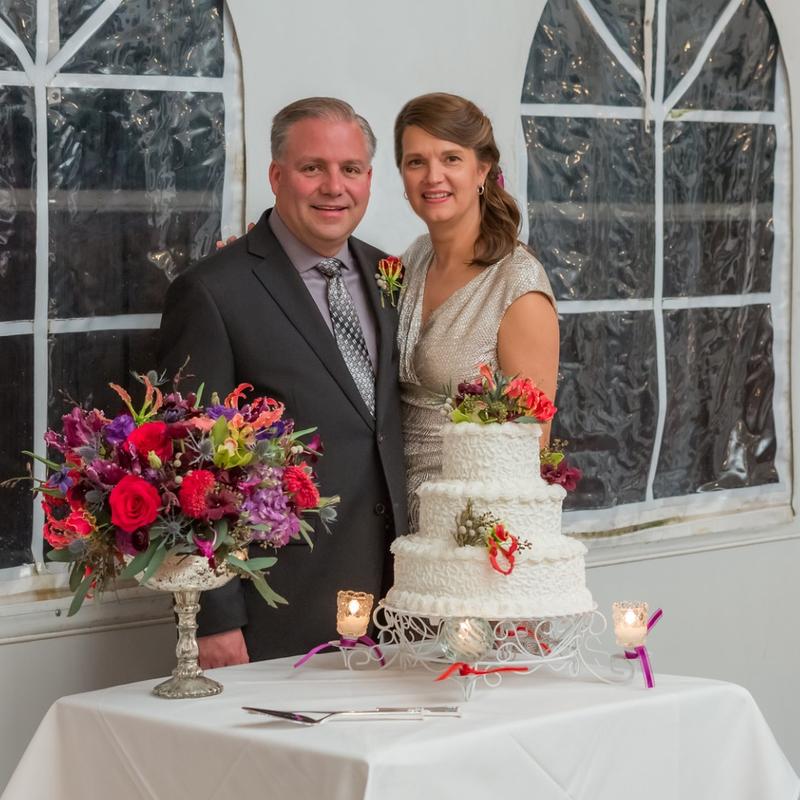} &
\includegraphics[width=0.42\linewidth, height=0.21\textheight, keepaspectratio]{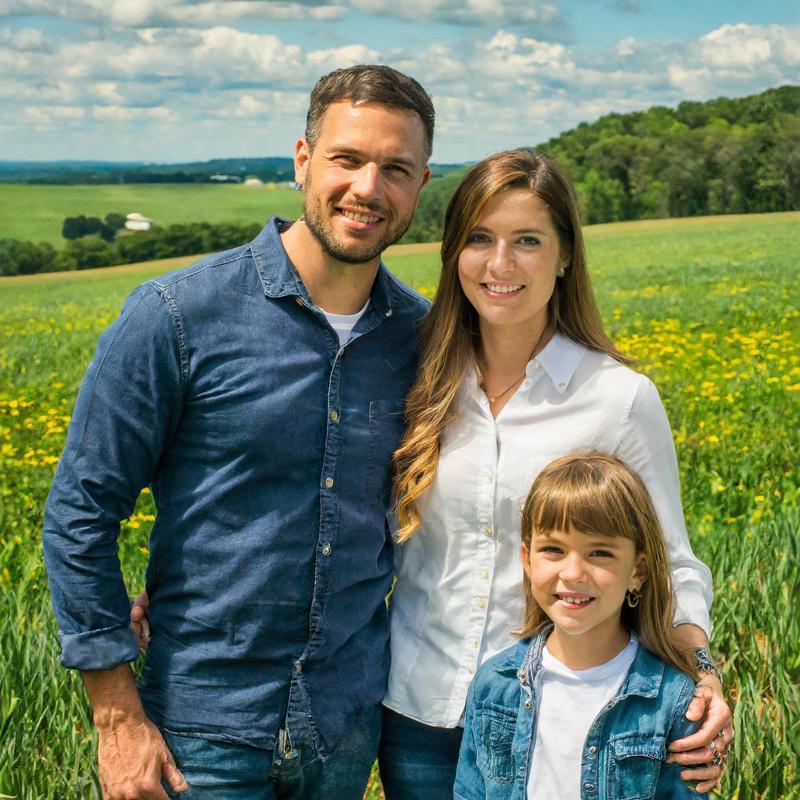} \\
\includegraphics[width=0.42\linewidth, height=0.21\textheight, keepaspectratio]{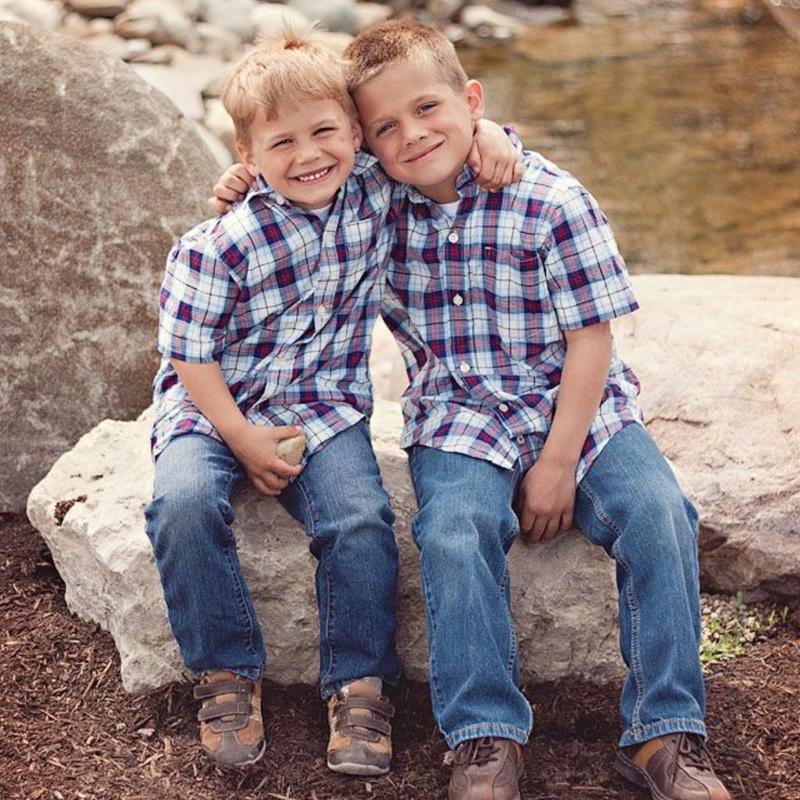} &
\includegraphics[width=0.42\linewidth, height=0.21\textheight, keepaspectratio]{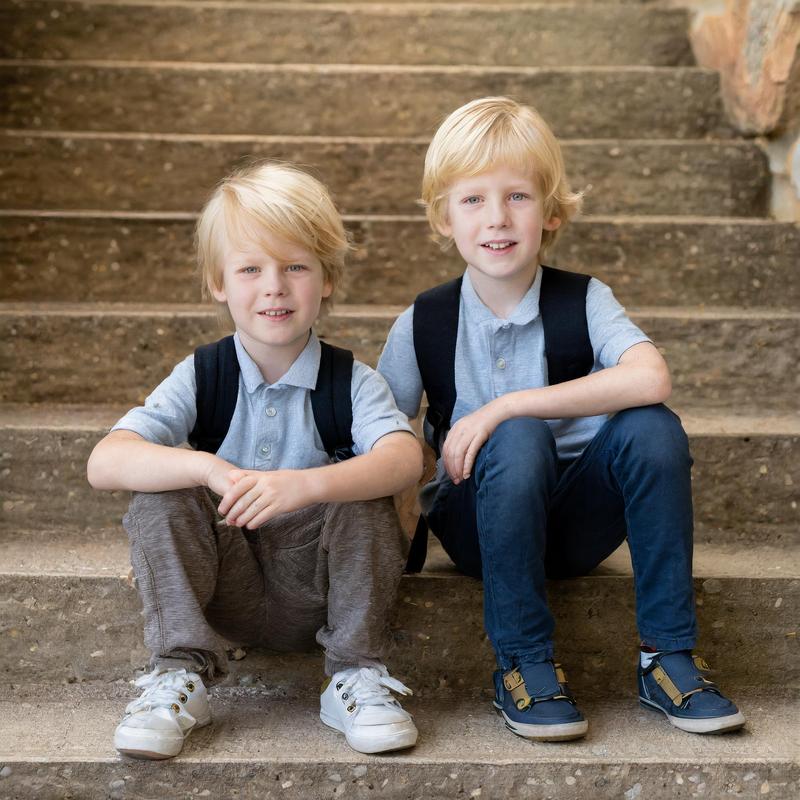} \\
\includegraphics[width=0.42\linewidth, height=0.21\textheight, keepaspectratio]{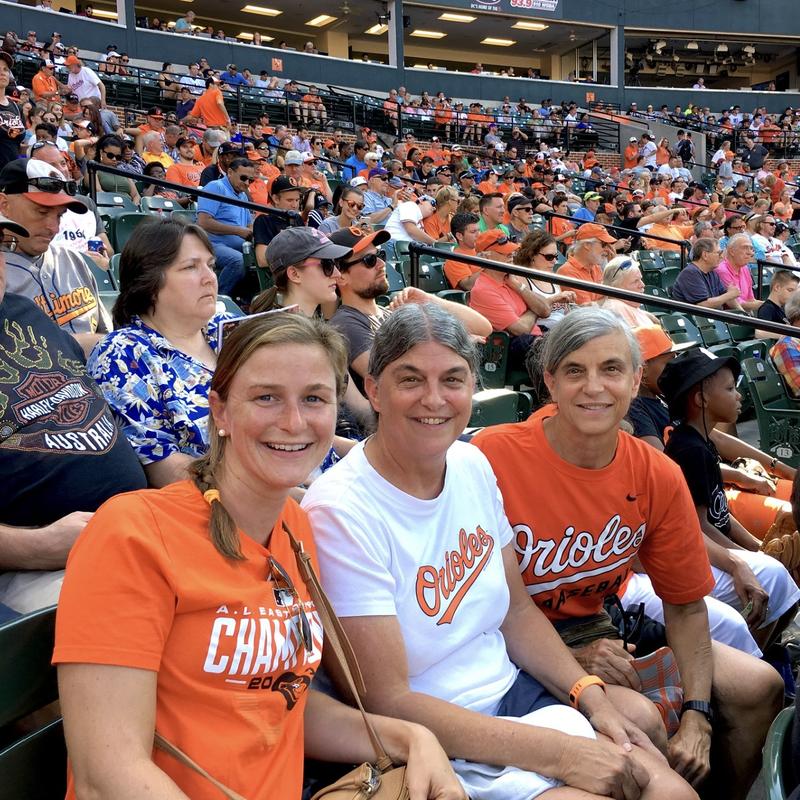} &
\includegraphics[width=0.42\linewidth, height=0.21\textheight, keepaspectratio]{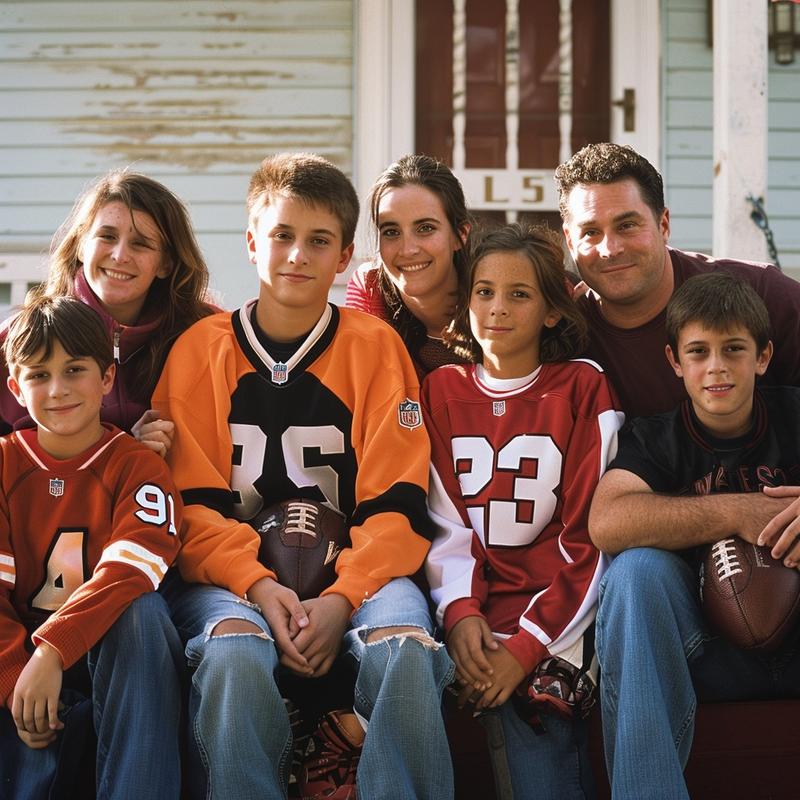} \\
\end{tabular}
\caption*{Pairs 29--32. Real photographs (left), matched AI-generated images (right).}
\end{figure}
\clearpage

\begin{figure}[h!]
\centering
\renewcommand{\arraystretch}{0.6}
\begin{tabular}{cc}
\includegraphics[width=0.42\linewidth, height=0.21\textheight, keepaspectratio]{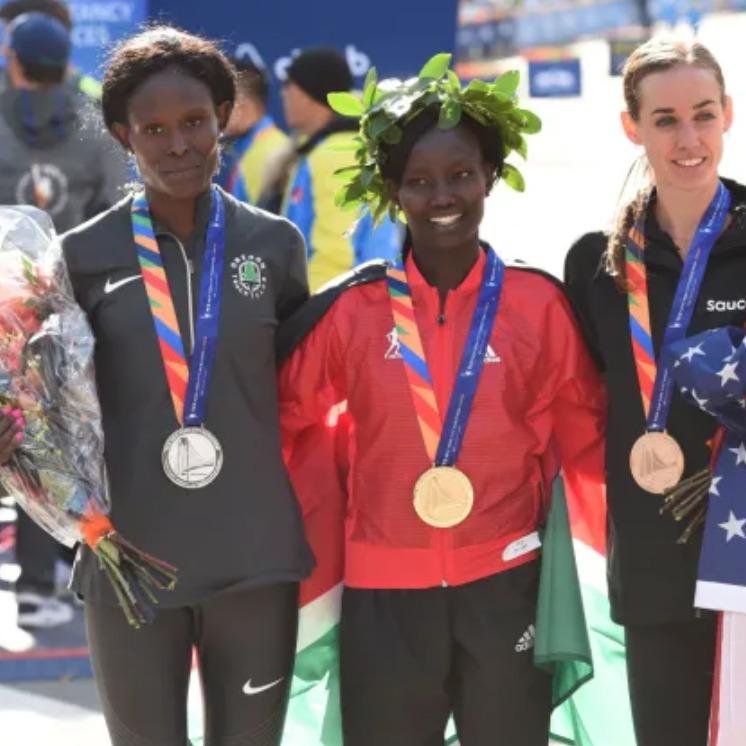} &
\includegraphics[width=0.42\linewidth, height=0.21\textheight, keepaspectratio]{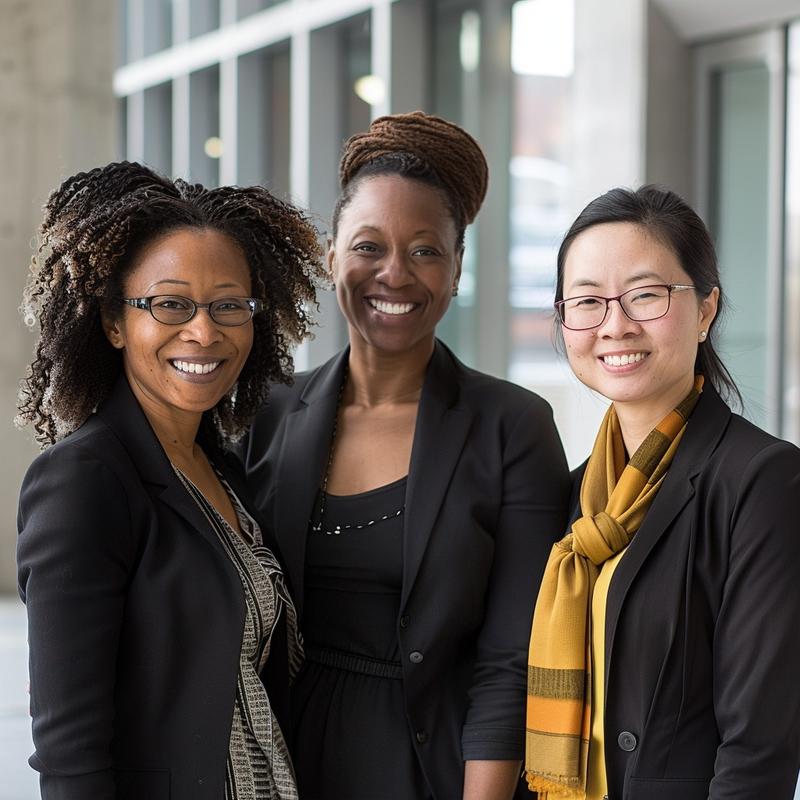} \\
\includegraphics[width=0.42\linewidth, height=0.21\textheight, keepaspectratio]{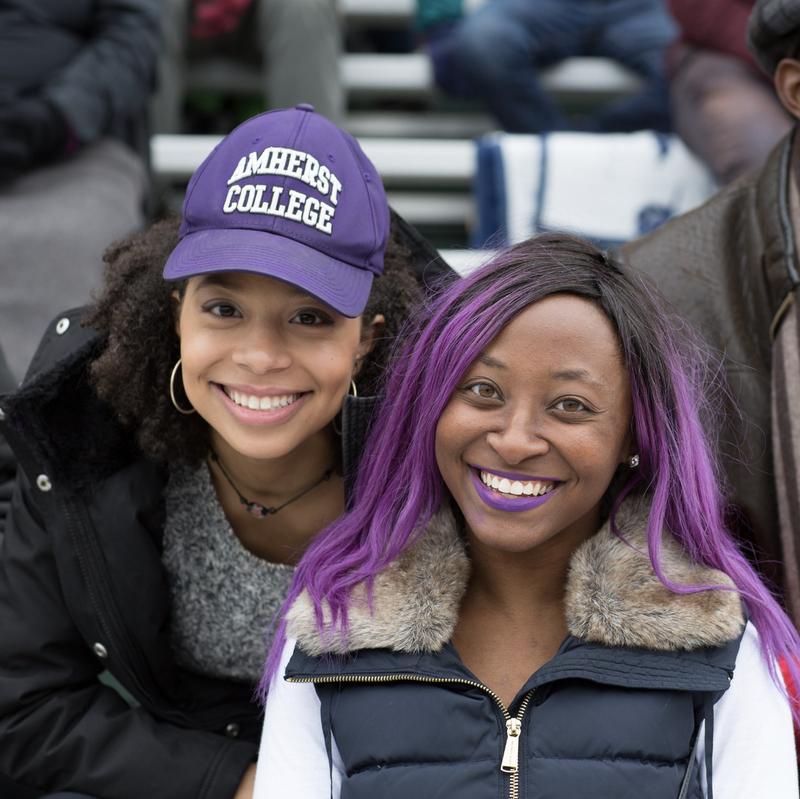} &
\includegraphics[width=0.42\linewidth, height=0.21\textheight, keepaspectratio]{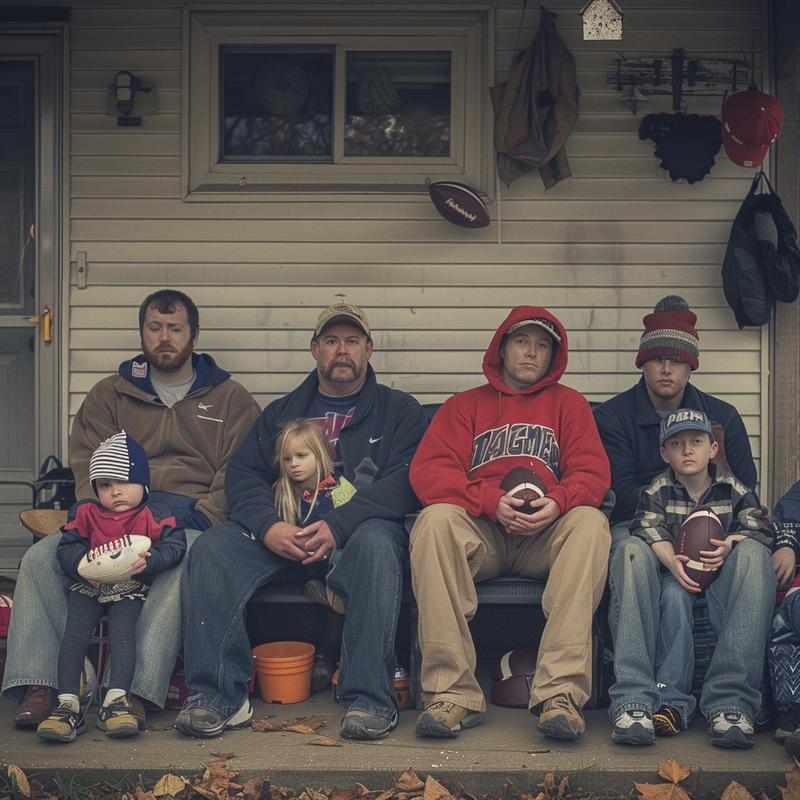} \\
\includegraphics[width=0.42\linewidth, height=0.21\textheight, keepaspectratio]{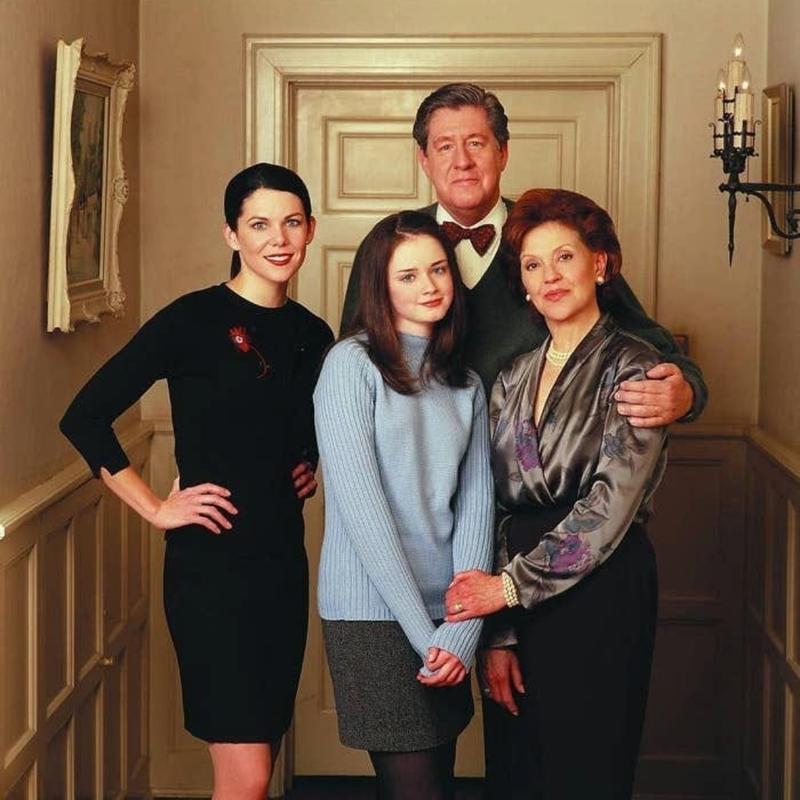} &
\includegraphics[width=0.42\linewidth, height=0.21\textheight, keepaspectratio]{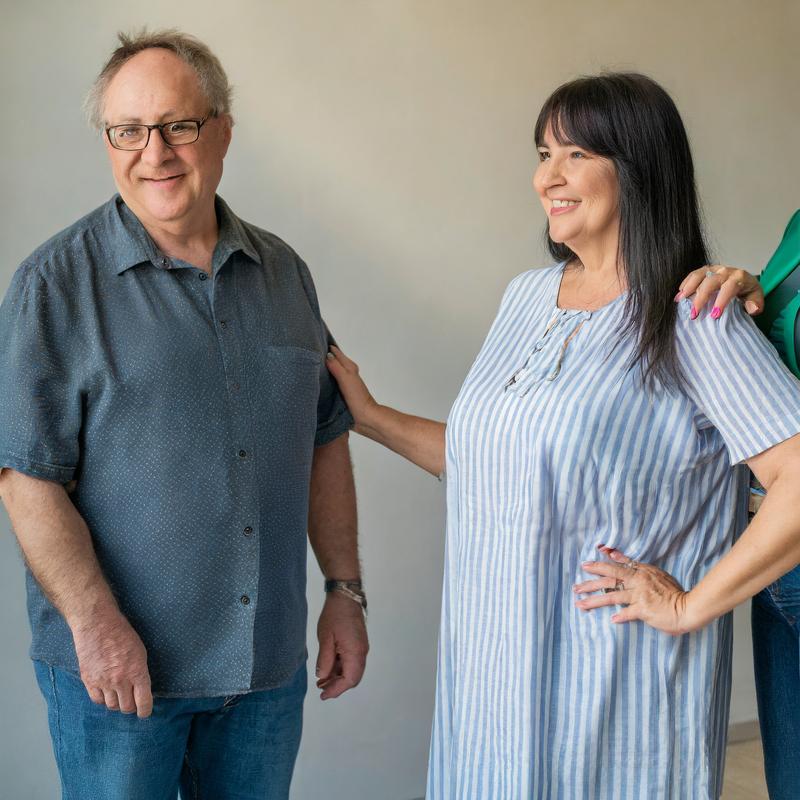} \\
\includegraphics[width=0.42\linewidth, height=0.21\textheight, keepaspectratio]{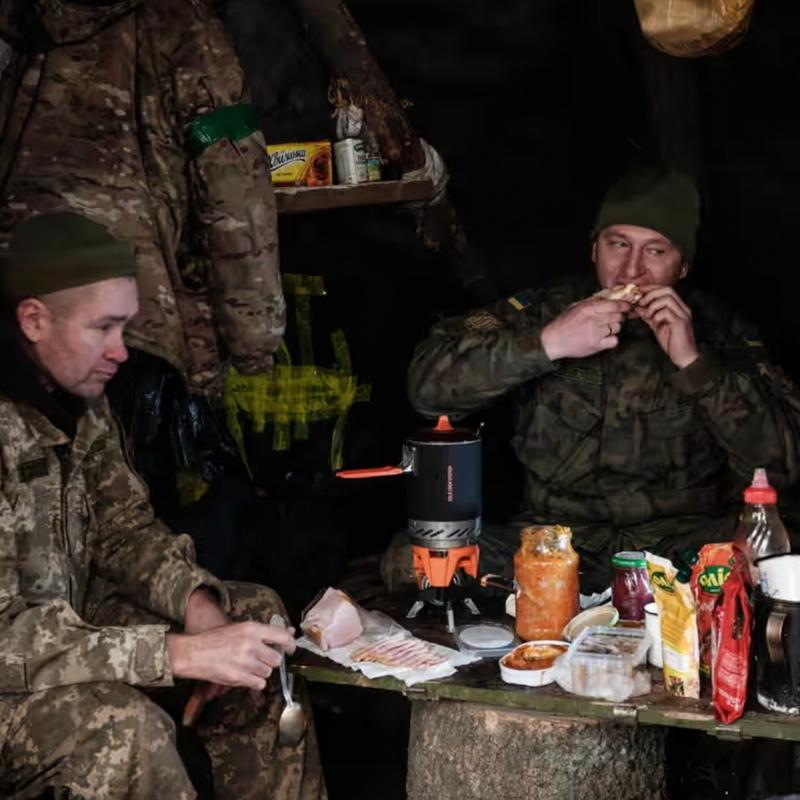} &
\includegraphics[width=0.42\linewidth, height=0.21\textheight, keepaspectratio]{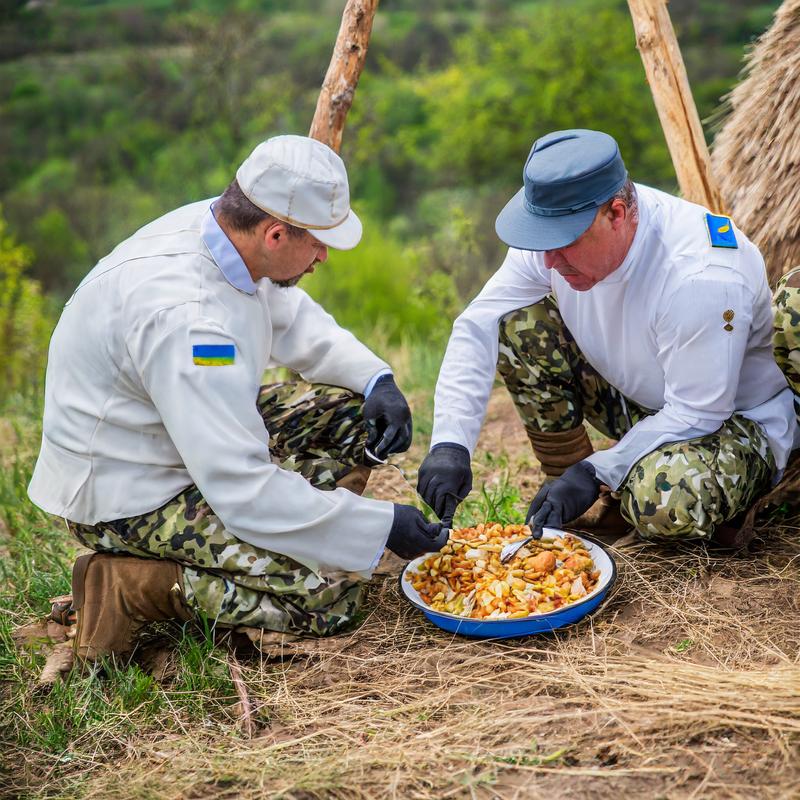} \\
\end{tabular}
\caption*{Pairs 33--36. Real photographs (left), matched AI-generated images (right).}
\end{figure}
\clearpage

\begin{figure}[h!]
\centering
\renewcommand{\arraystretch}{0.6}
\begin{tabular}{cc}
\includegraphics[width=0.42\linewidth, height=0.21\textheight, keepaspectratio]{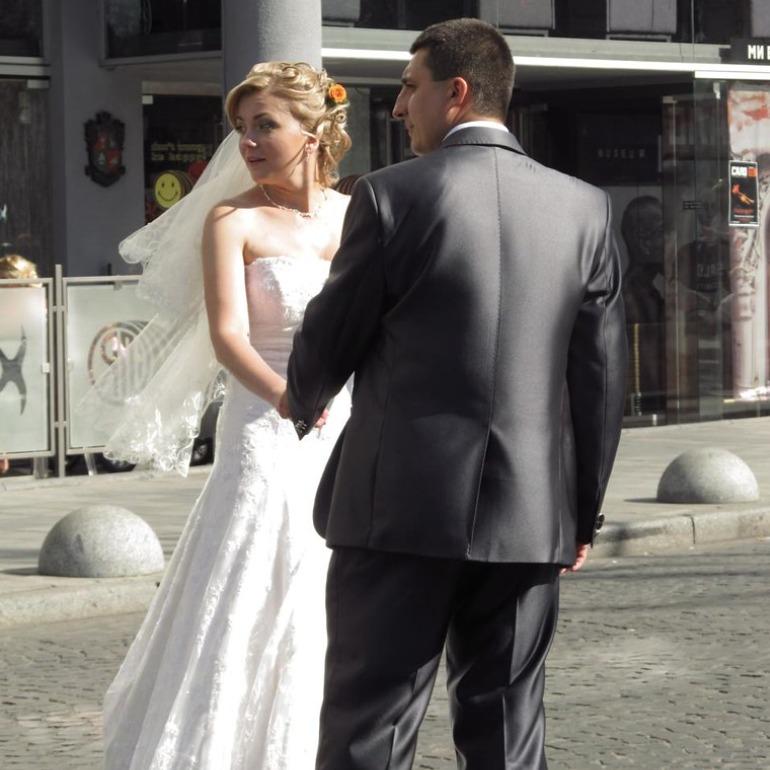} &
\includegraphics[width=0.42\linewidth, height=0.21\textheight, keepaspectratio]{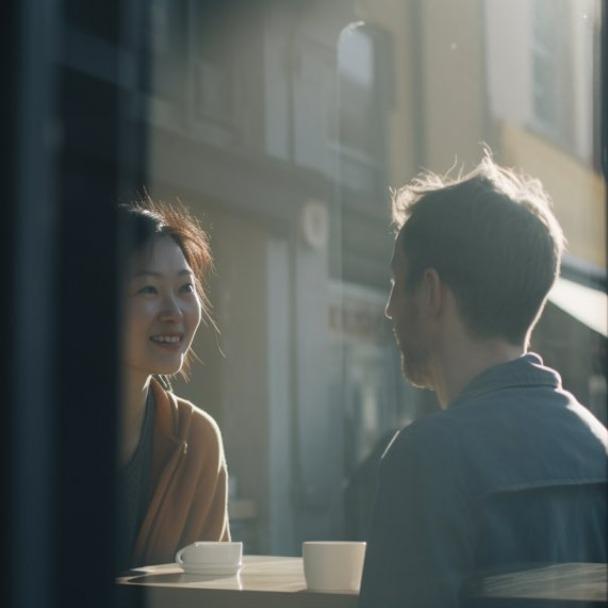} \\
\includegraphics[width=0.42\linewidth, height=0.21\textheight, keepaspectratio]{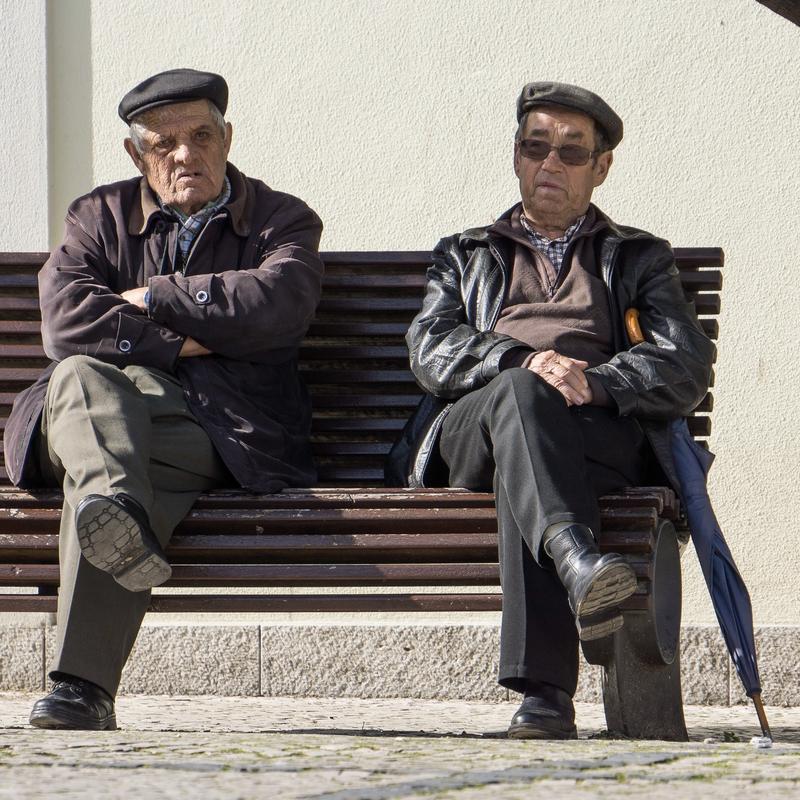} &
\includegraphics[width=0.42\linewidth, height=0.21\textheight, keepaspectratio]{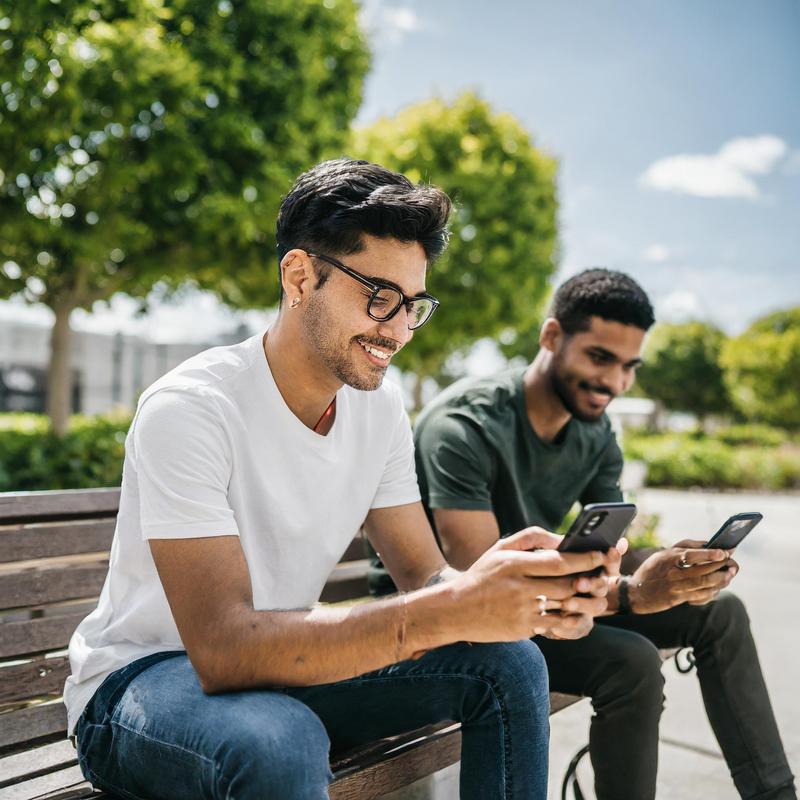} \\
\includegraphics[width=0.42\linewidth, height=0.21\textheight, keepaspectratio]{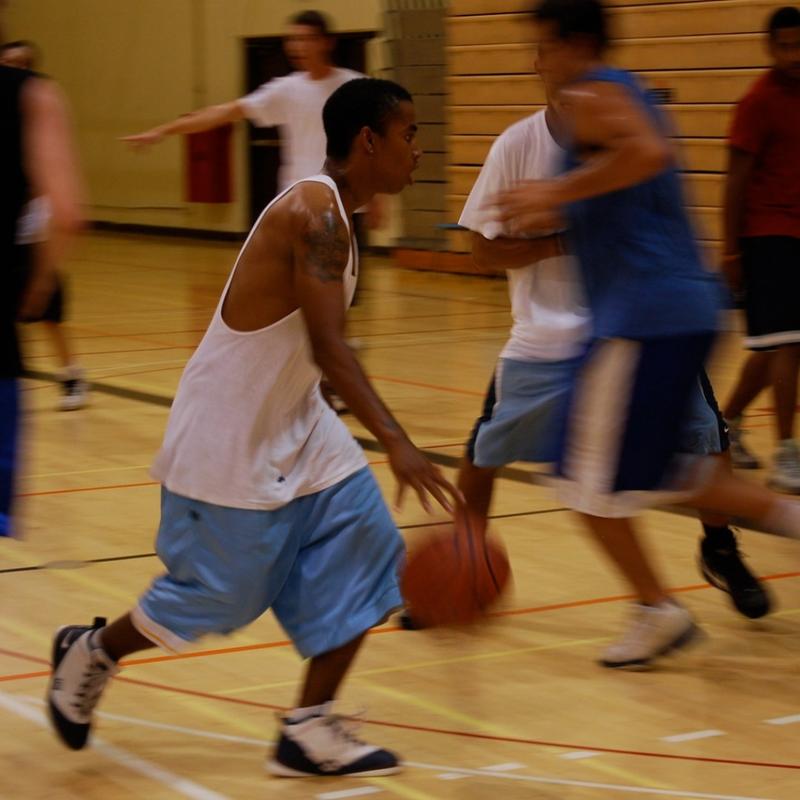} &
\includegraphics[width=0.42\linewidth, height=0.21\textheight, keepaspectratio]{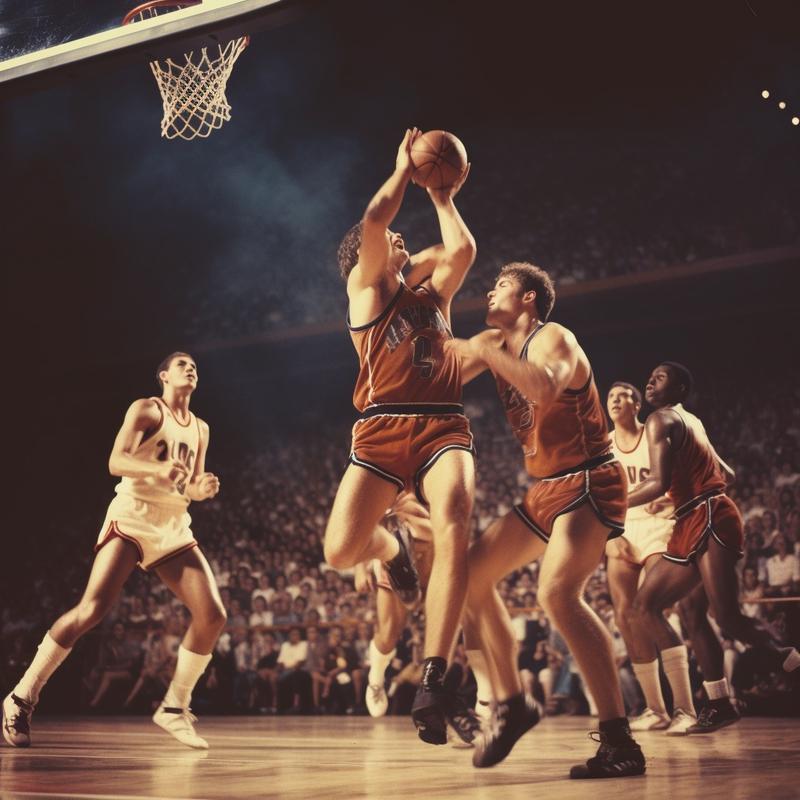} \\
\includegraphics[width=0.42\linewidth, height=0.21\textheight, keepaspectratio]{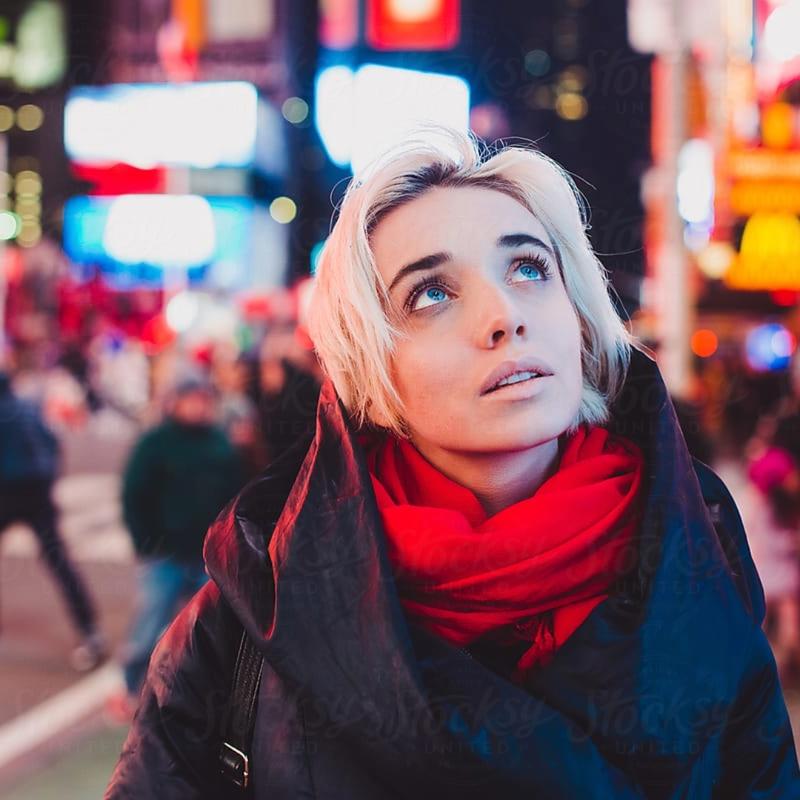} &
\includegraphics[width=0.42\linewidth, height=0.21\textheight, keepaspectratio]{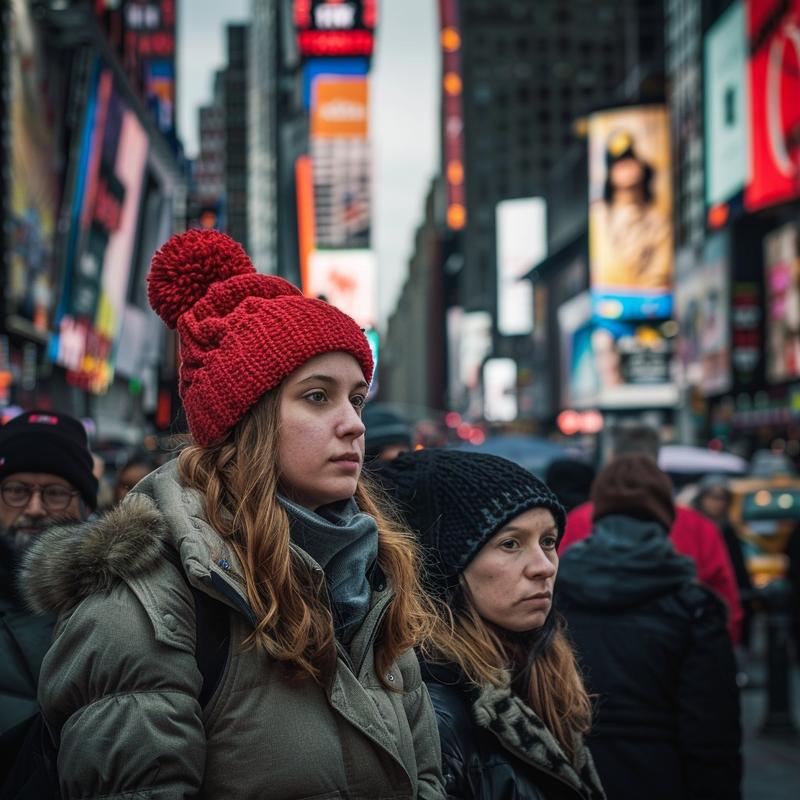} \\
\end{tabular}
\caption*{Pairs 37--40. Real photographs (left), matched AI-generated images (right).}
\end{figure}
\clearpage

\begin{figure}[h!]
\centering
\renewcommand{\arraystretch}{0.6}
\begin{tabular}{cc}
\includegraphics[width=0.42\linewidth, height=0.21\textheight, keepaspectratio]{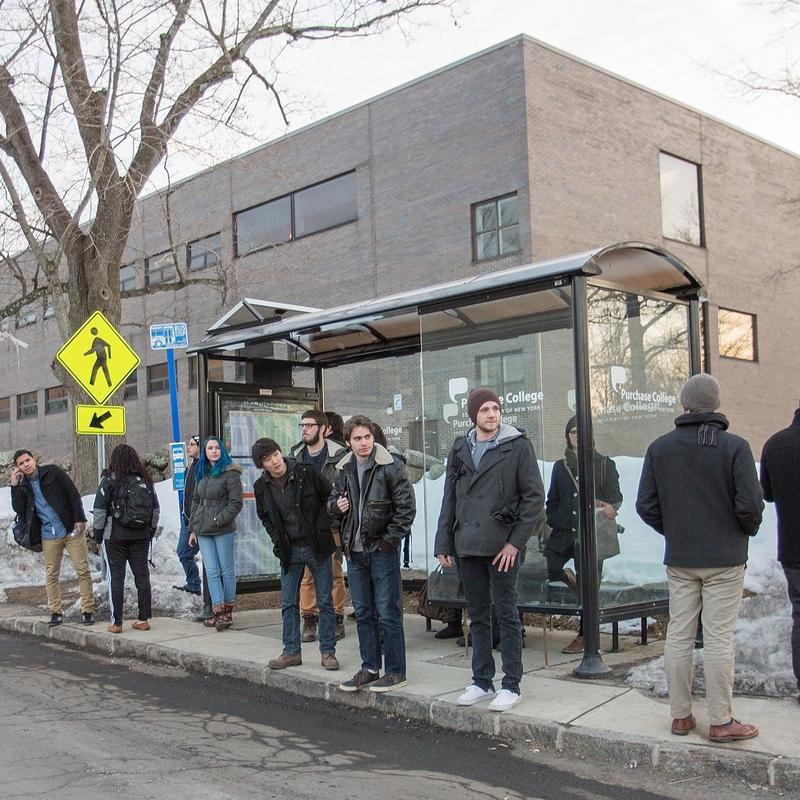} &
\includegraphics[width=0.42\linewidth, height=0.21\textheight, keepaspectratio]{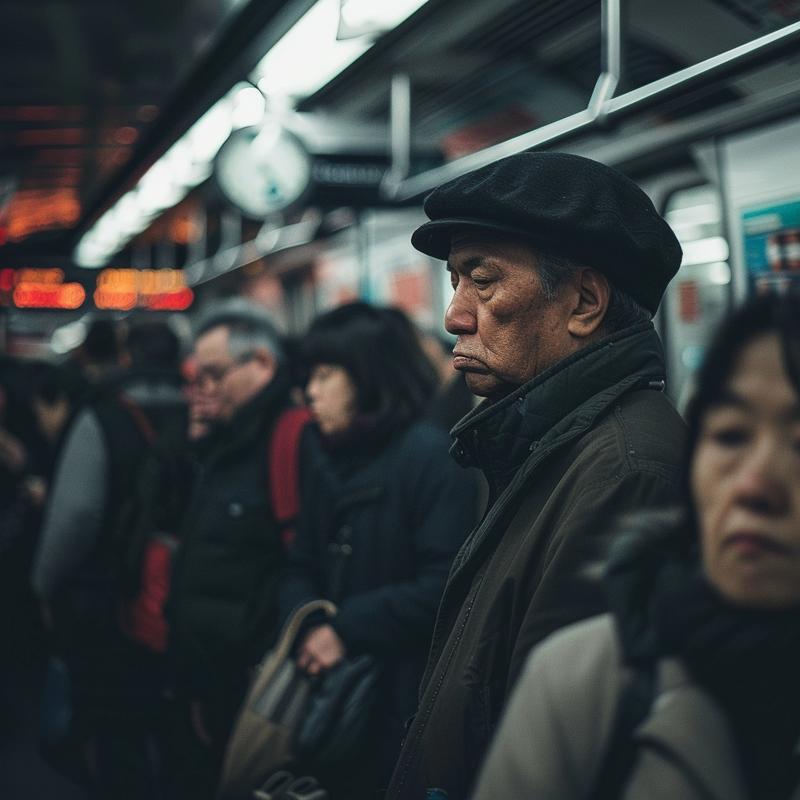} \\
\includegraphics[width=0.42\linewidth, height=0.21\textheight, keepaspectratio]{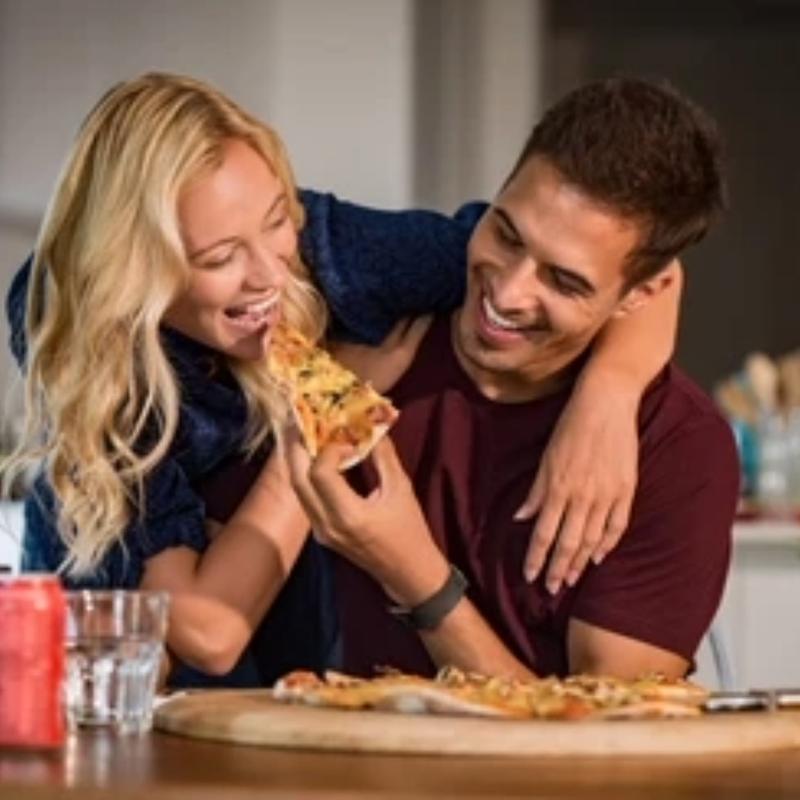} &
\includegraphics[width=0.42\linewidth, height=0.21\textheight, keepaspectratio]{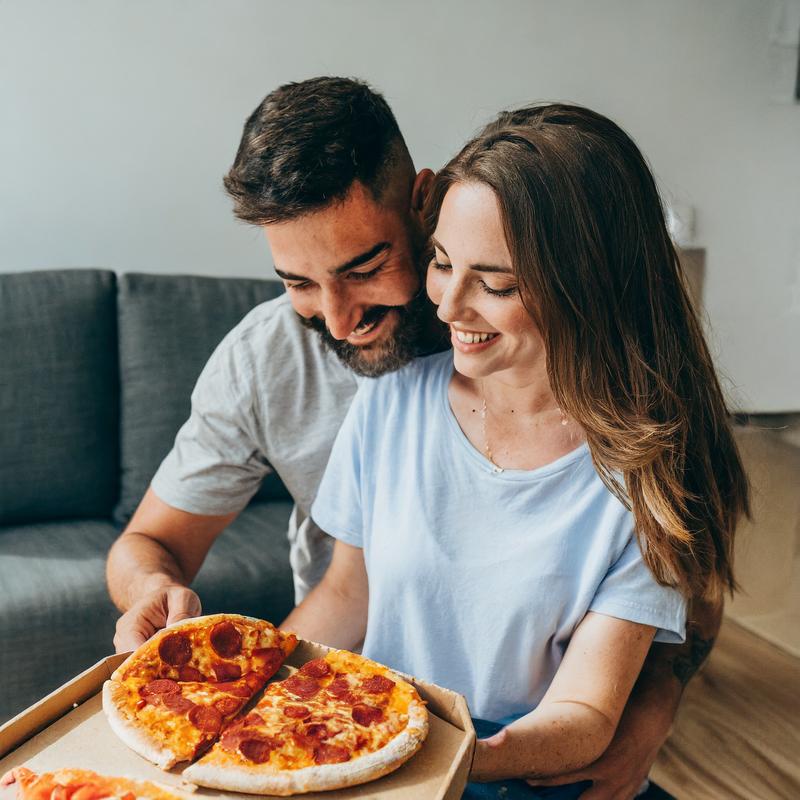} \\
\includegraphics[width=0.42\linewidth, height=0.21\textheight, keepaspectratio]{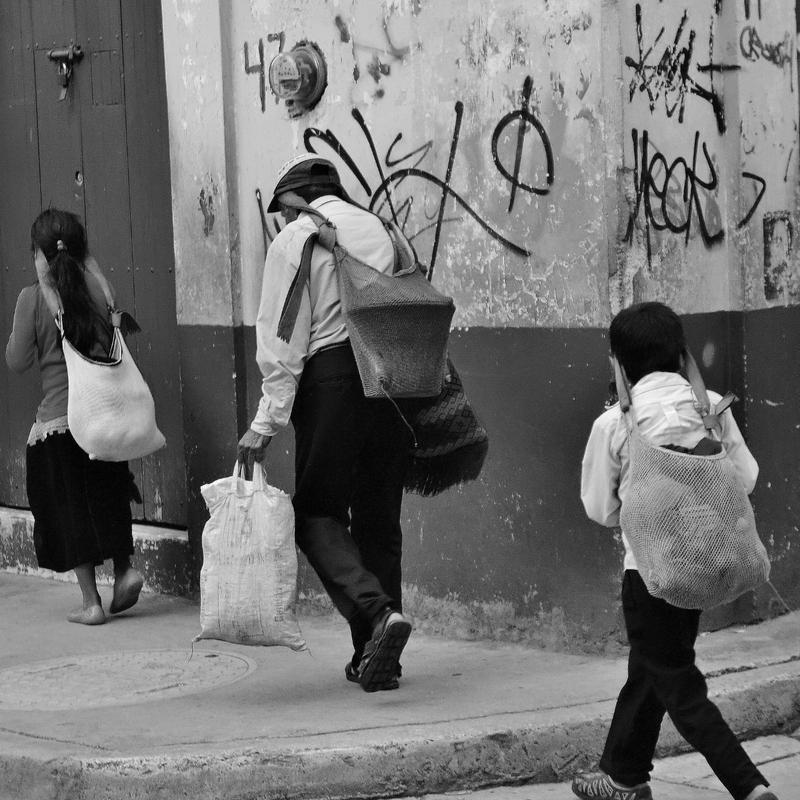} &
\includegraphics[width=0.42\linewidth, height=0.21\textheight, keepaspectratio]{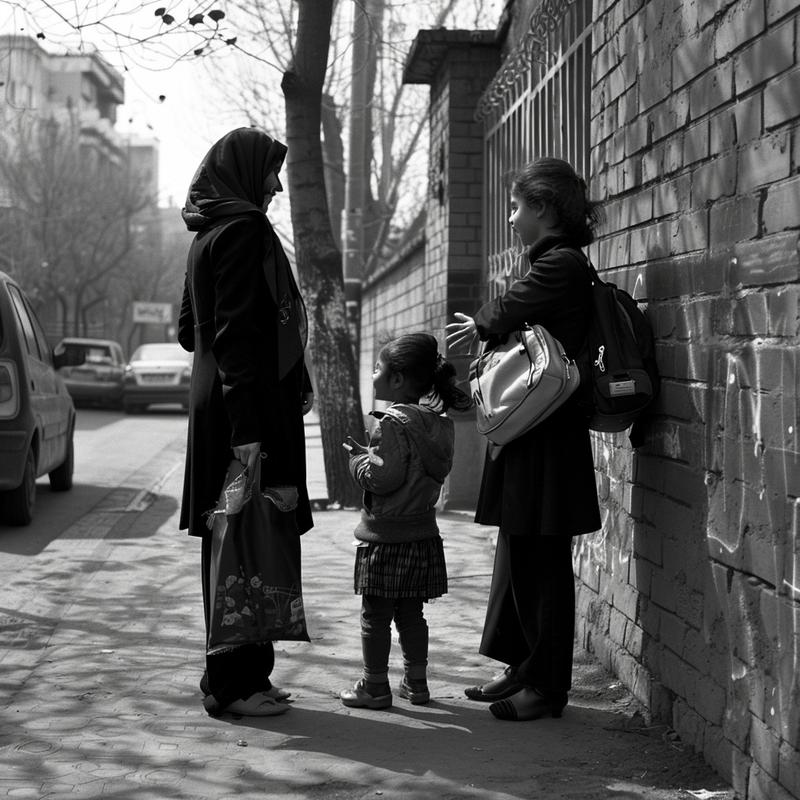} \\
\includegraphics[width=0.42\linewidth, height=0.21\textheight, keepaspectratio]{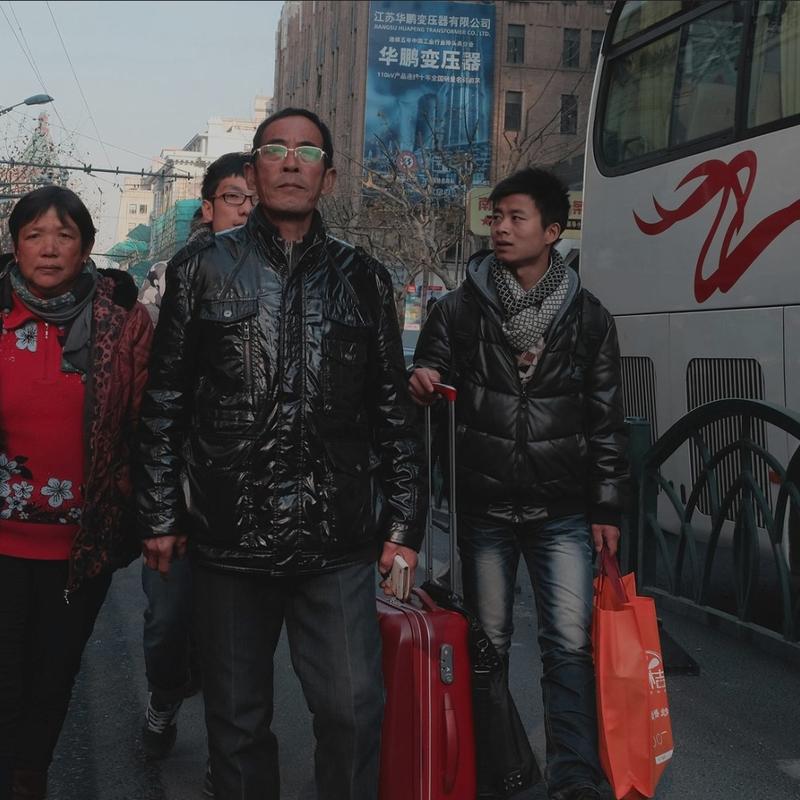} &
\includegraphics[width=0.42\linewidth, height=0.21\textheight, keepaspectratio]{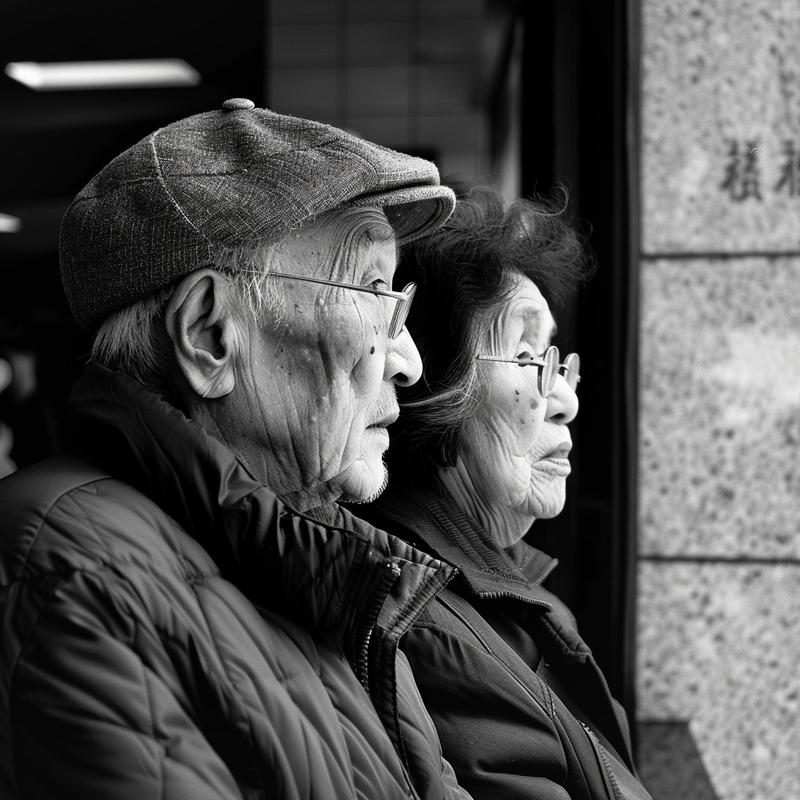} \\
\end{tabular}
\caption*{Pairs 41--44. Real photographs (left), matched AI-generated images (right).}
\end{figure}
\clearpage

\begin{figure}[h!]
\centering
\renewcommand{\arraystretch}{0.6}
\begin{tabular}{cc}
\includegraphics[width=0.42\linewidth, height=0.21\textheight, keepaspectratio]{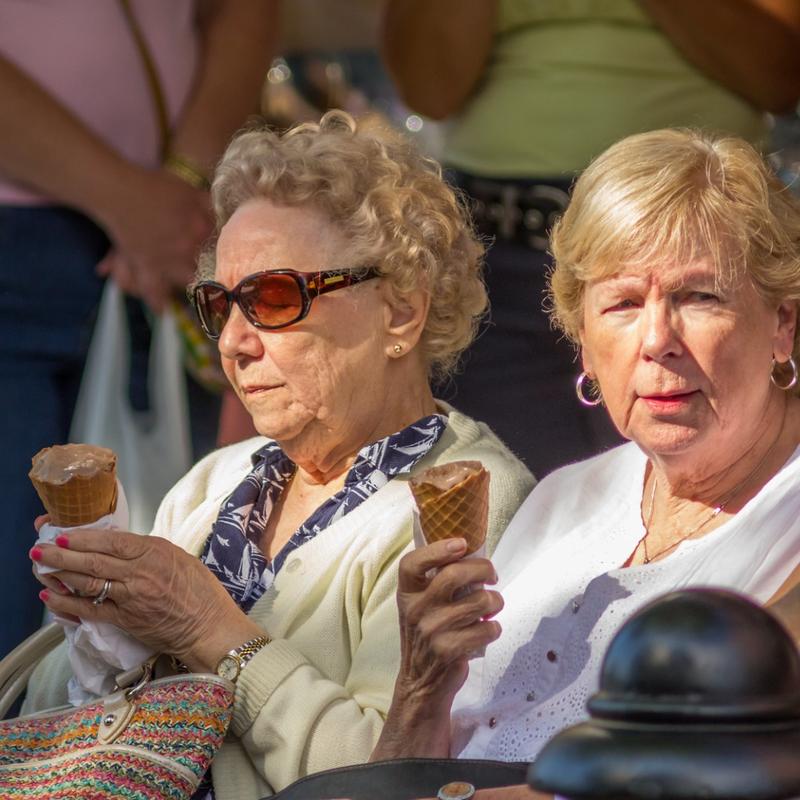} &
\includegraphics[width=0.42\linewidth, height=0.21\textheight, keepaspectratio]{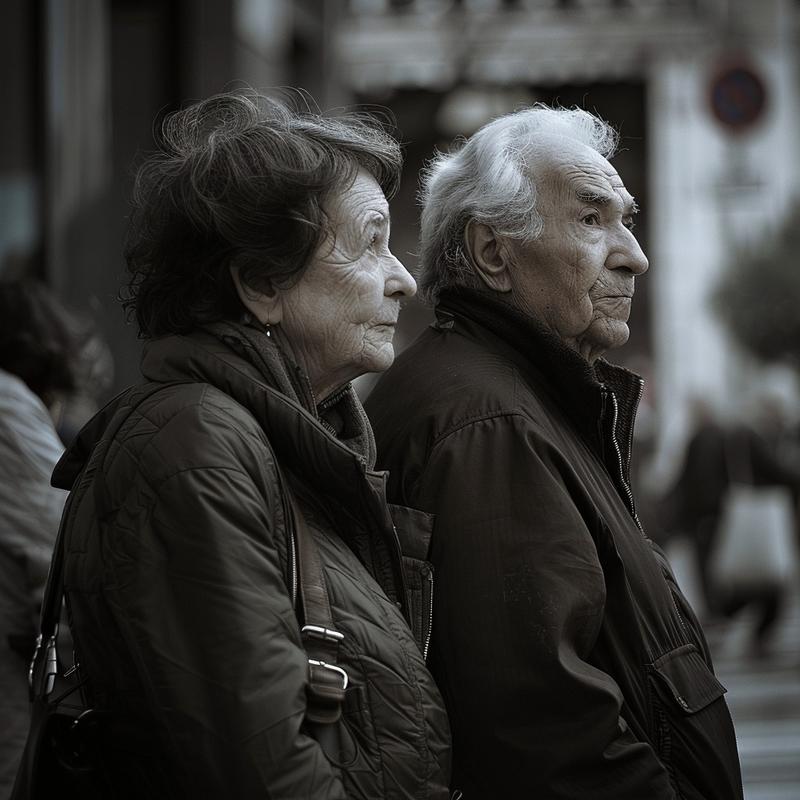} \\
\includegraphics[width=0.42\linewidth, height=0.21\textheight, keepaspectratio]{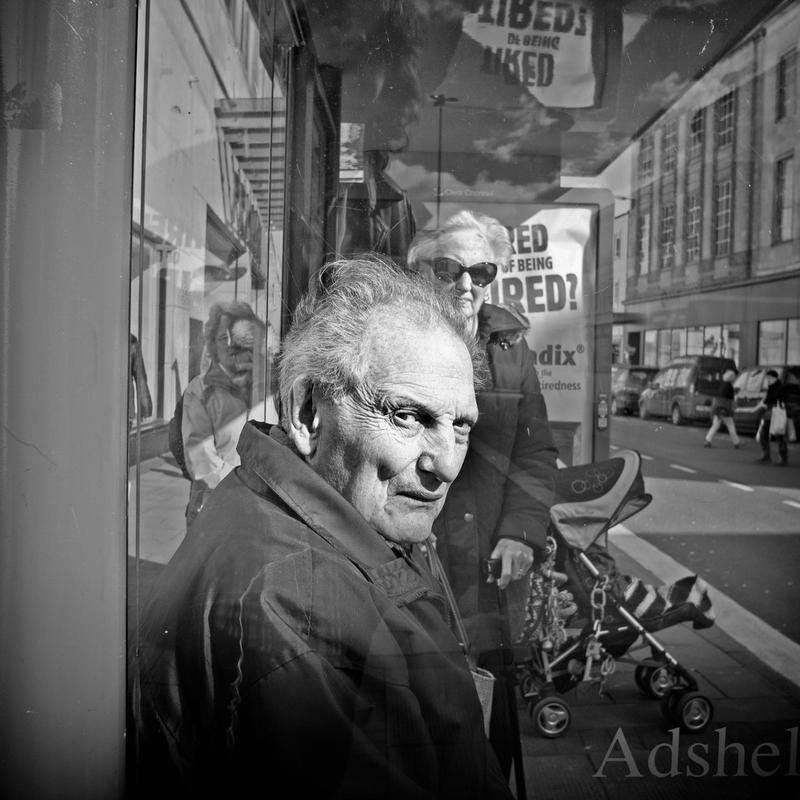} &
\includegraphics[width=0.42\linewidth, height=0.21\textheight, keepaspectratio]{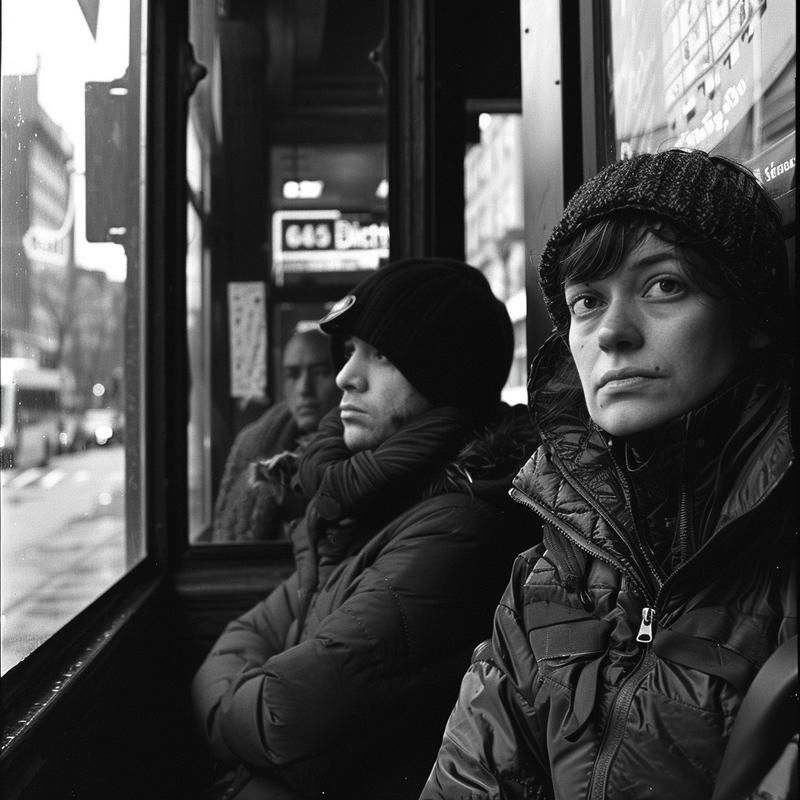} \\
\includegraphics[width=0.42\linewidth, height=0.21\textheight, keepaspectratio]{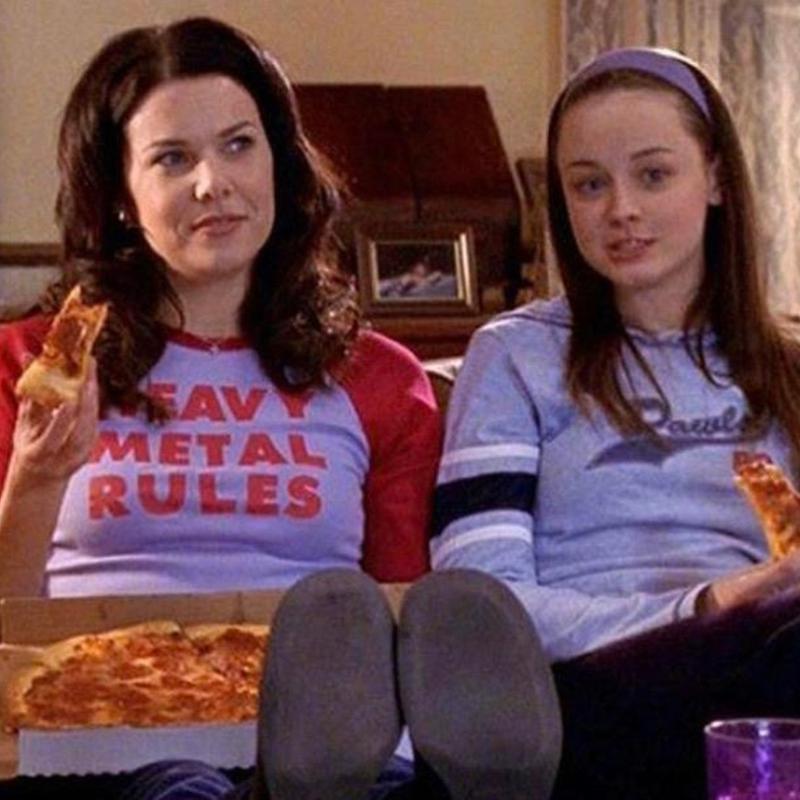} &
\includegraphics[width=0.42\linewidth, height=0.21\textheight, keepaspectratio]{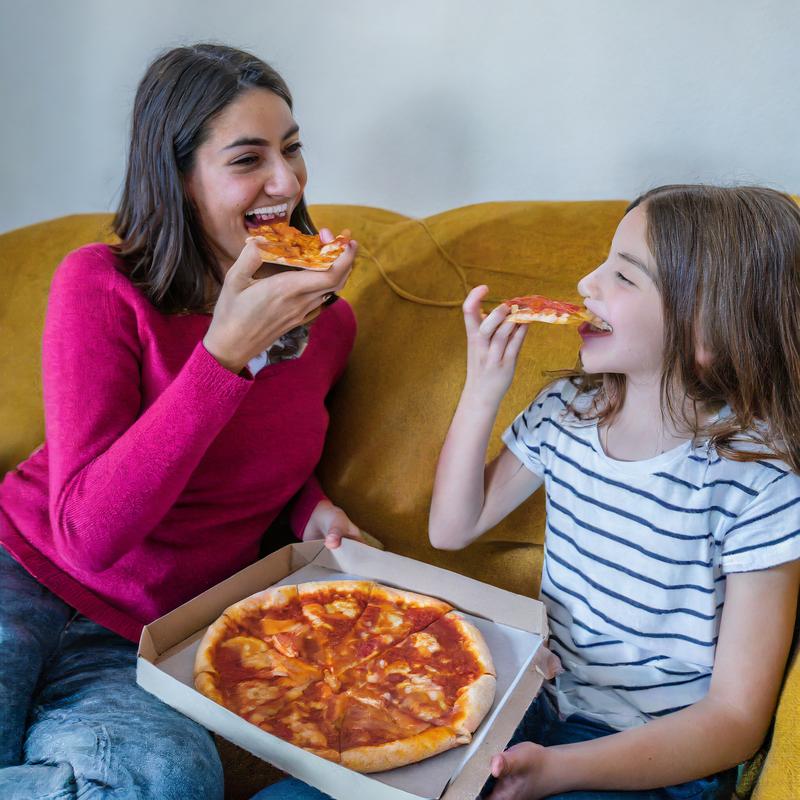} \\
\end{tabular}
\caption*{Pairs 45--47. Real photographs (left), matched AI-generated images (right).}
\end{figure}
\clearpage

\begin{figure}[h!]
\centering
\renewcommand{\arraystretch}{0.6}
\begin{tabular}{ccc}
\includegraphics[width=0.30\linewidth, height=0.22\textheight, keepaspectratio]{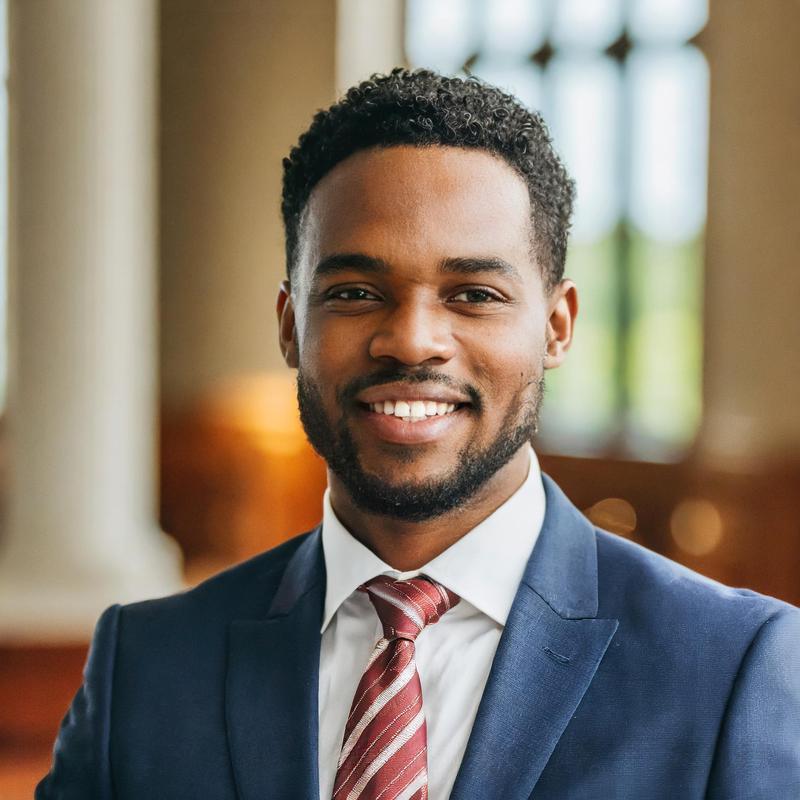} & \includegraphics[width=0.30\linewidth, height=0.22\textheight, keepaspectratio]{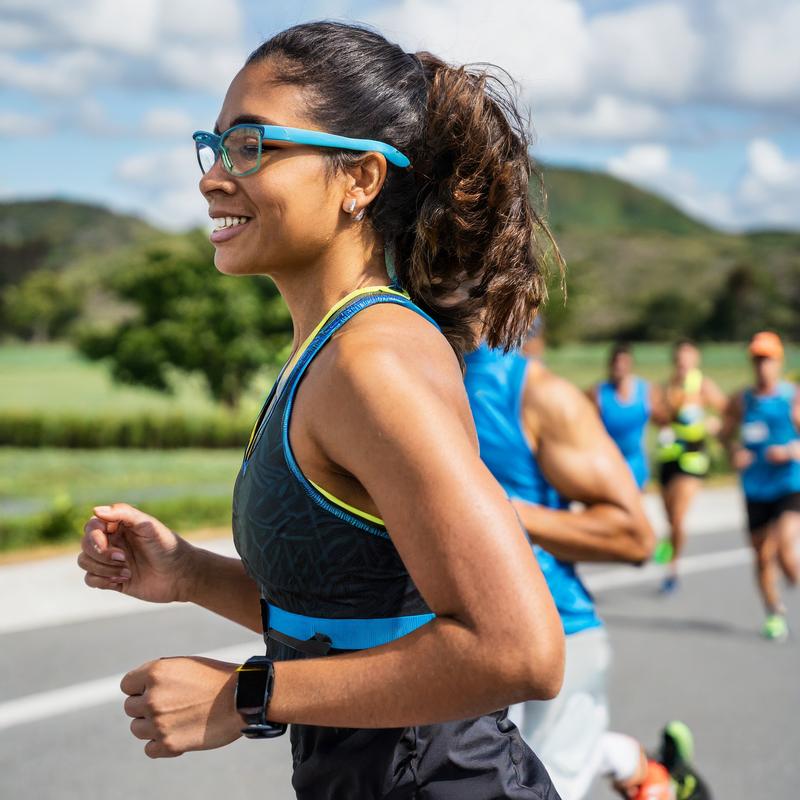} & \includegraphics[width=0.30\linewidth, height=0.22\textheight, keepaspectratio]{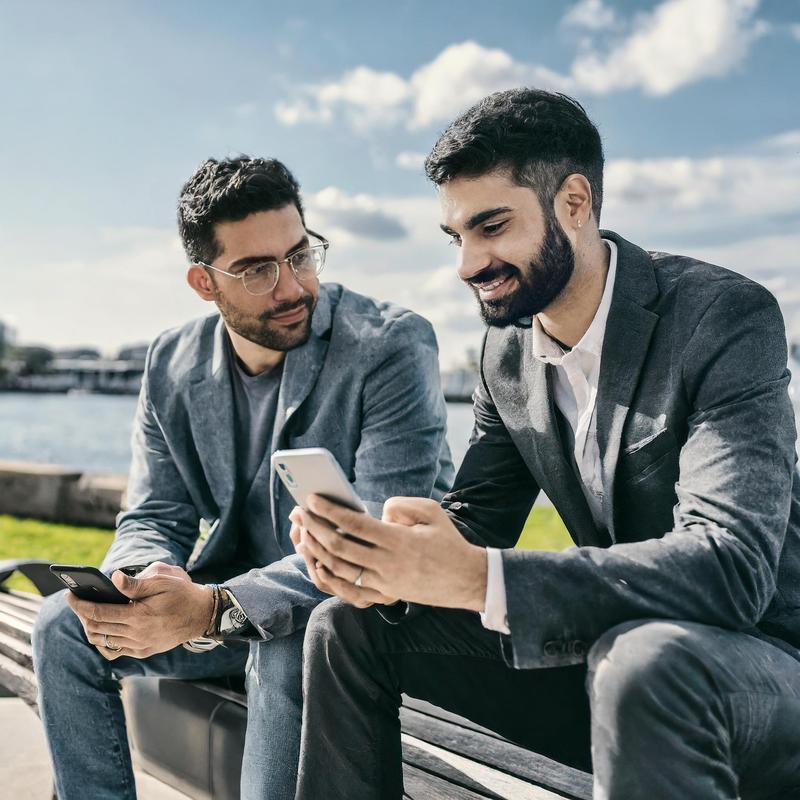} \\
\end{tabular}
\caption*{Additional AI-generated images.}
\end{figure}
\end{document}